\documentclass[aip,jcp,reprint,twocolumn,superscriptaddress]{revtex4-2}

\usepackage{amssymb,amsmath,amsthm,bbm,bbold}
\usepackage{physics}
\usepackage{mathtools,dsfont}
\usepackage{makecell}
\usepackage{graphicx}
\usepackage{hyperref}
\usepackage[arrow, matrix, curve]{xy}
\usepackage{bm}
\newcommand{\rme}{\mathrm{e}}
\usepackage{enumitem}
\usepackage{scalerel,stackengine}
\usepackage[normalem]{ulem}
\usepackage{color}

\newcommand\equalhat{\mathrel{\stackon[1.5pt]{=}{\stretchto{\scalerel*[\widthof{=}]{\wedge}{\rule{1ex}{3ex}}}{0.5ex}}}}

\newcommand{\F}{\mathcal{F}}
\def\ZZ{\mathbbm{Z}}
\def\RR{\mathbbm{R}}
\def\CC{\mathbbm{C}}

\newcommand{\g}{\gamma}
\newcommand{\G}{\Gamma}
\newcommand{\gt}{\tilde{\gamma}}

\newcommand{\Ftce}{\widetilde{\mathcal{F}}_{\CC}^{(e)}}
\newcommand{\Ftcp}{\widetilde{\mathcal{F}}_{\CC}^{(p)}}
\newcommand{\Fce}{\mathcal{F}_{\CC}^{(e)}}
\newcommand{\Fcp}{\mathcal{F}_{\CC}^{(p)}}
\newcommand{\Fre}{\mathcal{F}_{\RR}^{(e)}}
\newcommand{\Frp}{\mathcal{F}_{\RR}^{(p)}}

\usepackage[usenames,dvipsnames]{xcolor}
\hypersetup{colorlinks=true, linkcolor=BrickRed, urlcolor=blue!50!black, citecolor=blue!50!black}

\begin{document}
\title{Refining and relating fundamentals of functional theory}

\author{Julia Liebert}
\affiliation{Department of Physics, Arnold Sommerfeld Center for Theoretical Physics,
Ludwig-Maximilians-Universit\"at M\"unchen, Theresienstrasse 37, 80333 M\" unchen, Germany}
\affiliation{Munich Center for Quantum Science and Technology (MCQST), Schellingstrasse 4, 80799 M\"unchen, Germany}

\author{Adam Yanis Chaou}
\affiliation{Department of Physics, Arnold Sommerfeld Center for Theoretical Physics,
Ludwig-Maximilians-Universit\"at M\"unchen, Theresienstrasse 37, 80333 M\" unchen, Germany}
\affiliation{Munich Center for Quantum Science and Technology (MCQST), Schellingstrasse 4, 80799 M\"unchen, Germany}
\affiliation{Dahlem Center for Complex Quantum Systems and Institut f\"ur Physik,
Freie Universit\"at Berlin, Arnimallee 14, 14195 Berlin, Germany}

\author{Christian Schilling}
\email{c.schilling@lmu.de}
\affiliation{Department of Physics, Arnold Sommerfeld Center for Theoretical Physics,
Ludwig-Maximilians-Universit\"at M\"unchen, Theresienstrasse 37, 80333 M\" unchen, Germany}
\affiliation{Munich Center for Quantum Science and Technology (MCQST), Schellingstrasse 4, 80799 M\"unchen, Germany}

\date{\today}
\pacs{}

\begin{abstract}
To advance the foundation of one-particle reduced density matrix functional theory (1RDMFT) we refine and relate some of its fundamental features and underlying concepts. We define by concise means the scope of a 1RDMFT, identify its possible natural variables and explain how symmetries could be exploited. In particular, for systems with time-reversal symmetry, we explain why there exist six equivalent universal functionals, prove concise relations among them and conclude that the important notion of $v$-representability is relative to the scope and choice of variable. All these fundamental concepts are then comprehensively discussed and illustrated for the Hubbard dimer and its generalization to arbitrary pair interactions $W$. For this, we derive by analytical means the pure and ensemble functionals with respect to both the real- and complex-valued Hilbert space. The comparison of various functionals allows us to solve the underlying $v$-representability problems analytically and the dependence of its solution on the pair interaction is demonstrated. Intriguingly, the gradient of each universal functional is found to always diverge repulsively on the boundary of the domain. In that sense, this key finding emphasizes the universal character of the fermionic exchange force, recently discovered and proven in the context of translationally-invariant one-band lattice models.
\end{abstract}

\maketitle

\section{Introduction\label{sec:intro}}
\begin{figure*}[htb]
\includegraphics[width=1.0\linewidth]{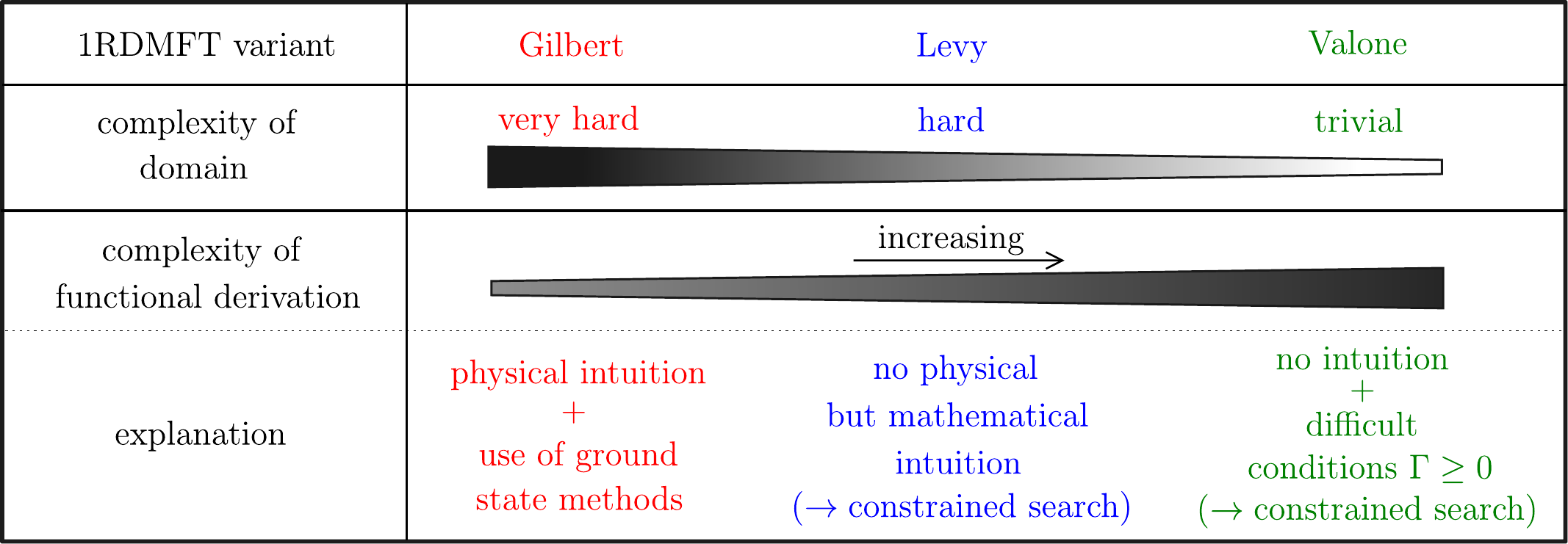}
\caption{Picking one's poison in 1RDMFT (qualitative consideration): By changing the employed variant of 1RDMFT (Gilbert, Levy or Valone), the complexity is interchanged to some degree between the functional's domain and the task of deriving functional approximations (see text for more details). \label{fig:table}}
\end{figure*}

The $v$-representability problem plays a pivotal role in functional theory, especially from a historical point of view: The Hohenberg-Kohn theorem\cite{HK, GD95, Capelle07} proves the existence of a universal functional on the set of exactly those densities which correspond to pure ground states. The same holds true for Gilbert's generalization\cite{Gilbert} to non-local external potentials $v$ and the corresponding domain of one-particle reduced density matrices (1RDMs). Particularly the problem of understanding which 1RDMs are $v$-representable has been perceived as too complex and the relaxation of the functional's domain to the set of $N$-representable 1RDMs was vital for the development of 1RDM-functional theory (1RDMFT)\cite{LE79,V80}. Yet, the relaxation of the domain first to pure $N$-representable\cite{LE79} and then to ensemble $N$-representable 1RDMs\cite{V80} comes at a cost which has been underestimated so far. As it is illustrated in Fig.~\ref{fig:table}, reducing the complexity of the functional's domain in turn increases the difficulty of deriving functional approximations. To be more specific, by resorting to Levy's pure state 1RDMFT one includes in the constrained search formalism unphysical $N$-particle quantum states which never occur in nature as ground states. Hence, resorting to our intuition about ground state physics becomes less effective and fitting approaches need to be extended beyond the solution of ground state problems. Circumventing then according to Valone the resulting highly intricate pure state $N$-representability constraints (generalized Pauli constraints) \cite{KL06, AK08, Kly} necessitates the implementation of non-linear positivity conditions on the $N$-particle ensemble states. In particular, compelling evidence has recently been provided that the complexity of the generalized Pauli constraints is merely shifted from the functional's domain to the universal functional itself\cite{Schilling2018}.
These unpleasant consequences of reducing the domain's complexity and the recent development of machine learning techniques call for a more thorough assessment of the original 1RDMFT approach by Gilbert with an emphasis on the $v$-representability problem and its complexity.

The main goal of this work is to elaborate on the $v$-representability problem and its relation to other fundamental features and concepts in 1RDMFT. Accordingly, we complement all the recent theoretical investigations of 1RDMFT \cite{CP04, C05, RP11, P12, WK15, BCG15, Schilling2018, GUL18, Schilling2019,GWK19, Cioslowski19, C20b,C20a,G20,C153-20, M21,Schilling21, LCLS21,DSKBR22, SYNF22,SG22, GBM22, GBM22-2} and hope that our insights could guide the intense development of novel functionals and their implementations \cite{Kamil2016, BCEG17, SKB17, SB18, MTP18, Mitxelena18, Piris19,Mitxelena20, Piris21-1, Piris21-2, Su21a, YFS21, Kooi22,SBTMPK22, RMGV22, LYDC22, LKO22, AFS22, DSKBR23}.
To achieve this, as a first key achievement, we introduce the so-called \emph{scope} of a functional theory. This novel concept will be vital for our general understanding since it identifies a functional  variable in a concise way. By focussing then on time-reversal symmetric Hamiltonians we make a crucial observation with far-reaching consequences: the notion of $v$-representability is \emph{relative} and depends, in complete analogy to 1RDMFT, on the scope, the variable and the optional reductions of the constrained search to pure and real-valued quantum states. By recalling a well-known geometric interpretation of the Legendre-Fenchel transformation, we establish a fruitful connection between the notion of $v$-representability and the form of the universal functional. This clearly demonstrates that several crucial concepts in 1RDMFT are connected. In order to discuss and illustrate all these fundamental concepts, we then solve by analytical means the Hubbard dimer and its generalization to arbitrary pair-interactions. The former has been widely used in DFT and 1RDMFT to illustrate conceptual aspects and test functionals for larger lattice systems\cite{Pastor2000,Neck2001, Lopez2002, Requist2008, Pastor2011a, Pastor2011b, Fuks2013, Fuks2014, Carrascal2015, Cohen2016, Kamil2016,Deur17, Deur18,Schilling2018, Deur19, DVR21, V-bachelor}. In particular, we show that 1RDMs that are not $v$-representable with respect to real-valued Hamiltonians indeed become $v$-representable if a complex-valued Hilbert space is considered.
The comparison of our work to previous ones\cite{Neck2001,Cohen2016} also demonstrates that the scope of questions in 1RDMFT that allow for analytical and thus fully conclusive answers has been underestimated so far.

The paper is structured as follows. In Sec.~\ref{sec:RDMFTfound} we refine and relate important conceptual aspects of 1RDMFT and in particular provide a comprehensive discussion of $v$-representability. All these fundamental aspects are then illustrated and discussed for the Hubbard dimer in Sec.~\ref{sec:HubbardDimer} and its generalization to arbitrary pair-interaction in Sec.~\ref{sec:generic}.

\section{Foundational aspects of 1RDMFT}\label{sec:RDMFTfound}

In this section, we introduce in detail the conceptual aspects of 1RDMFT required for the analytic study of the Hubbard dimer model in Sec.~\ref{sec:HubbardDimer} and its generalization in Sec.~\ref{sec:generic}. On the one hand, this means to recall well-known concepts and on the other hand to refine them and to introduce new ones. A prime example for the latter will be the definition of the \emph{scope} of a functional theory and a rigorous argument which identifies the related natural variables.

\subsection{Pure and ensemble universal functionals \label{sec:intro1RDMFT}}

In order to keep our work self-contained, we first recap Levy's\cite{LE79} pure and Valone's\cite{V80} ensemble 1RDMFT and introduce some notation that is used throughout the paper. The $N$-fermion Hilbert space $\mathcal{H}_N \equiv \wedge^N\mathcal{H}_1$ has (complex) dimension $D=\left(\begin{smallmatrix}d\\N\end{smallmatrix}\right)$, where $\mathcal{H}_1$ is the underlying $d$-dimensional complex one-fermion Hilbert space. The set of all $N$-fermion density operators $\Gamma$ on $\mathcal{H}_N$ is denoted by $\mathcal{E}^N$. By definition, $\Gamma\in \mathcal{E}^N$ is self-adjoint, positive semidefinite and $\Tr_N\small[\Gamma] = 1$. The boundary points of the compact and convex set $\mathcal{E}^N$ are given by all those density operators which are not strictly positive, i.e., at least one of their eigenvalues vanishes.
Moreover, the extremal elements of $\mathcal{E}^N$ are given by the idempotent density operators $\Gamma^2 = \Gamma$ which constitute the set $\mathcal{P}^N$ of all pure $N$-fermion density operators. Then, according to the Krein-Milman theorem\cite{Krein1940},
$\mathcal{E}^N$ is the convex hull of its extremal elements. The sets of corresponding one-particle reduced density operators $\gamma$ are obtained by tracing out $N-1$ fermions of the elements $\Gamma$ of the respective sets of $N$-fermion density operators,
\begin{eqnarray}
\mathcal{P}^1_N &\equiv & N\Tr_{N-1}[\mathcal{P}^N] \label{eq:P1N}\\
\mathcal{E}^1_N &\equiv& N\Tr_{N-1}[\mathcal{E}^N]\,.\label{eq:E1N}
\end{eqnarray}
We refer to a 1RDM in $\mathcal{P}^1_N $ as being \emph{pure $N$-representable} and to those in $\mathcal{E}^1_N $ as being \emph{ensemble $N$-representable}.

As a first scientific accomplishment, we provide in the following a concise motivation and derivation of 1RDMFT. For this, we consider a fixed pair-interaction $W$ and introduce the corresponding (affine) class of total Hamiltonians of the form
\begin{equation}\label{eq:Hgeneral}
H(h) \equiv h + W\,,
\end{equation}
which are parameterized by the one-particle Hamiltonian $h$. The latter takes in 1st quantization the form
\begin{equation}\label{eq:hvsh1}
h \equiv h(h_1) \equiv h_1 \otimes \mathds{1}^{\otimes {N-1}}+ \ldots + \mathds{1}^{\otimes {N-1}} \otimes h_1
\end{equation}
with some suitable $h_1$ acting on $\mathcal{H}_1$.
As a novel scientific concept, we interpret the class \eqref{eq:Hgeneral} as the \emph{scope} of the resulting functional theory. This scope could be reduced by restricting $h$ to a subspace, e.g., by considering only operators $h$ that exhibit certain additional symmetries.

If we do not restrict $h$, the full 1RDM $\gamma$ represents the conjugate variable in a natural and mathematical concise sense: In virtue of the Riesz representation theorem applied to the linear map
\begin{eqnarray}\label{eq:Riesz}
h_1 \mapsto \Tr_N[h(h_1) \G] &\equiv & \langle h(h_1),\G\rangle_{\mathrm{HS},N} \nonumber \\
&=& \Tr_1[h_1 \g] \equiv \langle h_1,\g\rangle_{\mathrm{HS},1}\,,
\end{eqnarray}
the 1RDM $\g$ follows as the \emph{unique} `Riesz vector' for which the equality  $\Tr_N[h(h_1) \G]=\langle h_1,\g\rangle_{\mathrm{HS},1}$ holds for all $h_1$. Here, we introduced the Hilbert-Schmidt inner product $\langle A,B\rangle_{\mathrm{HS},m}\equiv \Tr_m[A^\dagger B]$ on the space of linear operators acting on $\mathcal{H}_1^{\otimes m}$.
An appealing aspect of our novel and mathematically concise reasoning is that it identifies the simplest reduced state of $\G$ which is still sufficient for calculating the expectation values of any one-particle Hamiltonian $h$. It is also worth noticing, that any restriction of the vector space of all one-particle Hamiltonians $h$ to a subspace would directly yield via \eqref{eq:Riesz} a new and simpler conjugate variable with fewer degrees of freedom than the full 1RDM. For instance, if we restricted to $h\equiv h(v) \equiv t+v $ for some fixed kinetic energy operator $t$ and variable external potential $v$, this general reasoning would yield the particle density as conjugate variable. Moreover,  the reduced state/conjugate variable identified via the Riesz representation theorem has the number of degrees of freedom as the variable $h$, reduced by one because of the normalization of quantum states.

After having chosen an interaction $W$ in \eqref{eq:Hgeneral} (e.g., the Coulomb pair-interaction), the ground state 1RDM and the ground state energy follow from the Levy-Lieb constrained search \cite{LE79, V80, L83},
\begin{eqnarray}\label{eq:Levy}
E(h) &=& \min_{\Gamma}\,\mathrm{Tr}_N\left[(h + W)\Gamma\right]\nonumber\\\
&=& \min_{\gamma}\,\left[\mathrm{Tr}_1[h_1\gamma] + \min_{\Gamma\mapsto\gamma}\mathrm{Tr}_N[W\Gamma]\right]\nonumber\\
&\equiv &  \min_{\gamma}\,\left[\mathrm{Tr}_1[h_1\gamma] + \mathcal{F}(\gamma)\right]\,.
\end{eqnarray}
This in turn defines the universal functional $\mathcal{F}(\gamma)$ and thus establishes a 1RDMFT. The minimizations in Eq.~\eqref{eq:Levy} can either refer to all states $\Gamma\in \mathcal{E}^N$ or just the pure states $\Gamma\in \mathcal{P}^N$. This immediately leads to the distinction of the universal pure/Levy functional $\mathcal{F}^{(p)}$,
\begin{equation}
\mathcal{F}^{(p)}(\gamma) = \min_{\mathcal{P}^N\ni\Gamma\mapsto\gamma}\Tr_{N}[W\Gamma]\,,
\end{equation}
with (non-convex) domain $\mathcal{P}^1_N$, and the universal ensemble/Valone functional $\mathcal{F}^{(e)}$
on the convex domain $\mathcal{E}^1_N$. Intriguingly, $\mathcal{F}^{(e)}$ and $\mathcal{F}^{(p)}$ are related through\cite{Schilling2018}
\begin{equation}\label{eq:FpconvFe}
\mathcal{F}^{(e)} = \mathrm{conv}\left(\mathcal{F}^{(p)}\right)\,,
\end{equation}
where $\mathrm{conv}(\cdot)$ denotes the lower convex envelope. As it has been outlined in the introduction, each of the two 1RDMFT variants has relative advantages and disadvantages. In the development of functional approximations, it is a matter of preference whether one would like shift a part of the complexity of the ground state problem from the universal functional into the functional's domain or not.

\subsection{Optional reductions: Real ($\RR$) versus complex ($\CC$) \label{sec:real-vs-complex}}

In this section, we present another key result of our work. First, we recall that time-reversal symmetric systems could be described by real-valued quantum states. We then explain that this symmetry effectively simulates a binary degree of freedom, which in turn introduces in pure state 1RDMFT a certain degree of mixedness through the constrained search formalism. As a consequence, the choice of a natural variable is not unique and the same holds true for the definition of the universal functional.

Quantum systems with time-reversal symmetry are of central importance in physics and chemistry. Most common applications of 1RDMFT so far
even consider Hamiltonians $H$ which exhibit a \emph{conventional} time-reversal symmetry $T$\footnote{Although most applications of 1RDMFT in quantum chemistry so far restrict to time-reversal symmetric Hamiltonians, there are quite a few relevant systems which break that symmetry. The latter include the systems with external magnetic fields and velocity dependent forces in general\cite{TCPH22}, and chiral cavities\cite{Huebner21}.}, i.e., $[H,T]=0$ and $T^2 =1$. For the class of all such Hamiltonians on a given Hilbert space $\mathcal{H}$, one can construct a (non-unique) basis $\mathcal{B}$ of time-reversal invariant orthonormal states with respect to which every $H$ takes the form of a real-valued matrix \cite{HGK19}. Consequently, the energy minimization in the Ritz variational principle  can then be restricted to pure or ensemble density matrices which are real-valued with respect to $\mathcal{B}$.
If we do not like to restrict to real-valued quantum states, we can interpret any pure state $\ket{\Psi}\in \mathcal{H} \cong \CC^D$ as a spinor-like object of the form
\begin{equation}\label{eq:spinor}
\ket{\Psi}\equalhat\begin{pmatrix}
a\ket{\Psi_r}\\
b\ket{\Psi_i}
\end{pmatrix}\,
\end{equation}
with $a^2+b^2=1$, $a,b\in\RR$. The latter condition denotes the difference to the spin-$1/2$ degree of freedom, where in general $a, b\in\CC$.
In Eq.~\eqref{eq:spinor}, $\ket{\Psi_r}$ and $\ket{\Psi_i}$ denote the real and imaginary part defined with respect to the reference basis $\mathcal{B}$. The identification  \eqref{eq:spinor} is nothing else than a $\mathcal{B}$-induced isomorphism for $\CC^D \cong \RR^D  \otimes \RR^2$. Accordingly, we can define the linear map $\Tr_{\RR^2}[\cdot]$ that traces out the degree of freedom which corresponds to the complex-valuedness of the state $\ket{\Psi}$, i.e.~$\Tr_{\RR^2}[\ket{\Psi}\!\bra{\Psi}]=a^2\ket{\Psi_r}\!\bra{\Psi_r}+b^2\ket{\Psi_i}\!\bra{\Psi_i}$. This finally leads to the observation that pure states in $\CC^D$ can be described by real density matrices on $\RR^D$ of rank less or equal to 2.
\begin{figure*}[htb]
\includegraphics[width=1.0\linewidth]{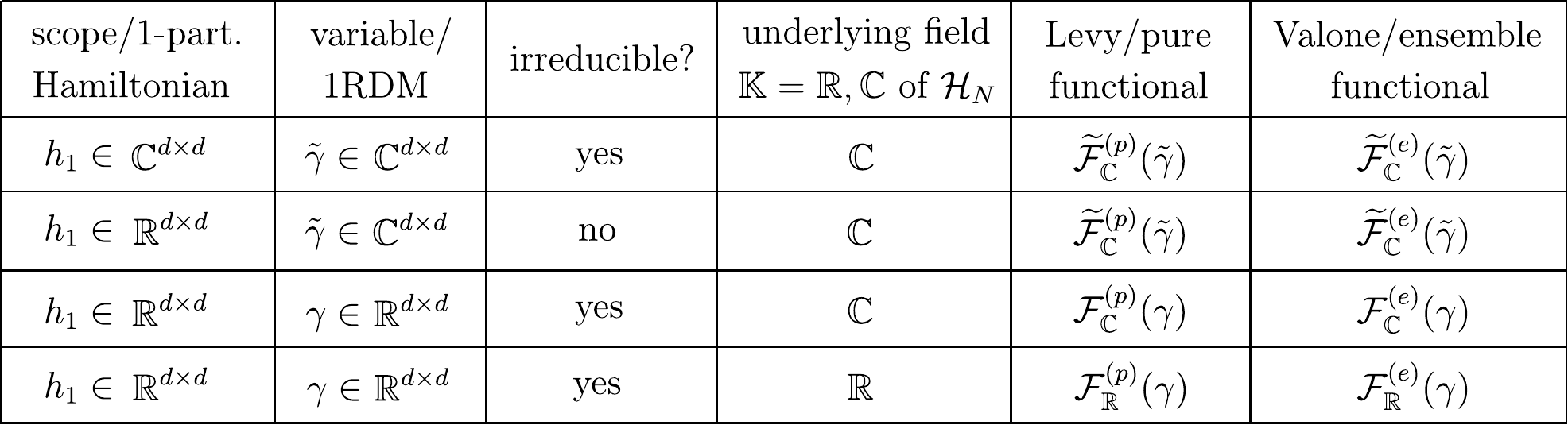}
\caption{Overview of some of our conceptual key results: Optional choices can be made in case of real-valued Hamiltonians (conventional time-reversal symmetry): First, one may follow the paradigm of irreducibility and reduce the functional's variable from the full complex-valued 1RDM $\gt \in \CC^{d \times d}$ to $\g \equiv \Re (\gt) \in \RR^{d \times d}$ and second one may then restrict in addition the constrained search to real-valued $N$-particle quantum states. The relation among the six possible functionals is explained in Fig.~\ref{fig:functionals-general}. \label{fig:choices}}
\end{figure*}

In our case of $N$-fermion quantum systems with conventional time-reversal symmetry, we may apply the above reasoning to both the one-particle Hamiltonian $h_1$ (here as an operator on the one-particle Hilbert space $\mathcal{H}_1$) and to the Hamiltonian and its individual parts acting on the $N$-fermion Hilbert space $\mathcal{H}_N \equiv \wedge^N\mathcal{H}_1$. In practise, these two applications can be made compatible, and in particular the former one implies the latter: For the class of all one-particle Hamiltonians $h_1$ on $\mathcal{H}_1$ we introduce an orthonormal  reference basis $\mathcal{B}_1 =\{\ket{\varphi_j}\}_{j=1}^d$ with respect to which all $h_1$ take the form of real-valued matrices. The basis $\mathcal{B}_1$ then induces the orthonormal reference basis $\mathcal{B}_N$ of Slater determinants with respect to which the total Hamiltonian $H$ and its parts $h\equiv h(h_1)$, $W$ are real-valued.  Actually, the latter would also be true for any basis $\mathcal{B}'_N$  whose elements are real-valued linear combinations of Slater determinants (e.g., spin-configuration states).

The consequences of restricting 1RDMFT to Hamiltonians $H(h)$ with conventional time-reversal symmetry are then twofold. First, as it has been explained in Sec.~\ref{sec:intro1RDMFT}, this restriction of $h_1$ and $h$, respectively, allows one to reduce the natural variable from the full 1RDM $\gt \in \CC^{d\times d}$ to its real part which we denote in the following by
\begin{equation}\label{eq:realgamma}
\g \equiv \Re(\gt) \in \RR^{d \times d}\quad ,\,\mbox{w.r.t.}~\mathcal{B}_1\,,
\end{equation}
where $\mathcal{B}_1$ denotes a suitable reference basis for $\mathcal{H}_1$, as described above. It is worth recalling here that the eigenvalues of the 1RDMs corresponding to non-degenerate ground states of time-reversally symmetric Hamiltonians are pairwise degenerate\cite{Smith66}. This mathematical implication thus played an important role in the analysis of the (quasi)pinning effect\cite{CM14,CS15}.
Second, if one follows the paradigm of irreducibility by resorting to $\g$ as the natural variable, one can in addition restrict the search space in the constrained search \eqref{eq:Levy} to density matrices which are real-valued with respect to $\mathcal{B}_N$.
These two reductions of 1RDMFT are optional and by realizing them or not, and by employing either the Levy/pure or Valone/ensemble variant one could choose among six possible universal functionals. We list all of them together with their defining characteristics in Fig.~\ref{fig:choices}. There, the first column explains whether the class of one-particle Hamiltonians respects conventional time-reversal symmetry or not, the second one indicates the potential reduction of the variable according to Eq.~\eqref{eq:realgamma} and the fourth one the optional choice of restricting the constrained search to real-valued states. The convention of the six universal functionals is as follows. If the functional depends on the full, complex-valued 1RDM $\gt$, we add a tilde ($\widetilde{\mathcal{F}}$) and otherwise not. The reference in the constrained search \eqref{eq:Levy} to real- or complex-valued density matrices is indicated by the index $\RR/\CC$, and the one to pure or ensemble states
by the superscript $(p/e)$.


We conclude this section, by presenting another key result of our work. To be more specific, we discover and explain that all six universal functionals are related to each other.
According to the Levy-Lieb constrained search \eqref{eq:Levy}, $\mathcal{F}^{(p/e)}_{\CC}$ is related to $\widetilde{\mathcal{F}}^{(p/e)}_{\CC}$ through (recall Eq.~\eqref{eq:realgamma})
\begin{equation}\label{eq:Levy_FC}
\mathcal{F}^{(p/e)}_{\CC}(\g) \equiv \min_{\Im(\gt)}\widetilde{\mathcal{F}}^{(p/e)}_{\CC}(\gt)\,.
\end{equation}
Hence, $\mathcal{F}^{(p/e)}_{\CC}$ can be obtained directly from $\widetilde{\mathcal{F}}^{(p/e)}_{\CC}$ through a minimization with respect to the imaginary part of $\gt$.
Moreover, the relation
\begin{equation}
\mathcal{F}_{\CC}^{(p)}(\g)\leq \mathcal{F}_{\RR}^{(p)}(\g)
\end{equation}
follows immediately from \eqref{eq:Levy} as well.
\begin{figure}[htb]
\frame{\includegraphics[width=1.0\linewidth]{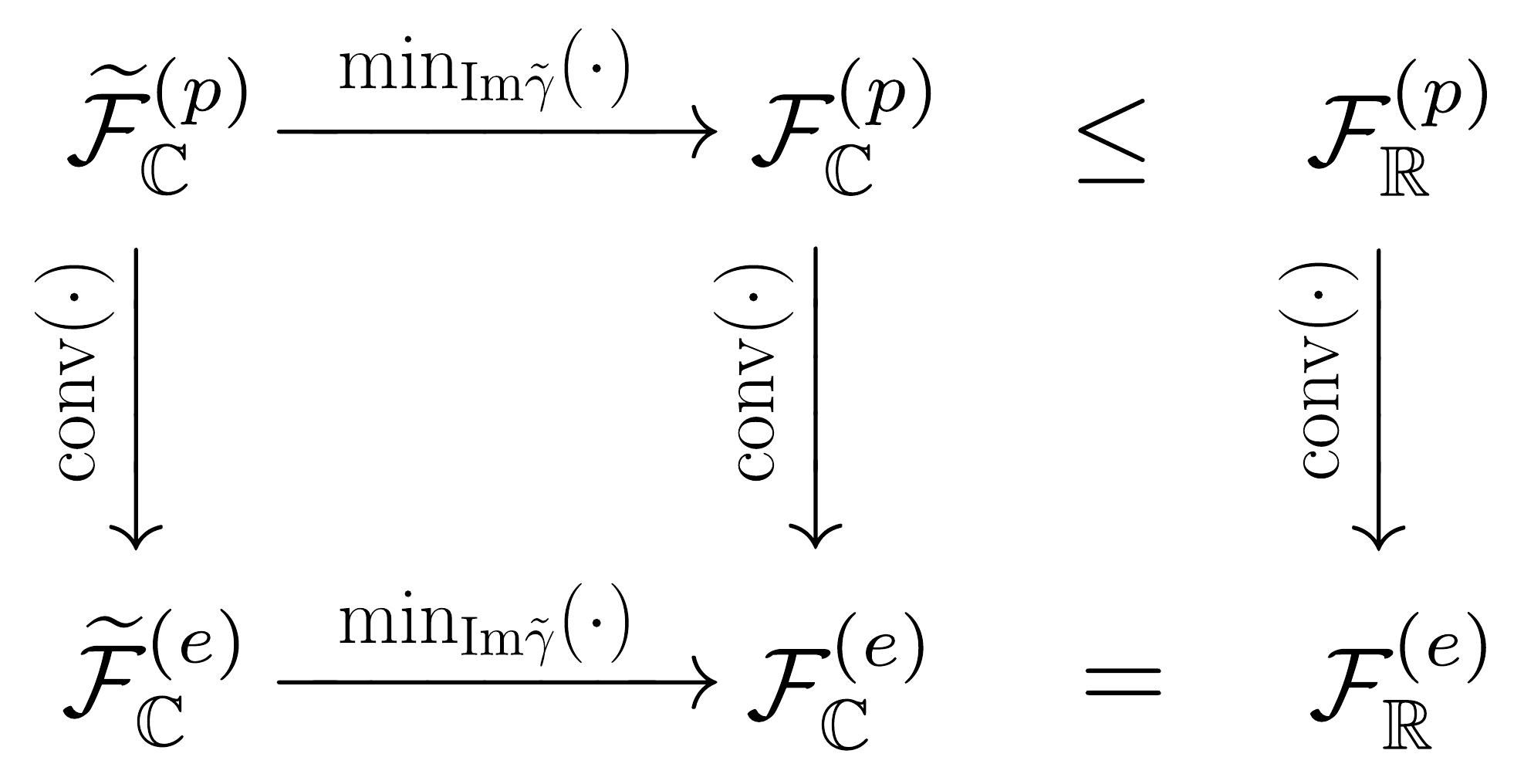}}
\caption{Illustration of the relations between the six universal functionals introduced in the text and listed in Fig.~\ref{fig:choices}. They hold for any system of arbitrary size which exhibits conventional time-reversal symmetry (see text for more details).  \label{fig:functionals-general}}
\end{figure}
Last but not least, the three functionals $\Ftcp, \Fcp, \Frp$ are related to their Valone/ensemble-partner functionals $\Ftce, \Fce, \Fre$ through the lower convex envelop\cite{Schilling2018}. This in turn implies that
\begin{equation}
\Fce(\g) = \Fre(\g).
\end{equation}
Various relations among the six functionals are illustrated in Fig.~\ref{fig:functionals-general}.
In Secs.~\ref{sec:HubbardDimer} and \ref{sec:generic} we will derive these functionals for the specific Hubbard dimer with on-site and generic interactions, respectively.

\subsection{General discussion of the $v$-representability problem\label{sec:v-rep-general}}

\subsubsection{Variants of $v$-representability problem}\label{sec:v-rep-var}
The discussion in the previous section \ref{sec:real-vs-complex} implies that also the concept of $v$-representability is relative: It refers to a pre-defined scope of a functional theory and the choice of a corresponding variable. To explain this absolutely vital aspect of our work, we consider the sequence,
\begin{equation}\label{eq:seq}
H \mapsto \ket{\Psi} \mapsto \g\,,
\end{equation}
where $H$ is some Hamiltonian on a fixed Hilbert space $\mathcal{H}$, $\ket{\Psi}$ its ground state and $\g$ --- in the most general context --- just some reduced information of $\ket{\Psi}$ which is obtained by applying a fixed linear map to $\ket{\Psi}\!\bra{\Psi}$. Obviously, the set of $\g$ that one can reach by varying $H$ in \eqref{eq:seq} over a certain subset $S$ of all Hermitian operators on $\mathcal{H}$
depends on the choice of $S$. In a similarly obvious fashion this sought-after set depends on the precise definition of $\g$, e.g., whether the latter is the 1RDM or particle density (in case of a system of $N$-identical particles) or more generally just the expectation values of a collection of distinctive observables. A comment is in order here concerning Hamiltonians in $S$ with degenerate ground states. For them one would just extend the notion \eqref{eq:seq} by considering all sequences \eqref{eq:seq} involving any possible ground states $\ket{\Psi}$ in the degenerate subspace. This resembles also the important fact that the original theorems due to Hohenberg-Kohn \cite{HK, Kohn83} in DFT and Gilbert \cite{Gilbert} in 1RDMFT  extend straightforwardly to degenerate pure ground states\cite{GD95, Capelle07}.

\begin{figure}[htb]
\frame{\includegraphics[width=0.8\linewidth]{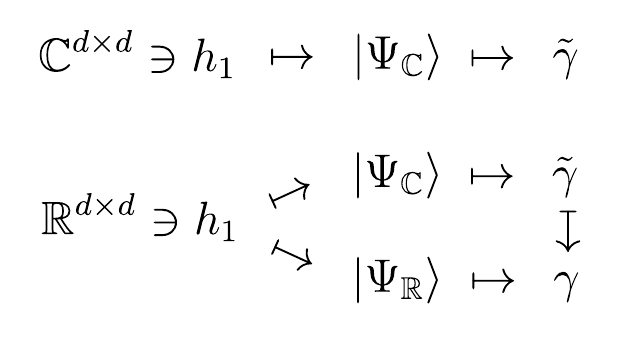}}
\caption{Our novel and more systematic perspective on 1RDMFT reveals that $v$-representability is a \emph{relative} concept: it refers to the \emph{scope} (set of variables $h_1$) and the choice of the corresponding \emph{conjugate variable}. For time-reversal invariant systems this yields in the context of 1RDMFT four variants, $\CC^{d\times d}\!\ni \!h_1 \mapsto \gt$, $\RR^{d\times d}\!\ni\! h_1\mapsto \gt$ and $\RR^{d\times d}\! \ni\! h_1\mapsto \g$, where the latter may involve complex- ($\ket{\Psi_{\CC}}$) or real-valued ($\ket{\Psi_{\RR}}$) $N$-fermion ground states.  \label{fig:v-rep}}
\end{figure}

As another key result of this work, the application of these general considerations to 1RDMFT for Hamiltonians \eqref{eq:Hgeneral} with a (conventional) time-reversal symmetry leads to four different meaningful notions of \emph{pure state $v$-representability}. As it is illustrated in Fig.~\ref{fig:v-rep}, the chosen scope might be either the complex- or real-valued one-particle Hamiltonians $h$. In the case of the latter, one may restrict the 1RDM $\gt$ to its  real-part $\g$ according to \eqref{eq:realgamma} and then optionally consider only real-valued $N$-fermion states in the constrained search formalism \eqref{eq:Levy}. This again demonstrates that the notion of pure state $v$-representability is a relative concept.
Moreover, in analogy to DFT (see, e.g., Refs.~\onlinecite{Helgaker22, Penz22}) one may even allow for \emph{mixed }ground states in \eqref{eq:seq} for degenerate Hamiltonians.  In turn, this yields in the same fashion as for pure state $v$-representability four notions of \emph{ensemble state $v$-representability}. In principle, the resulting eight sets could be denoted by $\widetilde{\mathcal{V}}_{p/e}^{\CC/\RR}$, $\mathcal{V}_{p/e}^{\CC/\RR}$, where the tilde indicates that the set refers to $\gt$ rather than $\g$. By definition, 1RDMs $\gt\in \widetilde{\mathcal{V}}_{p/e}^{\CC/\RR}$, and analogously for $\gamma\in \mathcal{V}_{p/e}^{\CC/\RR}$, are referred to as being real/complex pure/ensemble $v$-representable. 
In Secs.~\ref{sec:vrepr-dimer} and \ref{sec:vrepr-gen}, however, we will simplify those symbols to just $\mathcal{V}_{p/e}$ since it will be always clear from the context to which of the eight sets we are actually referring to.

\subsubsection{Relation between $v$-representability and universal functional}\label{sec:v-rep-f}
\begin{figure}[htb]
\centering
\includegraphics[width=0.85\linewidth]{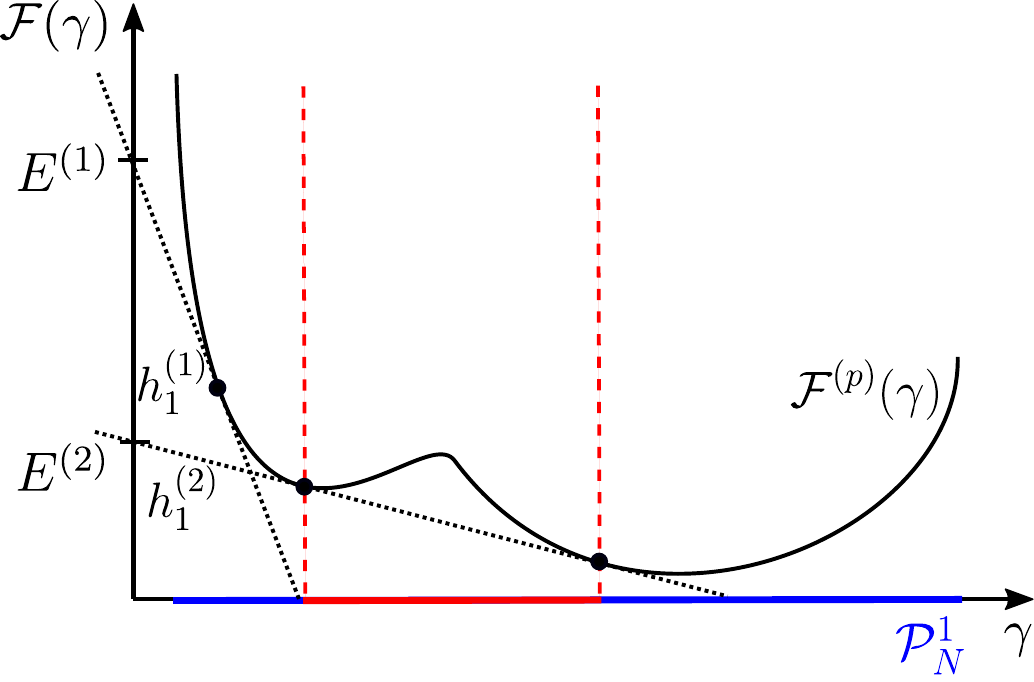}
\caption{Schematic illustration of the energy minimization and pure state $v$-representability for the pure universal functional $\mathcal{F}^{(p)}$ which is defined on the set $\mathcal{P}^1_N$ (red and blue). The red area between the two red dashed lines depicts the set of non pure state $v$-representable 1RDMs (see text for more details).\label{fig:F_schematic}}
\end{figure}

We first recall that according to the last line of Eq.~\eqref{eq:Levy}
the calculation of the ground state energy $E(h)$ through 1RDMFT can be interpreted (up to minus signs) as a Legendre-Fenchel transformation of the universal functional\cite{Schilling2018}. This and all the following comments made in this section are equally valid for all universal functionals shown in Fig.~\ref{fig:choices} and thus we introduce the simplified symbol $\mathcal{F}^{(p/e)}$ to represent any of them. As it is illustrated in Fig.~\ref{fig:F_schematic}, there is a simple geometric interpretation of the Legendre-Fenchel transformation and the calculation of $E(h)$, respectively\cite{Schilling2018}: The underlying minimization means nothing else than shifting the hyperplane defined by $\Tr_1[h_1\g]=const$ upwards until it touches the graph of the functional $\mathcal{F}^{(p/e)}$. The corresponding intercept with the vertical axis coincides (up to a minus sign) with the ground state energy $E(h)$ and the horizontal coordinate of the touch point is the corresponding ground state 1RDM. In case the graph of $\mathcal{F}^{(p/e)}$ is not convex or contains flat parts, there are corresponding $h_1$
leading to more than one touch point. This in turn means that the corresponding Hamiltonian $H(h)$ has a degenerate ground state space and thus can lead according to the sequence \eqref{eq:seq} to more than one ground state 1RDM. For instance, in the exemplary case of $h_1=h_1^{(1)}$ the ground state 1RDM is unique, whereas for $h_1^{(2)}$ the hyperplane touches $\mathcal{F}^{(p)}$ at two distinct points indicated by the black dots.  In particular, all 1RDMs between the two red dashed lines are not pure state $v$-representable since they cannot be obtained as touch points with the graph of $\mathcal{F}^{(p)}$  for any choice of the one-particle Hamiltonian. Accordingly, the notion of pure state $v$-representability is strongly linked to the form and more specifically the non-convexity of the pure functional $\mathcal{F}^{(p)}$: A 1RDM $\g$ is pure state $v$-representable if and only if the pure functional and its lower convex envelop (the corresponding ensemble functional) coincide at that point $\g$, i.e., $\mathcal{F}^{(p)}(\g)=\mbox{conv}\big(\mathcal{F}^{(p)}\big)(\g)$. In particular, this also means that the existence of not pure state $v$-representable sets of 1RDMs is tightly bound to the presence of ground state degeneracies\cite{Schilling2018}.

The same reasoning actually applies also in the context of ensemble 1RDMFT, yet with one crucial difference. Since the
ensemble functional $\F^{(e)}$ is convex any 1RDM in the interior of the domain $\mathcal{E}^1_N$ is ensemble state $v$-representable. This is in striking contrast to 1RDMs on the boundary  of $\mathcal{E}^1_N$. For instance, for arbitrary translationally-invariant one-band lattice models\cite{Schilling2019} the so-called fermionic exchange force (or Bose-Einstein condensation force for bosons\cite{Benavides20, LS21, M21}) repels the 1RDM from the boundary $\partial\mathcal{E}^1_N$ (and $\partial\mathcal{P}^1_N$ in pure 1RDMFT). Hence, 1RDMs on the boundary of the functional's domain are not pure/ensemble state $v$-representable, except for the non-generic case of a vanishing prefactor of the exchange force. Although there is little doubt that these implications are valid also for non-translationally invariant models --- compelling evidence follows from the work in Ref.~\cite{PhysRevA.96.052312, Schilling_2020, MSGLS20-2}--- no rigorous proof has been found so far. It will therefore be one of the crucial contributions of our work to confirm the existence of the fermionic exchange force for the class of all generalized Hubbard dimer models.

In summary, it is exactly the close relation between convexity of the universal functional and pure state $v$-representability that will play a central role for the discussion of the $v$-representability problem for the Hubbard dimer in Secs.~\ref{sec:HubbardDimer} and \ref{sec:generic}.

\section{Hubbard dimer with on-site interaction - singlet subspace\label{sec:HubbardDimer}}

In the following we provide an analytical discussion of the six universal functionals introduced in Sec.~\ref{sec:real-vs-complex} for the Hubbard dimer with on-site interaction. In particular, to complement the results obtained in Ref.~\onlinecite{Cohen2016}, we derive analytically the functional $\mathcal{F}_{\CC}^{(p)}$, namely by exploiting geometric aspects of the set of density matrices. Finally, we discuss the relations among the functionals and relate our findings to the concept of $v$-representability, as it has been outlined in Sec.~\ref{sec:vrepr-dimer}.

\subsection{Recap of the derivation of $\mathcal{F}_{\RR}^{(p)}$\label{sec:functional-onsite-real}}

To keep our paper self-contained, we recap in this section the derivation of the universal functional $\mathcal{F}^{(p)}_\RR$ for real-valued wave function as it was already derived, e.g., in Refs.~\onlinecite{Pastor2011a,Cohen2016}. Furthermore, similar concepts will be required in Sec.~\ref{sec:functional-onsite-complex} to derive the universal functional for complex-valued wave functions by analytical means.

The anisotropic Hubbard dimer with on-site interaction is described by the Hamiltonian
\begin{eqnarray}\label{eq:H}
H &=& -t\sum_{\sigma=\uparrow, \downarrow}(c_{1\sigma}^\dagger c_{2\sigma} + c_{2\sigma}^\dagger c_{1\sigma}) + \sum_{\sigma=\uparrow, \downarrow}(\epsilon_1 n_{1\sigma} + \epsilon_2 n_{2\sigma}) \nonumber\\
&\quad &+ U(n_{1\uparrow}n_{1\downarrow} + n_{2\uparrow}n_{2\downarrow})\,,
\end{eqnarray}
where the first term describes hopping with strength $t$ of electrons with spin $\sigma=\uparrow,\downarrow$ between the left and right site and $U$ is the on-site interaction. Moreover, $n_{i\sigma} = c_{i\sigma}^\dagger c_{i\sigma}$ denotes the occupation number operator which involves the fermionic annihilation (creation) operators $c_{i\sigma}$ ($c^\dagger_{i\sigma}$) acting on site $i=1,2$. Note that we skip the commonly used hat symbol on the operators $n_{i\sigma}$ since we will not use the same symbol for the operators and their expectation values. Consequently, to relate this to the context of 1RDMFT, the last term in Eq.~\eqref{eq:H} describes the \emph{fixed} pair-interaction $W$ and the first two terms constitute the \emph{variable} one-particle Hamiltonian $h$.

\begin{figure}[tb]
\includegraphics[width=0.65\linewidth]{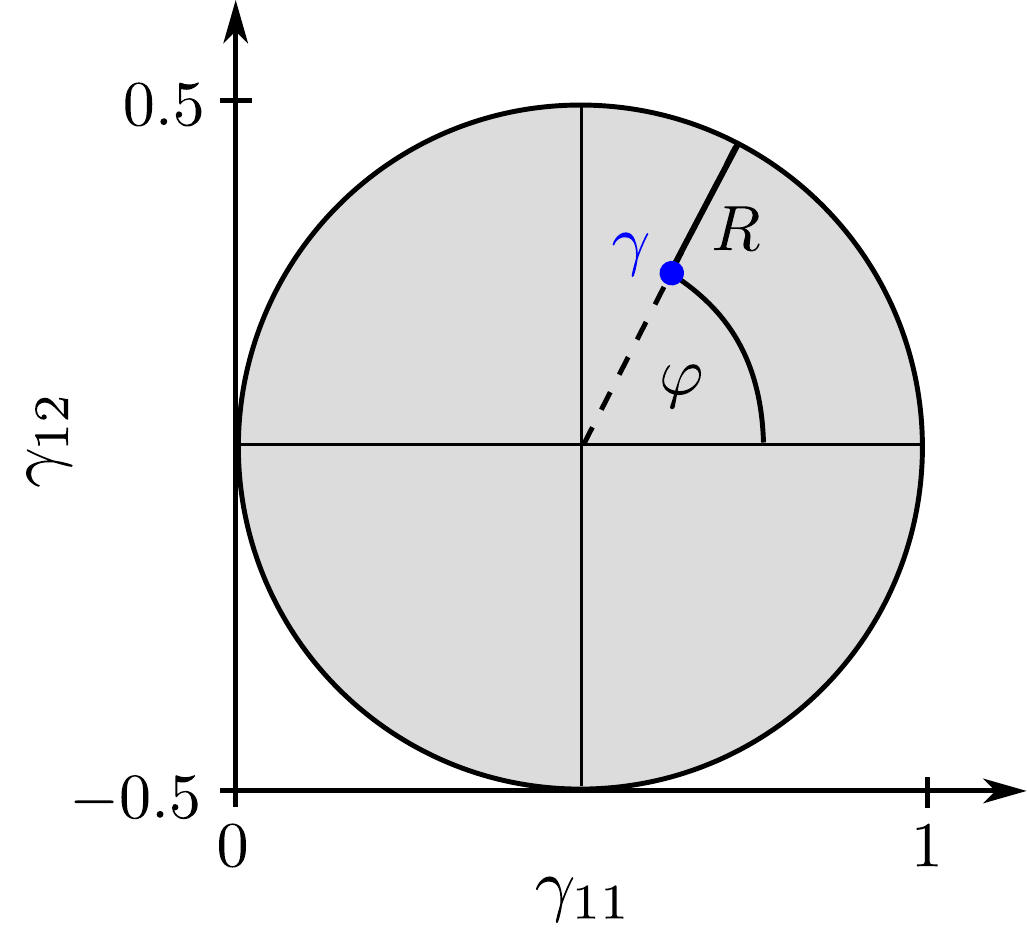}
\caption{Illustration of the set $\mathcal{P}^1_2 = \mathcal{E}^1_2$ of pure and ensemble $N$-representable real-valued 1RDMs $\g =(\g_{11},\g_{12})$ and their polar coordinates $R$ and $\varphi$. \label{fig:disk}}
\end{figure}

In the following, we consider the asymmetric Hubbard dimer, which means that we allow for an asymmetric external potential introduced by the second term in Eq.~\eqref{eq:H} which allows for $\epsilon_1\neq \epsilon_2$. At half-filling, $N=2$, we restrict the $N$-fermion Hilbert space $\mathcal{H}_N \equiv \wedge^N\mathcal{H}_1$ of dimension $D=\left(\begin{smallmatrix}d\\N\end{smallmatrix}\right)=6$ to the three-dimensional singlet subspace containing the ground state. In the singlet subspace, spin-symmetry furthermore implies that $\gt_{ij} \equiv\bra{i\uparrow}\gt\ket{j\uparrow} = \bra{i\downarrow}\gt\ket{j\downarrow}$, $i,j=1,2$. Thus, the 1RDM $\gt$ is block-diagonal with respect to the spin and the two non-vanishing blocks $\gt^\uparrow,\gt^\downarrow$ are equal, $\gt^\uparrow=\gt^\downarrow$. Therefore, we will effectively consider only one of them and denote it for the sake of simplicity by $\gt$ as well, where now $\Tr_1[\gt]=1$ instead of $\Tr_1[\gt]=2$. The same is assumed for the related $\g \equiv \Re(\gt)$.
Moreover, the two sets of 1RDMs defined in Eqs.~\eqref{eq:P1N} and \eqref{eq:E1N} are equal, that is $\mathcal{P}^1_2 = \mathcal{E}^1_2$\footnote{Although we will distinguish in the following carefully between different variants of 1RDMFT according to Sec.~\ref{sec:real-vs-complex} we will safely continue using the non-specific symbols $\mathcal{P}^1_N, \mathcal{E}^1_N$. It will namely be clear from the context to which variant they refer to.}. As an orthonormal reference basis $\mathcal{B} =\{\ket{\Phi_i}\}_{i=1}^3$ for the singlet subspace  we choose
\begin{eqnarray}\label{eq:basisstates}
\ket{\Phi_1} &=& c_{1\uparrow}^\dagger c_{1\downarrow}^\dagger\ket{0}\,, \\
\ket{\Phi_2} &=& c_{2\uparrow}^\dagger c_{2\downarrow}^\dagger\ket{0}\,,\nonumber \\
\ket{\Phi_3} &=& (c_{1\uparrow}^\dagger c_{2\downarrow}^\dagger - c_{1\downarrow}^\dagger c_{2\uparrow}^\dagger)\ket{0}/\sqrt{2}\,,\nonumber
\end{eqnarray}
where $\ket{0}$ denotes the vacuum state.
Due to the positivity condition of the 1RDM $\gt$, the set $\mathcal{P}^1_2 = \mathcal{E}^1_2$ is characterized by the condition\cite{Cohen2016}
\begin{equation}\label{eq:disk}
\left(\gt_{11}-\frac{1}{2}\right)^2 + |\gt_{12}|^2 \leq \frac{1}{4}\,.
\end{equation}
In the following, we interpret occasionally the 1RDM $\gt$ as a vector in $\RR^3$ with independent entries $(\gt_{11},\Re(\gt_{12}),\Im(\gt_{12}))$ and its real-part $\g \equiv \Re(\gt)$ as a vector in $\RR^2$ with independent entries $(\g_{11},\g_{12})=(\gt_{11},\Re(\gt_{12}))$.

If we deal with real-valued 1RDMs $\g$, their set is again described by relation \eqref{eq:disk}.
In that case this leads to the disk illustrated in Fig.~\ref{fig:disk}. Instead of Cartesian coordinates $(\gamma_{11}, \gamma_{12})$, we may also represent $\g$ in polar coordinates $R$ and $\varphi$. As it is illustrated in Fig.~\ref{fig:disk}, $R$ then denotes the distance of the 1RDM $\g$ to the boundary of the set $\mathcal{P}^1_2$ and $\varphi$ is the corresponding polar angle. Expressing the two independent matrix elements of the 1RDM, $\gamma_{11}$ and $\gamma_{12}$ in the polar coordinates $R, \varphi$ yields
\begin{eqnarray}\label{eq:polar-coord}
\gamma_{11} &=& [1+(1-2R)\mathrm{cos}(\varphi)]/2\nonumber \\
\gamma_{12} &=& (1-2R)\mathrm{sin}(\varphi)/2\,.
\end{eqnarray}

\begin{figure*}[htb]
\includegraphics[width=0.42\linewidth]{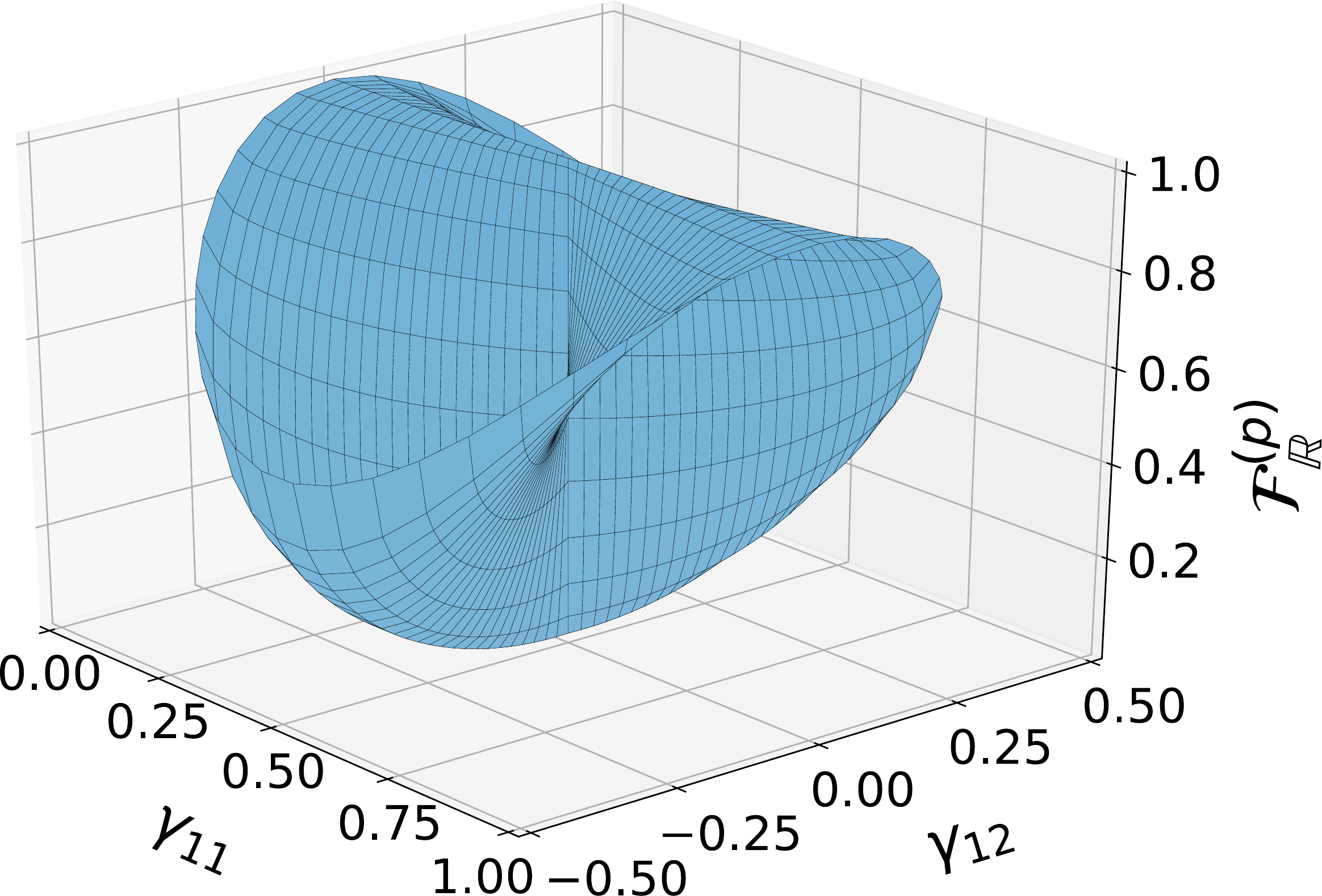}
\hspace{1.1cm}
\includegraphics[width=0.42\linewidth]{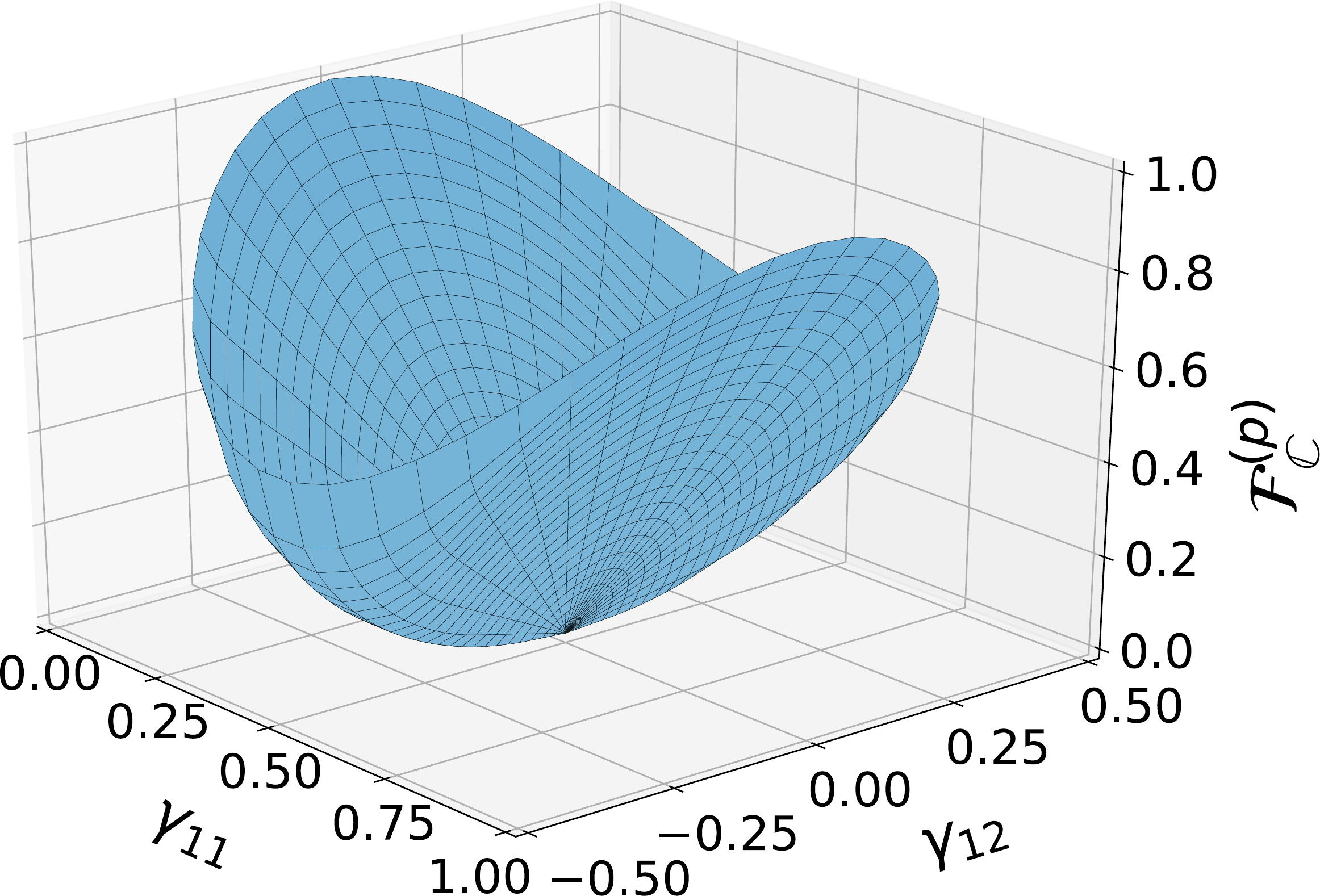}
\caption{The universal functional $\mathcal{F}_{\RR}^{(p)}$ (left) and $\mathcal{F}_{\CC}^{(p)}$ (right) for the Hubbard dimer with $U=1$. \label{fig:FR-FC}}
\end{figure*}

To recap the common 1RDMFT approach to the Hubbard dimer, we recall that the underlying Hamiltonian \eqref{eq:H}, in particular its one-particle term $h$, is time-reversal symmetric. As it is explained in Sec.~\ref{sec:real-vs-complex}, this allows one to first reduce the functional's variable from the full 1RDM $\gt$ to its real part $\g \equiv \Re(\gt)$ and then to restrict the constrained minimization of $\bra{\Psi}W\ket{\Psi}$ to pure states
\begin{equation}\label{eq:PsiB1}
\ket{\Psi} = a\ket{\Phi_1}+b\ket{\Phi_2}+c\ket{\Phi_3}
\end{equation}
with \emph{real} coefficients $a, b, c\in \RR$.
It is then a straightforward exercise to show that the corresponding functional $\mathcal{F}^{(p)}_{\RR}$ for the Hubbard dimer \eqref{eq:H} with \eqref{eq:PsiB1} is given by\footnote{References in the literature usually discuss repulsive interactions with $U\geq 0$ \cite{Pastor2011a, Cohen2016}. Nevertheless, the derivation of $\mathcal{F}_{\RR}^{(p)}$ can be extended to attractive interactions in a straightforward manner yielding the result in Eq.~\eqref{eq:FRR_1}}
\begin{eqnarray}\label{eq:FRR_1}
\lefteqn{\mathcal{F}_{\RR}^{(p)}(\gamma)}&&\\
&=& U\frac{(\gamma_{11}-\tfrac{1}{2})^2+\tfrac{1}{2}\gamma^2_{12} [1 - \mathrm{sgn}(U)\sqrt{1-4(\gamma_{11}-\tfrac{1}{2})^2-4\gamma^2_{12}}]}{(\gamma_{11}-\tfrac{1}{2})^2+\gamma^2_{12}}\,.\nonumber
\end{eqnarray}
We illustrate $\mathcal{F}_{\RR}^{(p)}$ in the left panel of Fig.~\ref{fig:FR-FC}.
Note that $\mathcal{F}_{\RR}^{(p)} = 0$ at $(\gamma_{11}, \gamma_{12}) = (1/2, 0)$ and thus the functional is lower semi-continuous\cite{R97} at this point. This is in particular relevant from a conceptual and mathematical point of view since this ensures that the minimum of the energy functional $\Tr_1[h_1\gamma] + \mathcal{F}^{(p)}_\RR(\g)$ is attained for every choice of $h_1$. As a consequence of Eq.~\eqref{eq:FpconvFe} \cite{Schilling2018}, the knowledge of \eqref{eq:FRR_1} is sufficient to construct, at least in principle, the ensemble functional $\mathcal{F}^{(e)}_{\RR}$, namely as the lower convex envelope of $\mathcal{F}^{(p)}_\RR$. Yet, it is worth noticing that the numerical calculation of a lower convex envelope in larger dimensions represents a rather involved problem.

\subsection{Analytic derivation of universal functionals for complex-valued wave functions\label{sec:functional-onsite-complex}}

In the following, we derive first the functional $\widetilde{\mathcal{F}}^{(p)}_{\CC}(\gt)$ and afterwards the two functionals $\mathcal{F}_{\CC}^{(p/e)}(\g)$ for the Hubbard dimer with on-site interaction described by the Hamiltonian in Eq.~\eqref{eq:H}.
The functional $\widetilde{\mathcal{F}}^{(p)}_{\CC}$ is obtained by minimizing $\bra{\Psi}W\ket{\Psi}$ over all pure $N$-fermion states of the form in Eq.~\eqref{eq:PsiB1}
with complex coefficients $a, b, c \in \CC$ and $\ket{\Phi_1}$, $\ket{\Phi_2}$ and $\ket{\Phi_3}$ are given by Eq.~\eqref{eq:basisstates}.
For the crucial expectation value of the interaction $W= U(n_{1\uparrow}n_{1\downarrow} + n_{2\uparrow}n_{2\downarrow})$ one finds
\begin{equation}
\Tr_N[W\ket{\Psi}\!\bra{\Psi}]= U(1-|c|^2)\,.
\end{equation}
Since $|c|$ turns out to be decoupled from the phase of $\gt_{12}$, it directly follows that $\widetilde{\mathcal{F}}_{\CC}^{(p)}$ is independent of the phase of $\gt_{12}$. Finally, one finds (see Appendix \ref{app:Ftilde})
\begin{widetext}
\begin{eqnarray}\label{eq:FCC_1main}
{\widetilde{\mathcal{F}}_{\CC}^{(p)}(\gt)} = U\frac{(\gt_{11}-\tfrac{1}{2})^2+\tfrac{1}{2}|\gt_{12}|^2[1 - \mathrm{sgn}(U)\sqrt{1-4(\gt_{11}-\tfrac{1}{2})^2-4|\gt_{12}|^2}]}{(\gt_{11}-\tfrac{1}{2})^2+|\gt_{12}|^2}\,.
\end{eqnarray}
\end{widetext}
In particular, we thus observe that (recall $\gt_{11} \in \RR$)
\begin{equation}
\widetilde{\mathcal{F}}_{\CC}^{(p)}(\gt_{11},\gt_{12}) = \mathcal{F}_{\RR}^{(p)} (\gt_{11}, |\gt_{12}|)\,.
\end{equation}
This simply means that $\widetilde{\mathcal{F}}_{\CC}^{(p)}$ is related to the well-known result for  $\mathcal{F}_{\RR}^{(p)}$, one just needs to  replace $\gamma_{12}$ by $|\gt_{12}|$.
According to Eq.~\eqref{eq:FpconvFe}, the functional $\widetilde{\mathcal{F}}_{\CC}^{(p)}$ in Eq.~\eqref{eq:FCC_1main} determines  $\widetilde{\mathcal{F}}^{(e)}_{\CC}$. The latter namely follows as the lower convex envelop of the former. Thus, the remaining universal functionals to be calculated are $\mathcal{F}_{\CC}^{(p/e)}$.

To derive $\mathcal{F}^{(p)}_\CC$ according to Eq.~\eqref{eq:Levy_FC}, we first notice that each state \eqref{eq:PsiB1} with complex coefficients $a, b, c$ can be separated into its real and imaginary part (relative to the basis states \eqref{eq:basisstates}),
\begin{equation}
\ket{\Psi} = w_r\ket{\Psi_r} + i w_i\ket{\Psi_i}\,,
\end{equation}
with $w_r, w_i\in \RR$ and $w_r^2+w_i^2=1$.
Since only the imaginary part of $\ket{\Psi}\!\bra{\Psi}$ contributes to $\Im(\gt_{12})$ and only the real part
\begin{equation}\label{eq:GR2}
\Re(\ket{\Psi}\!\bra{\Psi}) = w_r^2 \ket{\Psi_r}\!\bra{\Psi_r}  + w_i^2\ket{\Psi_i}\!\bra{\Psi_i}
\end{equation}
to
$\Tr [\ket{\Psi}\!\bra{\Psi} W] = \bra{\Psi} W \ket{\Psi}$,
Eq.~\eqref{eq:Levy_FC} together with the definition of $\Ftcp$ yields
\begin{eqnarray}\label{eq:FC_rank2}
\mathcal{F}^{(p)}_\CC (\g) & =& \min_{\scriptsize \begin{array}{c}\G\mapsto \g,\\ \mathrm{rank}(\G)\leq 2 \end{array} }\Tr_N[W\G]\,.
\end{eqnarray}
Here, the minimization of the interaction energy is performed over all real-valued density operators $\G$ with rank of at most 2.

The goal is now to show that $\mathcal{F}^{(p)}_\CC$ \eqref{eq:FC_rank2} equals the ensemble functional $\mathcal{F}^{(e)}_\RR$. For this, we first recall from Sec.~\ref{sec:intro1RDMFT} that $\mathcal{F}^{(e)}_\RR$ is obtained by minimizing the \emph{linear} functional $\Tr_N[W\Gamma]$ over the convex and compact set
\begin{equation}\label{eq:E2gamma}
\mathcal{E}^2(\gamma) \equiv \left\{\Gamma\in \mathcal{E}^2\,|\,\Gamma\mapsto\gamma\right\}\,,
\end{equation}
of real-valued $2$-electron density operators $\G$ which map to the given 1RDM $\g$.
Since the partial trace $\Tr_{1}[\cdot]$ is linear, the set $\mathcal{E}^2(\gamma)$ can be interpreted as the intersection of the set $\mathcal{E}^2$ with the hyperplane described by $\Tr_1[\cdot]=\gamma$. As a result, the boundary points of $\mathcal{E}^2(\gamma)$ are of at most rank $D-1$  (because they are in particular also boundary points of $\mathcal{E}^2$).
Then, minimizing $\Tr_2[W\Gamma]$ over all $\Gamma\in \mathcal{E}^2(\gamma)$ means to shift a hyperplane whose normal vector is defined by the interaction $W$ (i.e., hyperplanes of constant interaction energy) in direction $-W$ until it touches the boundary of $\mathcal{E}^2(\gamma)$\cite{Schilling2018}. Consequently, to derive the ensemble functional $\mathcal{F}^{(e)}_\RR(\g)$ we can restrict the minimization over all density operators to those that are of at most rank $D-1$. Comparing this with Eq.~\eqref{eq:FC_rank2} reveals that the definitions of $\mathcal{F}^{(e)}_\RR$ and $\mathcal{F}^{(p)}_\CC$ coincide for $D=3$. Since we have $D=3$ for the asymmetric Hubbard dimer restricted to the singlet subspace, this observation finally leads to
\begin{equation}\label{eq:FCeqFe}
\mathcal{F}^{(p)}_\CC(\g) = \mathcal{F}^{(e)}_\RR (\gamma)\,.
\end{equation}
Thus, the universal functional $\mathcal{F}^{(p)}_\CC$ is equal to $\mathcal{F}^{(e)}_\RR$ and, therefore, also equals the lower convex envelope of $\mathcal{F}^{(p)}_\RR$.

The key result \eqref{eq:FCeqFe} already resembles an important conclusion of our work: Whenever one adds `unnecessary' degrees of freedom to $\mathcal{H}_N$ in the constrained search formalism of Levy one simulates effectively a certain degree of mixedness. Indeed, since the interaction $W$ does not depend on the extra degrees of freedom, one can trace them out again and obtain a mixed state on $\mathcal{H}_N$ due to the possible entanglement between the extra degrees of freedom and those of $\mathcal{H}_N$. In case of a sufficiently small-dimensional $N$-fermion Hilbert space (as in the case of the Hubbard dimer) this even yields the entire set of density operators and accordingly Valone's constrained search formalism.
We also would like to stress that the derivation of Eq.~\eqref{eq:FCeqFe} does not require any knowledge of the interaction under consideration and is merely based on the dimensionality of the singlet subspace and the geometry of the set of density operators. In particular this means that the proof of Eq.~\eqref{eq:FCeqFe} is equally valid for the generalized Hubbard dimer in Sec.~\ref{sec:generic}.


To obtain a closed form for the functional $\Fcp$ (and $\Fre$) one performs the minimization of $\widetilde{\mathcal{F}}^{(p)}_{\CC}(\gt)$ with respect to the imaginary part $\Im(\gt_{12})$. This elementary exercise then leads to
\begin{eqnarray}\label{eq:FCC_2}
\lefteqn{\mathcal{F}_{\CC}^{(p)}(\gamma_{11}, \g_{12})} && \\
&=& \begin{cases}
U(1 - 2\gamma_{11})\,,\!&\text{if } \gamma_{12}^2\leq \gamma_{11}(1-2\gamma_{11})\\
U(2\gamma_{11}-1)\,,\!&\text{if }\gamma_{12}^2\leq \gamma_{11}(3 - 2\gamma_{11})-1\\
\mathcal{F}^{(p)}_{\RR}\left(\gamma_{11}, \gamma_{12}\right)\,,\!&\text{otherwise}\\
\end{cases}\nonumber\,.
\end{eqnarray}
The above equation describes nothing else than the lower convex envelope of $\mathcal{F}^{(p)}_{\RR}$.
We plot the universal functional $\mathcal{F}_{\CC}^{(p)}$ in the right panel of Fig.~\ref{fig:FR-FC}.
The result in Eq.~\eqref{eq:FCC_2} was first deduced from numerical studies in Ref.~\onlinecite{Cohen2016}, yet without providing any analytical evidence. In contrast, our work has provided a complete analytic proof through Eq.~\eqref{eq:FCeqFe}.

\begin{figure}[htb]
\frame{\includegraphics[width=1.0\linewidth]{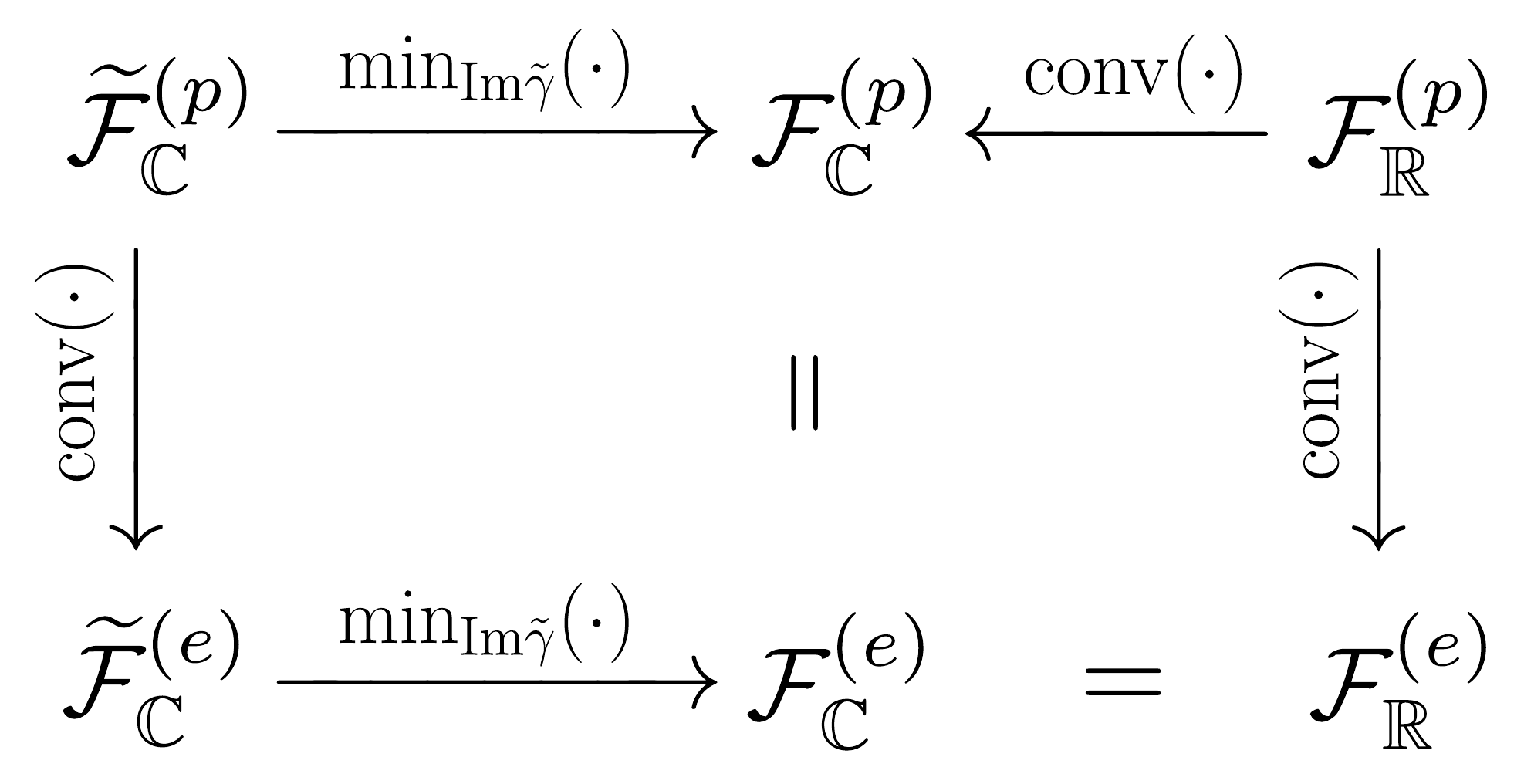}}
\caption{Illustrations of various relations between the six universal functionals for the Hubbard dimer and its generalizations (see text for more details). \label{fig:functionals}}
\end{figure}

We present the relations among various functionals derived in this section in Fig.~\ref{fig:functionals}.

\subsection{Discussion of v-representability \label{sec:vrepr-dimer}}

Equipped with the six universal functionals derived in the previous section, we now turn towards the $v$-representability problem and apply the general concepts introduced in Sec.~\ref{sec:v-rep-general} to the asymmetric Hubbard dimer defined in Eq.~\eqref{eq:H}. For this, we first recall the distinction between pure state and ensemble state $v$-representability from Sec.~\ref{sec:v-rep-general}.

\begin{figure}[htb]
\centering
\includegraphics[width=0.48\linewidth]{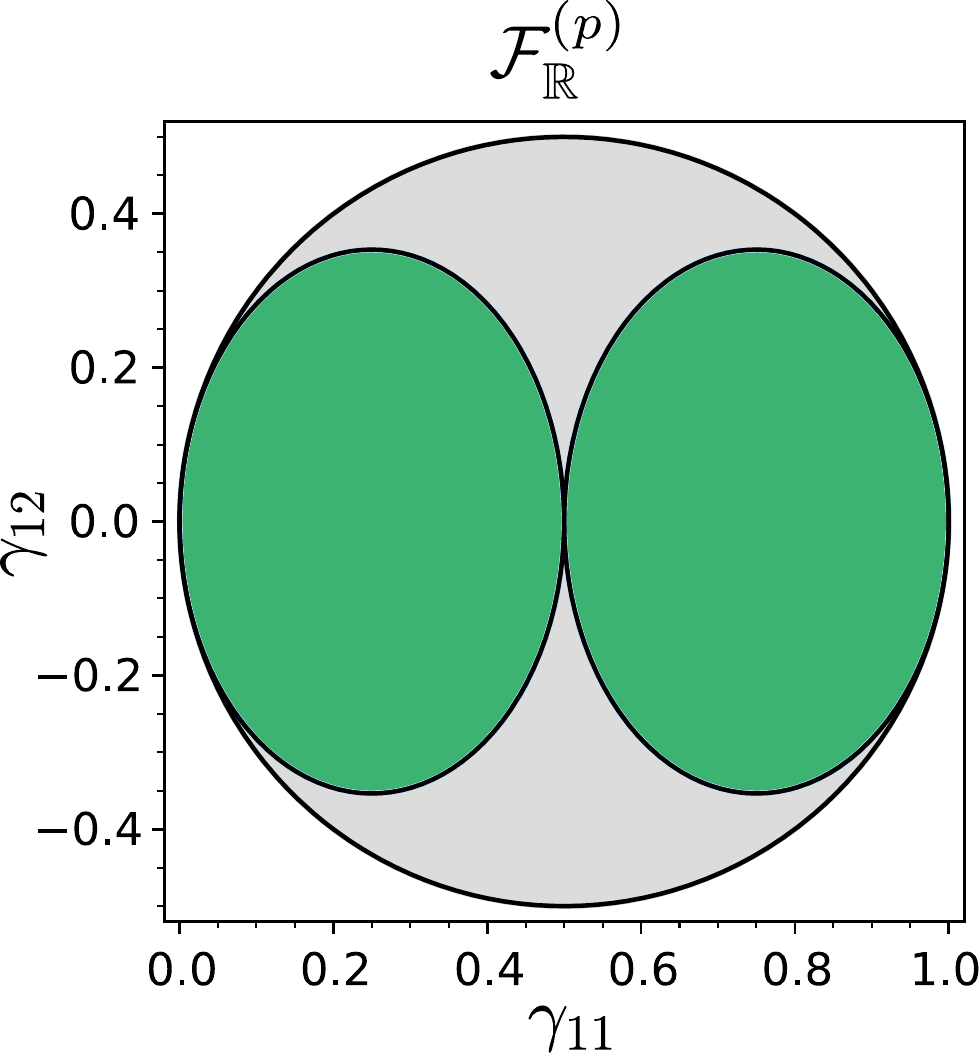}
\hspace{0.1cm}
\includegraphics[width=0.48\linewidth]{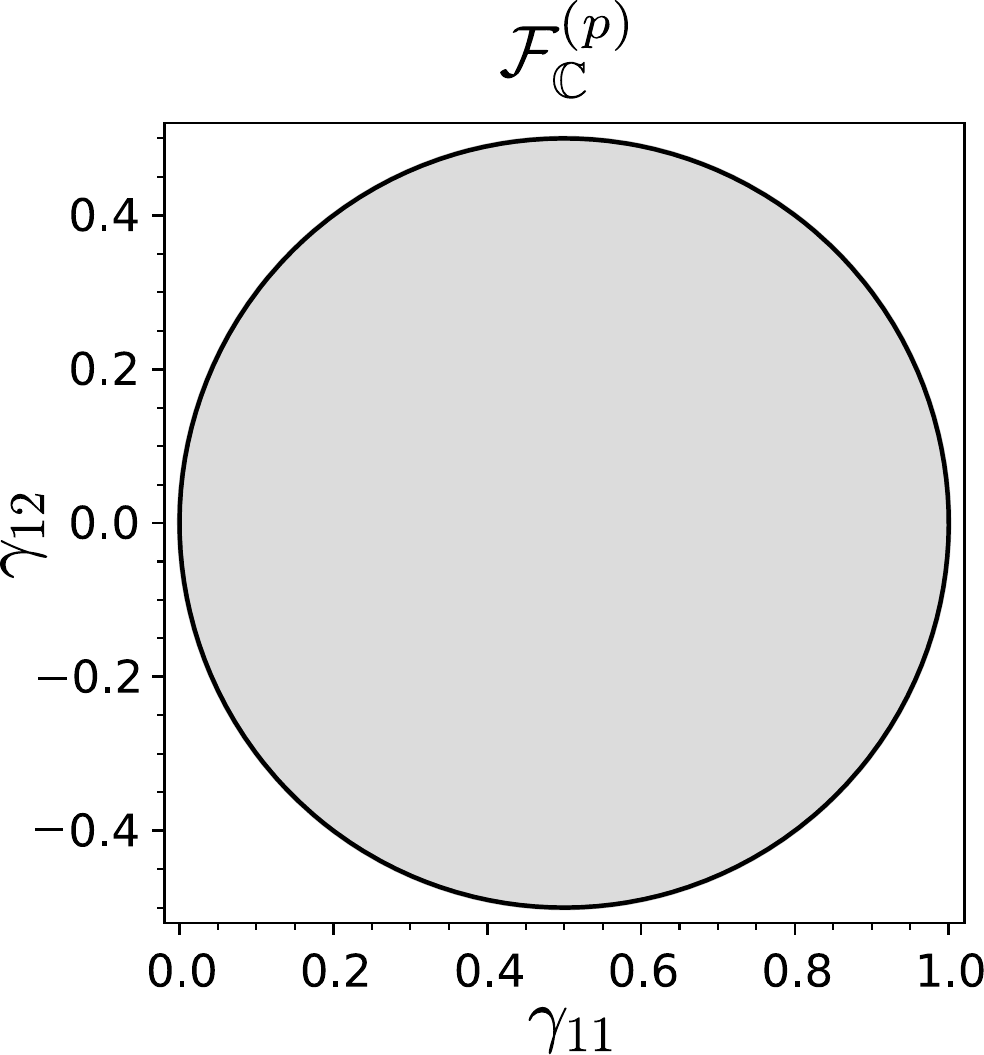}
\caption{Illustration of the non-$v$-representable subregions (green) of the set $\mathcal{P}^1_N$ (grey and green) for the two universal functionals $\mathcal{F}^{(p)}_{\RR}(\gamma)$ (left) and $\mathcal{F}^{(p)}_{\CC}(\g)$ (right). \label{fig:vrepr-dimer}}
\end{figure}

First, we focus on the pure state $v$-representability problem which we solve by comparing the graphs of $\mathcal{F}_{\RR}^{(p)}$ and $\Fcp=\mathcal{F}_{\RR}^{(e)}$. Thereby we complement the illustration of the non-$v$-representable regions for $\mathcal{F}_{\RR}^{(p)}$ in Ref.~\onlinecite{Cohen2016} with a comprehensive discussion of $v$-representability with respect to real-valued or complex-valued one-particle Hamiltonians $h$. By restricting to conventional time-reversal symmetric Hamiltonians, similar illustrations were obtained for the Anderson model\cite{Pastor2011b} and general interactions by numerical means\cite{Neck2001}.
In Fig.~\ref{fig:vrepr-dimer} we plot the functional's domain $\mathcal{P}^1_N$ and illustrate the set of 1RDMs which are not pure state $v$-representable in green for both $\mathcal{F}^{(p)}_\RR$ (left panel) and $\mathcal{F}^{(p)}_\CC$ (right panel).
The set of all pure state $v$-representable 1RDMs is shown in grey.
For $\mathcal{F}^{(p)}_{\RR}$ \eqref{eq:FRR_1}, there exist two solid green ellipses of 1RDMs $\g$ which are not pure state $v$-representable, whereas for $\mathcal{F}^{(p)}_{\CC}$ all $(\gt_{11}, \Re(\gt_{12}))$ (except the boundary) are pure state $v$-representable. It can be easily shown that the equation of the two ellipses for $\mathcal{F}^{(p)}_{\RR}$ is given by
\begin{equation}
2\g_{12}^2 +\left(2\left\vert\g_{11}-\frac{1}{2}\right\vert - \frac{1}{2}\right)^2 = \frac{1}{4}\,.
\end{equation}
Since the functional $\mathcal{F}^{(p)}_{\CC}$ is convex (recall Eq.~\eqref{eq:FCeqFe} and Fig.~\ref{fig:functionals}), all 1RDMs in the interior of the underlying domain are according to Sec.~\ref{sec:v-rep-f} indeed complex pure state $v$-representably. This is an important insight which demonstrates again that $v$-representability is a relative concept.

For both $\mathcal{F}^{(p)}_{\CC}$ and $\mathcal{F}^{(p)}_{\RR}$, all points on the boundary of $\mathcal{P}^1_N$, except the two points $(\g_{11}, \g_{12})= (0,0), (1, 0)$, are neither real nor complex pure state $v$-representable, as a direct result of the fermionic exchange force\cite{Schilling2019}. Since we are going to calculate this force for the asymmetric Hubbard dimer with generic interactions in Sec.~\ref{sec:force} containing \eqref{eq:H} as a special case, we skip its derivation here.

To discuss the last of the three pure functionals, we observe that the $\widetilde{\mathcal{F}}^{(p)}_\CC(\gamma)$ is not convex. This implies directly according to Sec.~\ref{sec:v-rep-f} that some of the complex pure state $N$-representable 1RDMs $\gt$ are not complex pure state $v$-representable. To clarify this aspect, let us now consider a 1RDM $\gamma$ which is not real pure state $v$-representable but complex pure state $v$-representable. Then, it follows that the 1RDM $\tilde{\gamma}\mapsto\gamma$ obtained from the minimization in Eq.~\eqref{eq:Levy_FC} has a non-zero imaginary part. 

We can explicitly determine this imaginary part by constructing the degenerate complex-valued $N$-particle state. In the case of $\F_\RR^{(p)}$ there exists an $h$ such that the two 1RDMs
\begin{equation}\label{eq:gammaellipse}
\gamma^{(1)} = \ket{2}\!\bra{2}\,,\quad \gamma^{(2)} =\frac{1}{2} \left(\ket{1}\!\bra{1}+\ket{2}\!\bra{2}\right)
\end{equation}
following from the quantum states $\ket{2\uparrow, 2\downarrow}$ and $\frac{1}{\sqrt{2}}(\ket{1\uparrow, 2\downarrow} - \ket{1\downarrow, 2\uparrow})$, respectively, correspond to the same ground state energy.
Therefore, also a superposition
\begin{equation}\label{eq:supo}
\ket{\Psi} = x \ket{2\uparrow, 2\downarrow}\pm \sqrt{1 - |x|^2}\frac{1}{\sqrt{2}}(\ket{1\uparrow, 2\downarrow} - \ket{1\downarrow, 2\uparrow})
\end{equation}
leads to the same ground state energy.
Since $x = |x|\rme^{i\varphi}\in \mathbb{C}$ with $|x|\in [0, 1]$ in Eq.~\eqref{eq:supo} we obtain
\begin{equation}\label{eq:gammasupo}
\begin{split}
\tilde{\gamma}_{11} &= \frac{1}{2} (1 - |x|^2)\\\
\tilde{\gamma}_{12} &= \pm \frac{1}{\sqrt{2}} x\sqrt{1-|x|^2} = \pm  \frac{1}{\sqrt{2}}|x|\rme^{i\varphi}\sqrt{1-|x|^2} \\\
&\equiv \mathrm{Re}(\tilde{\gamma}_{12}) + i\,\mathrm{Im}(\tilde{\gamma}_{12})\,,
\end{split}
\end{equation}
and similarly for the second ellipse. By varying $|x|$ and setting $\varphi=0$ one obtains the ellipses in the $(\gamma_{11},\gamma_{12})$ plane which define the non-$v$-representable 1RDMs in the case of $\F_\RR^{(p)}$. However, for $\varphi\neq 0$ we can reach any point $(\gamma_{11}, \gamma_{12})$ inside the ellipse such that \eqref{eq:gammasupo} is satisfied.

Since the three ensemble functionals $\widetilde{\mathcal{F}}_{\CC}^{(e)}, \mathcal{F}_{\CC}^{(e)}$ and $\mathcal{F}_{\RR}^{(e)}$ are convex, all 1RDMs in the interior of the respective functional's domain are ensemble $v$-representable as anticipated in Sec.~\ref{sec:v-rep-f}. In analogy to $\mathcal{F}_{\CC}^{(p)}$ and $\mathcal{F}_{\RR}^{(p)}$, the 1RDMs at the boundary of the functional's domain are not ensemble $v$-representable (except for $(\gamma_{11}, \gamma_{12})= (0,0), (1, 0)$) due to the fermionic exchange force.
As explained in Sec.~\ref{sec:v-rep-general}, the 1RDMs which are not pure but ensemble state $v$-representable correspond to degenerate ground states.

\section{Generalized Hubbard dimer-singlet subspace\label{sec:generic}}

Exact closed expressions for universal 1RDM-functionals of model systems such as the ordinary Hubbard dimer are quite rare but then frequently used to illustrate conceptual aspects of 1RDMFT\cite{Neck2001, Lopez2002, Requist2008, Pastor2011a, Cohen2016, Kamil2016, Schilling2018}. It is therefore one of the main achievements of this paper to derive analytically some of the universal functionals (in particular $\mathcal{F}_{\RR}^{(p)}$) for the Hubbard dimer with generalized pair-interactions $W$.  This will also allow to confirm conclusively that the subsets of non-$v$-representable 1RDMs strongly depend on the interaction $W$ between the particles as it has been proposed in Ref.~\onlinecite{Neck2001} based on numerical investigations.

As for the Hubbard dimer with on-site interaction, we choose as orthonormal reference basis the three states in Eq.~\eqref{eq:basisstates}. Then, the most general isotropic, reflection symmetric interaction reads
\begin{eqnarray}\label{eq:W_generic}
W &=& U\left(\ket{\Phi_1}\!\bra{\Phi_1} + \ket{\Phi_2}\!\bra{\Phi_2}\right)  + V\left(\ket{\Phi_1}\!\bra{\Phi_2} + \mathrm{h.c.}\right) \nonumber \\
& \quad &+ X\left(\ket{\Phi_1}\!\bra{\Phi_3} + \ket{\Phi_2}\!\bra{\Phi_3} + \mathrm{h.c.}\right)\,,
\end{eqnarray}
where $U, V, X\in \RR$. The term proportional to $\ket{\Phi_3}\!\bra{\Phi_3}$ can be discarded due to a possible overall shift of the total energy.
The first term in Eq.~\eqref{eq:W_generic} describes the Hubbard on-site interaction of strength $U$ as in Sec.~\ref{sec:HubbardDimer}. Note that we do not distinguish between repulsive and attractive on-site interactions. A direct coupling between the singlet basis states $\ket{\Phi_1}$ and $\ket{\Phi_2}$ is introduced by the second term in \eqref{eq:W_generic}, $V(c^\dagger_{1\uparrow}c^\dagger_{1\downarrow}\ket{0}\!\bra{0}c_{2\downarrow}c_{2\uparrow} +\mathrm{h.c.})$, and therefore it corresponds to a transition involving two fermions from site $i$ to site $j\neq i$ as it might occur for a compound of two fermions in a singlet state (e.g.~for  cobosons). The third term in \eqref{eq:W_generic} finally comprises all eight possible hopping processes of one fermion embedded in the two-fermion level. Therefore, it describes the energy cost or gain of transitions between a double and a single occupied site.

\subsection{Derivation of $\mathcal{F}_{\RR}^{(p)}$}

In contrast to the ordinary Hubbard dimer with on-site interaction discussed in the previous section, the universal functional $\widetilde{\mathcal{F}}^{(p)}_\CC$ for the generic reflection symmetric interaction \eqref{eq:W_generic} will in general depend on the phase of $\gt_{12}\in \CC$. Due to this additional degree of freedom, the constrained search for deriving $\widetilde{\mathcal{F}}^{(p)}_\CC$ cannot be performed analytically anymore.
Instead, we commence by deriving the pure universal functional $\mathcal{F}^{(p)}_\RR$.
Minimizing the expectation value of the interaction $W$ over all states of the form \eqref{eq:PsiB1} with real coefficients $a, b, c\in \RR$ yields in a straightforward manner the key result (see Appendix \ref{app:FR_gen})
\begin{eqnarray}\label{eq:FRR_polar}
&&\mathcal{F}_\RR^{(p)}(\gamma) = U + \sqrt{2}X(1-2R)\mathrm{sin}(\varphi) + \frac{1}{2}(V-U)\mathrm{sin}^2(\varphi) \nonumber \\
&& - \frac{1}{2}\sqrt{1-(1-2R)^2}\left\vert (V-U)\mathrm{sin}^2(\varphi) - 2V\right\vert\,,
\end{eqnarray}
where $R, \varphi$ are the polar coordinates introduced in Eq.~\eqref{eq:polar-coord}. According to Eq.~\eqref{eq:FRR_polar}, a non-zero $X$ only adds a tilt to the functional. Therefore, we choose $X=0$ in Table \ref{fig:FRR_arbitrary} to plot the universal functional $\mathcal{F}_\RR^{(p)}$ for $U=1$ and different values of $V$ in the left panel. Each row in Table \ref{fig:FRR_arbitrary} corresponds to a different value of $V$ as indicated on top of the plot of $\mathcal{F}_\RR^{(p)}$. Next to $\mathcal{F}_\RR^{(p)}$ we show the domain of the functional and illustrate the set of not real pure state $v$-representable 1RDMs in green, whereby the dashed lines depict the constant values of $\gamma_{11}$ and $\gamma_{12}$ for which we plot a 2D slice of the functional in the third and fourth column. The implications thereof with respect to the $v$-representability problem will be discussed in Sec.~\ref{sec:vrepr-gen}.

In order to derive $\Fcp$, we recall that our proof of Eq.~\eqref{eq:FCeqFe} was merely based on the geometry of quantum states and, thus, is independent of the interaction $W$ under consideration. Therefore, the six universal functionals for the generalized Hubbard dimer (recall Fig.~\ref{fig:choices}) obey the same relations among each other as for the ordinary Hubbard dimer with on-site interaction (see Fig.~\ref{fig:functionals}).
In particular, the functional $\mathcal{F}_{\CC}^{(p)}$ is given by
\begin{equation}\label{eq:FCeqFe2}
\mathcal{F}_{\CC}^{(p)}(\g) = \mathrm{conv}\left(\mathcal{F}_{\RR}^{(p)}(\g)\right)= \Fre(\g)\,.
\end{equation}
It is worth stressing here that, according to the proof of Eq.~\eqref{eq:FCeqFe} in Sec.~\ref{sec:HubbardDimer}, the simple relations between $\mathcal{F}^{(p)}_\CC$, $\mathcal{F}^{(p)}_\RR$ and $\Fre$ in Eq.~\eqref{eq:FCeqFe2} will typically not hold anymore for dimensions $D>3$. Nevertheless, they may still hold for specific systems with distinctive simplifying properties, as e.g., the Fermi-Hubbard model with $N=2$ electrons on an arbitrary number of lattice sites \cite{Kienesberger-MA}.

\begin{table*}[!]
\centering
\begin{tabular}{c  c  c  c}
 \thead{\textbf{Functional $\mathbf{\mathcal{F}_{\RR}^{(p)}(\gamma_{11}, \gamma_{12})}$}}& \thead{\textbf{Domain of $\mathbf{\mathcal{F}_{\RR}^{(p)}}$}}& \thead{\textbf{$\mathbf{\mathcal{F}_{\RR}^{(p)}}$ along\,\,\,\,\includegraphics[width=0.08\linewidth]{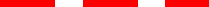}}} &\thead{\textbf{$\mathbf{\mathcal{F}_{\RR}^{(p)}}$ along\,\,\,\,\includegraphics[width=0.08\linewidth]{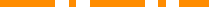}}}\\
\hline
 \thead{\includegraphics[width=0.24\linewidth]{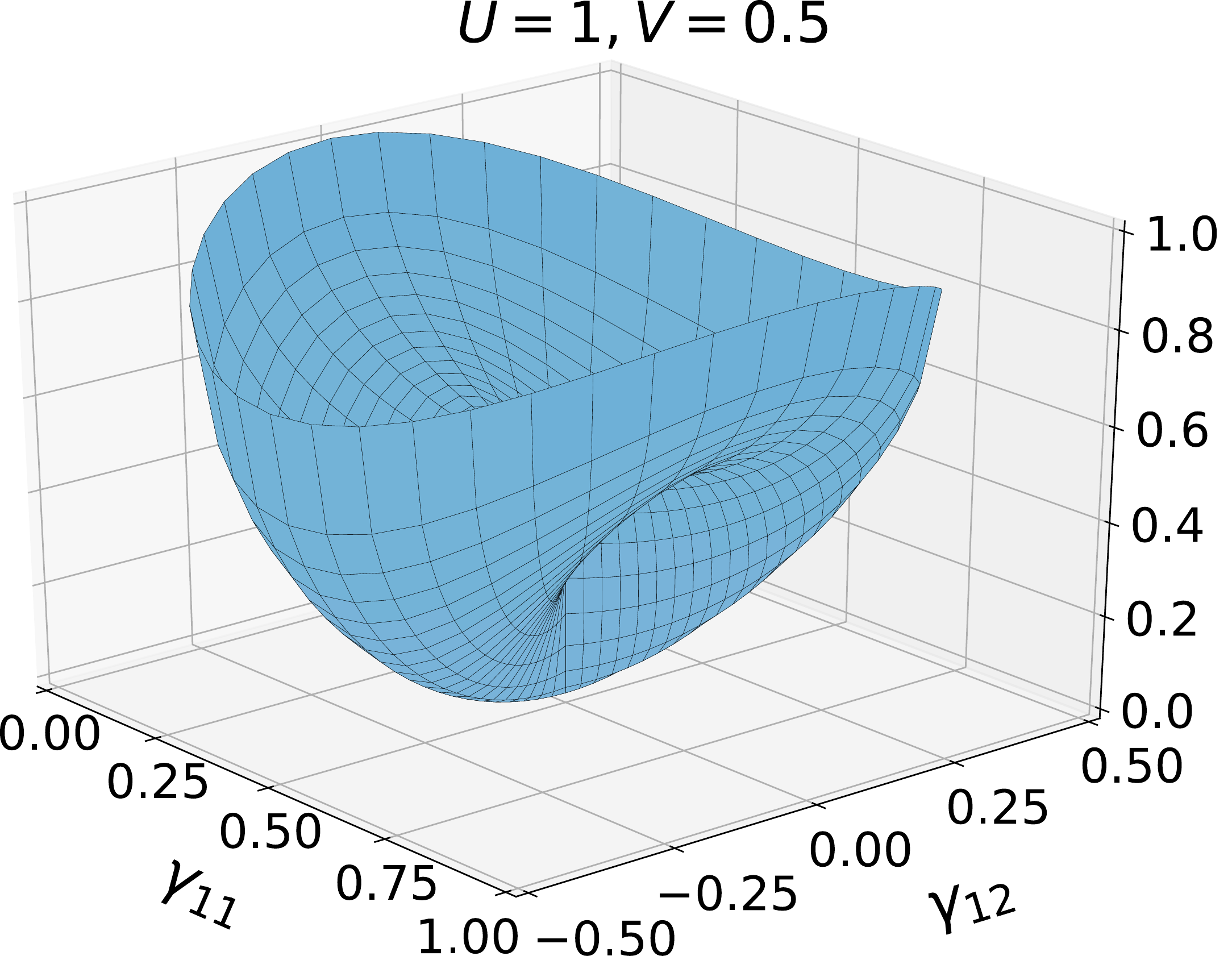}}& \thead{\includegraphics[width=0.20\linewidth]{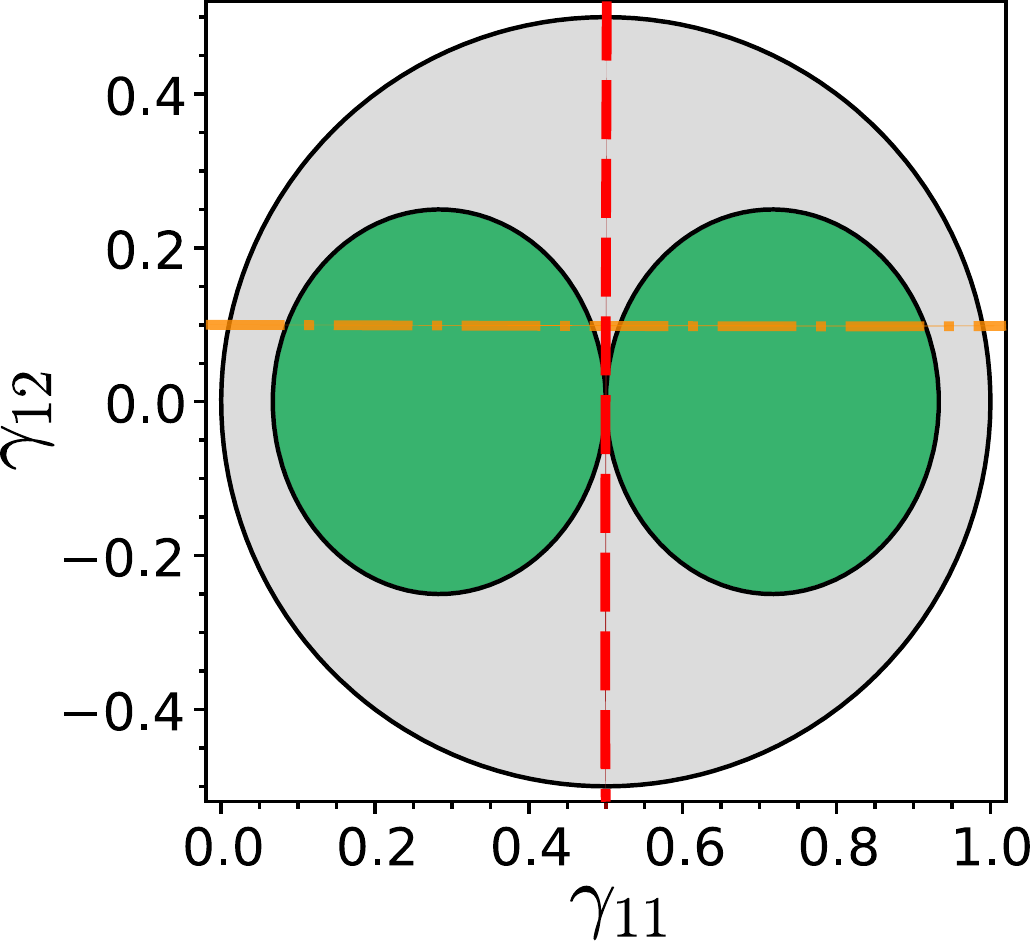}}& \thead{\includegraphics[width=0.24\linewidth]{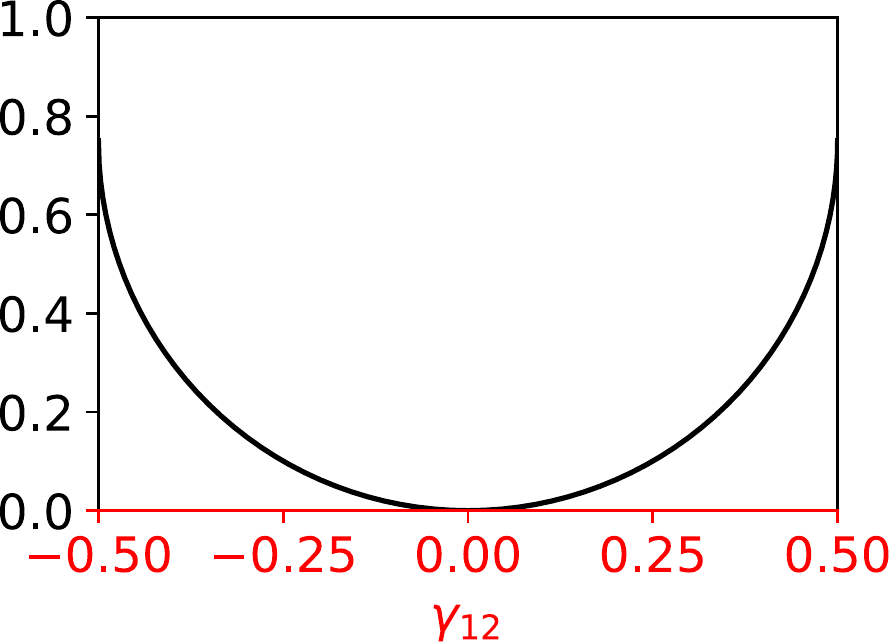}}&
 \thead{\includegraphics[width=0.24\linewidth]{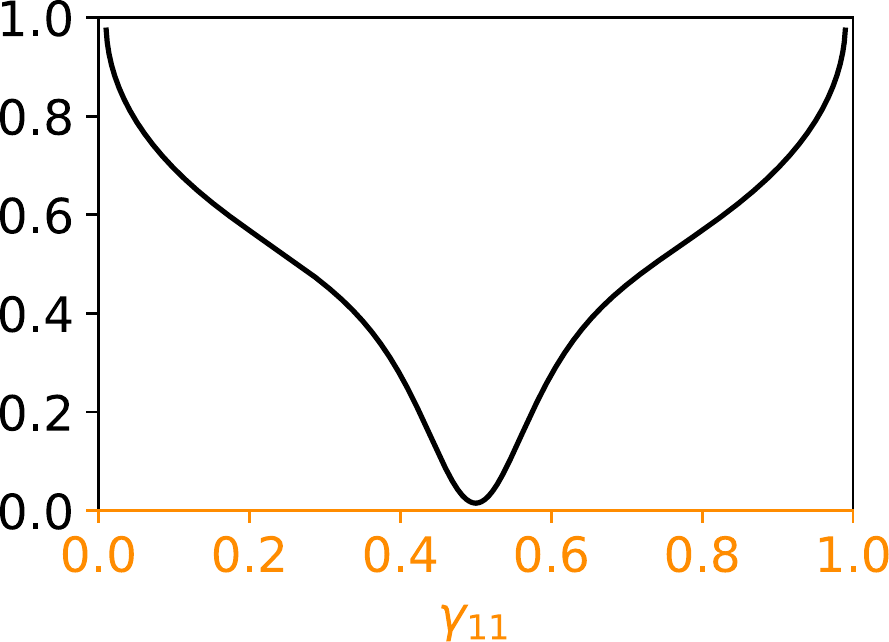}}\\
 \thead{\includegraphics[width=0.24\linewidth]{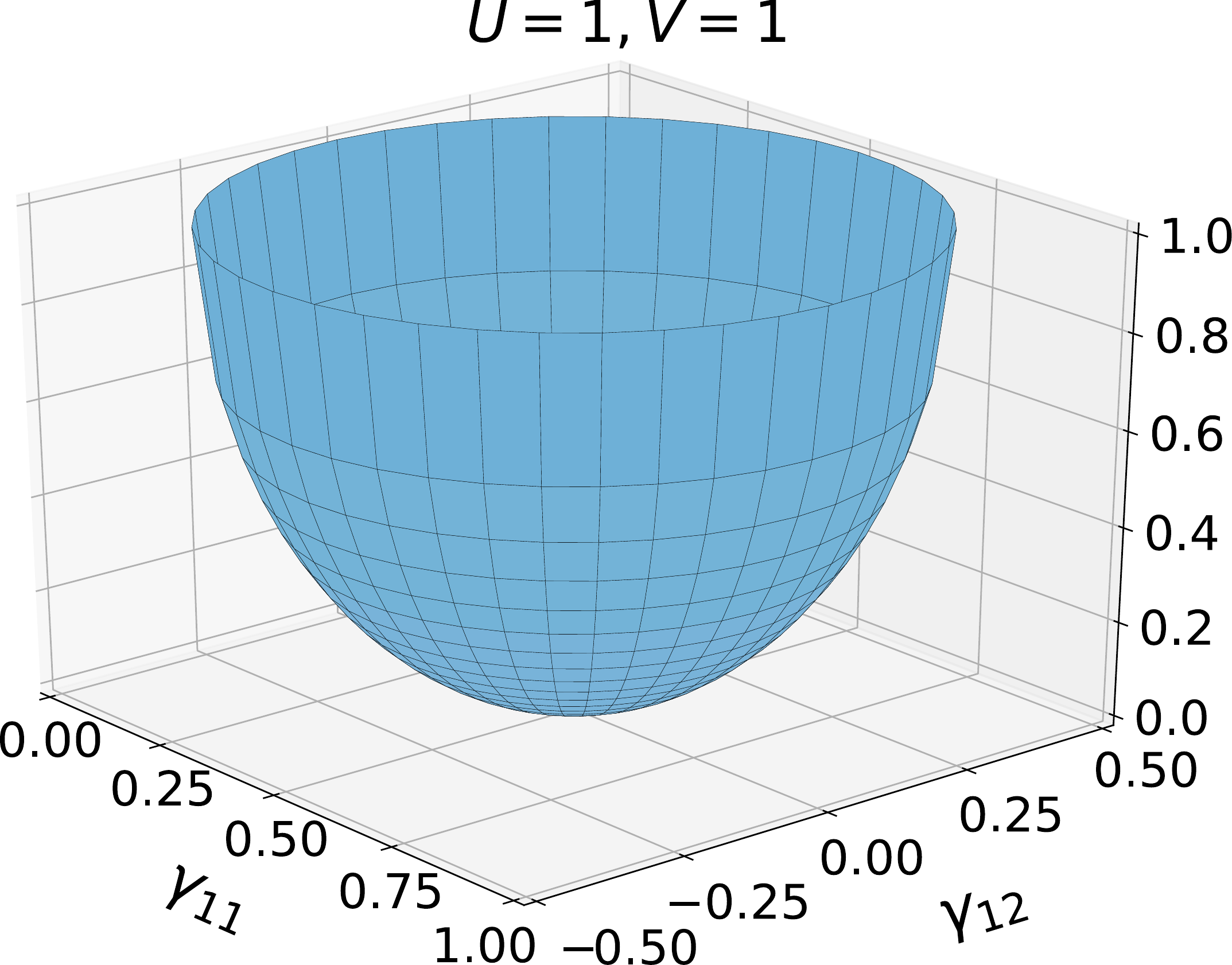}}&
\thead{\includegraphics[width=0.2\linewidth]{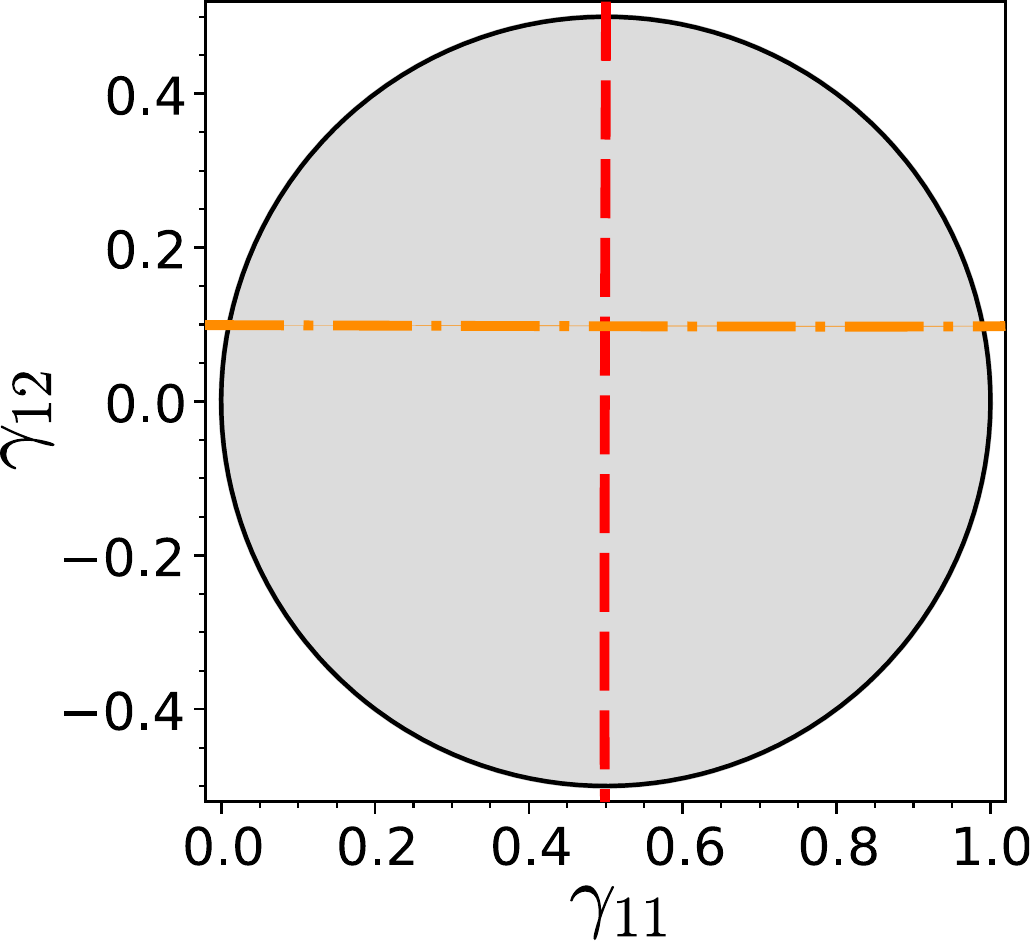}}&
\thead{\includegraphics[width=0.24\linewidth]{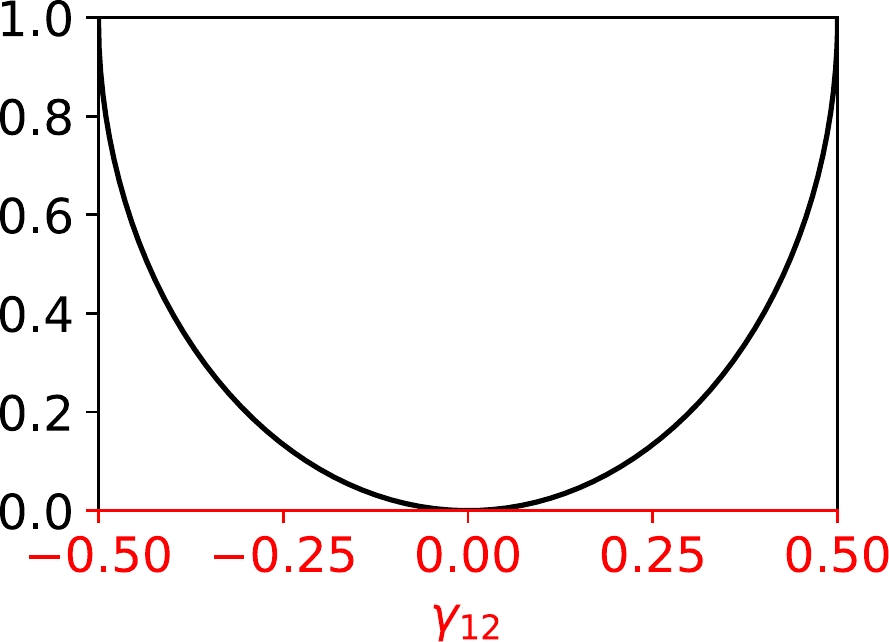}}&
\thead{\includegraphics[width=0.24\linewidth]{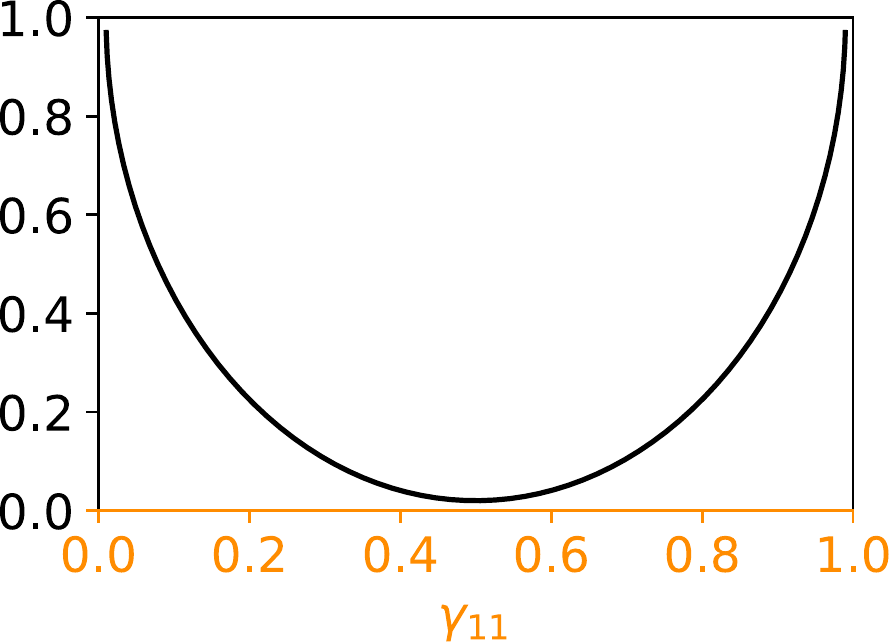}}\\
\thead{\includegraphics[width=0.24\linewidth]{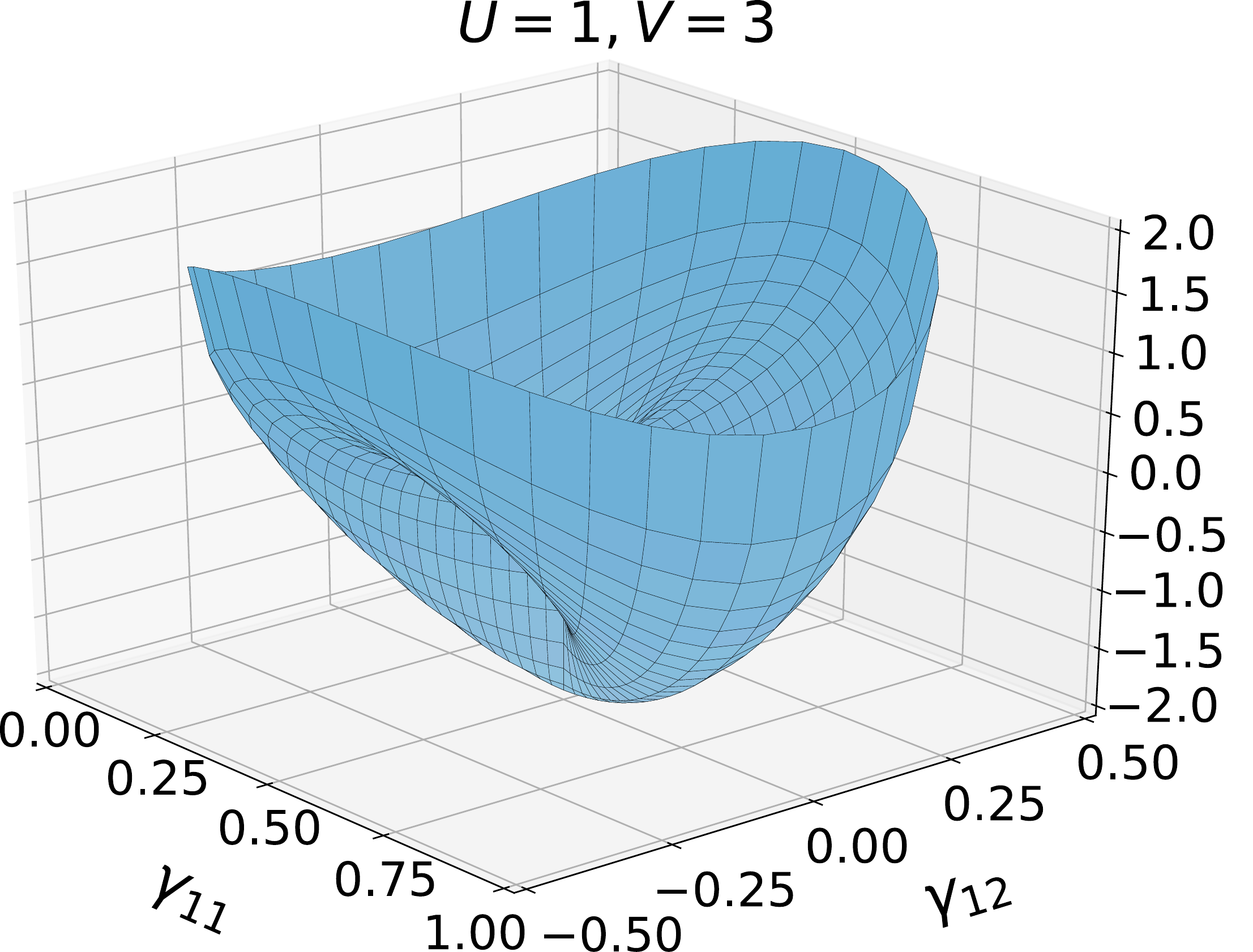}}&
\thead{\includegraphics[width=0.2\linewidth]{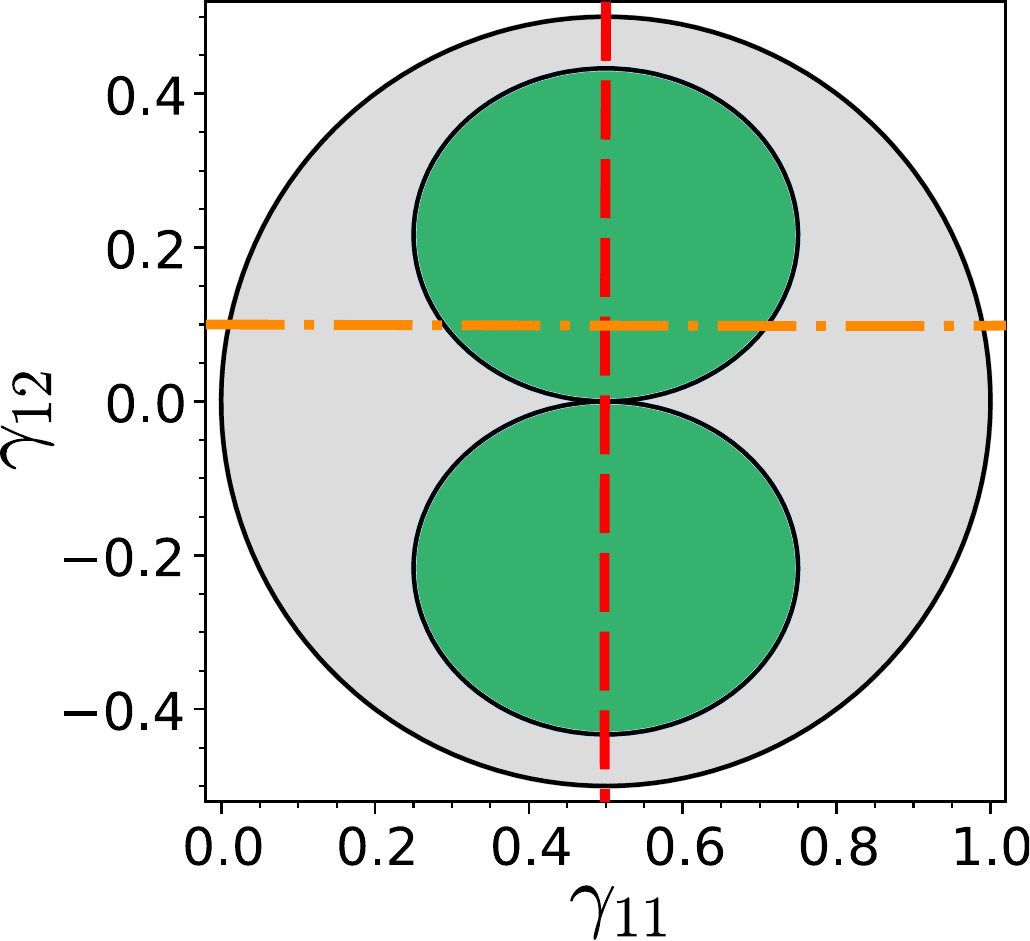}}&
\thead{\includegraphics[width=0.24\linewidth]{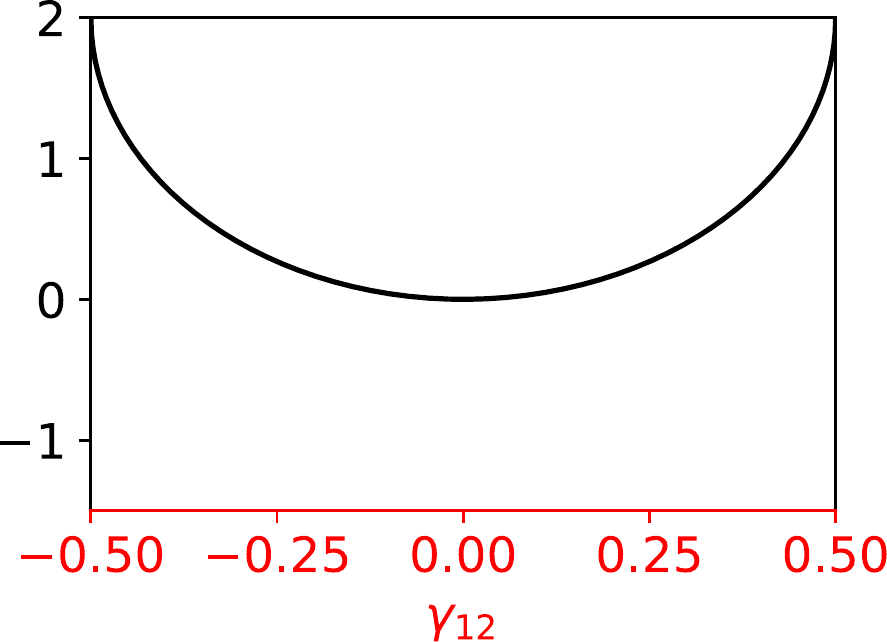}}&
\thead{\includegraphics[width=0.24\linewidth]{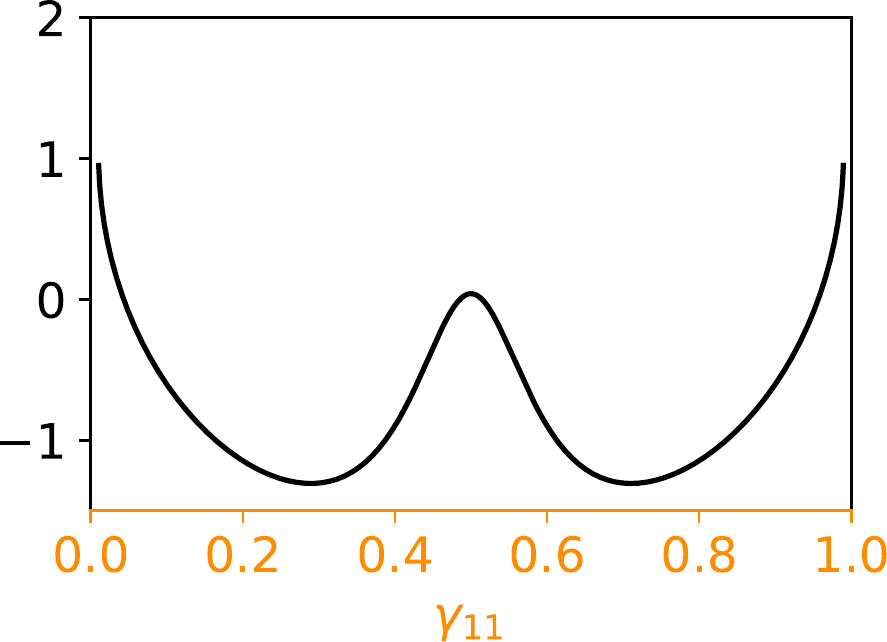}}\\
\thead{\includegraphics[width=0.24\linewidth]{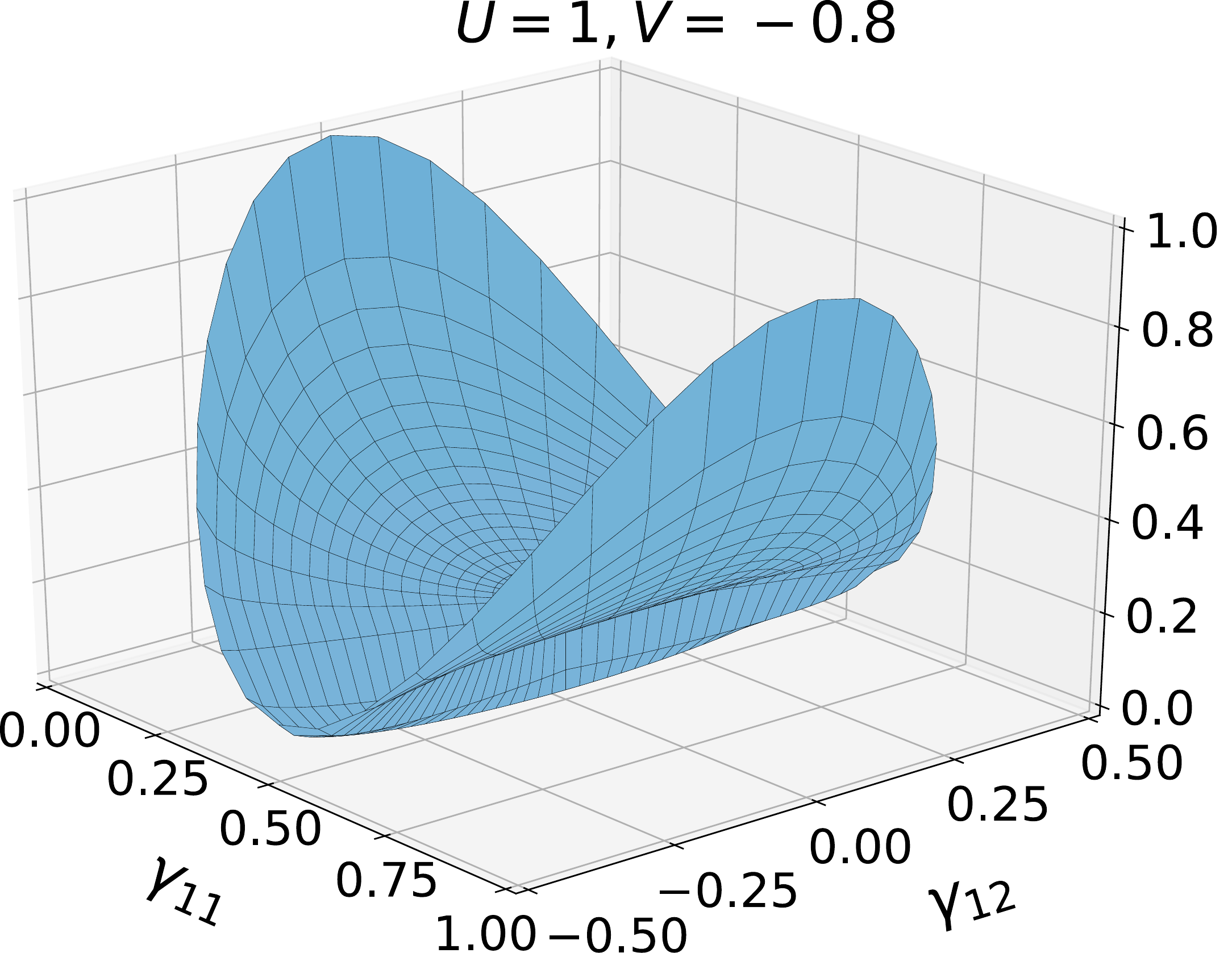}}&
\thead{\includegraphics[width=0.2\linewidth]{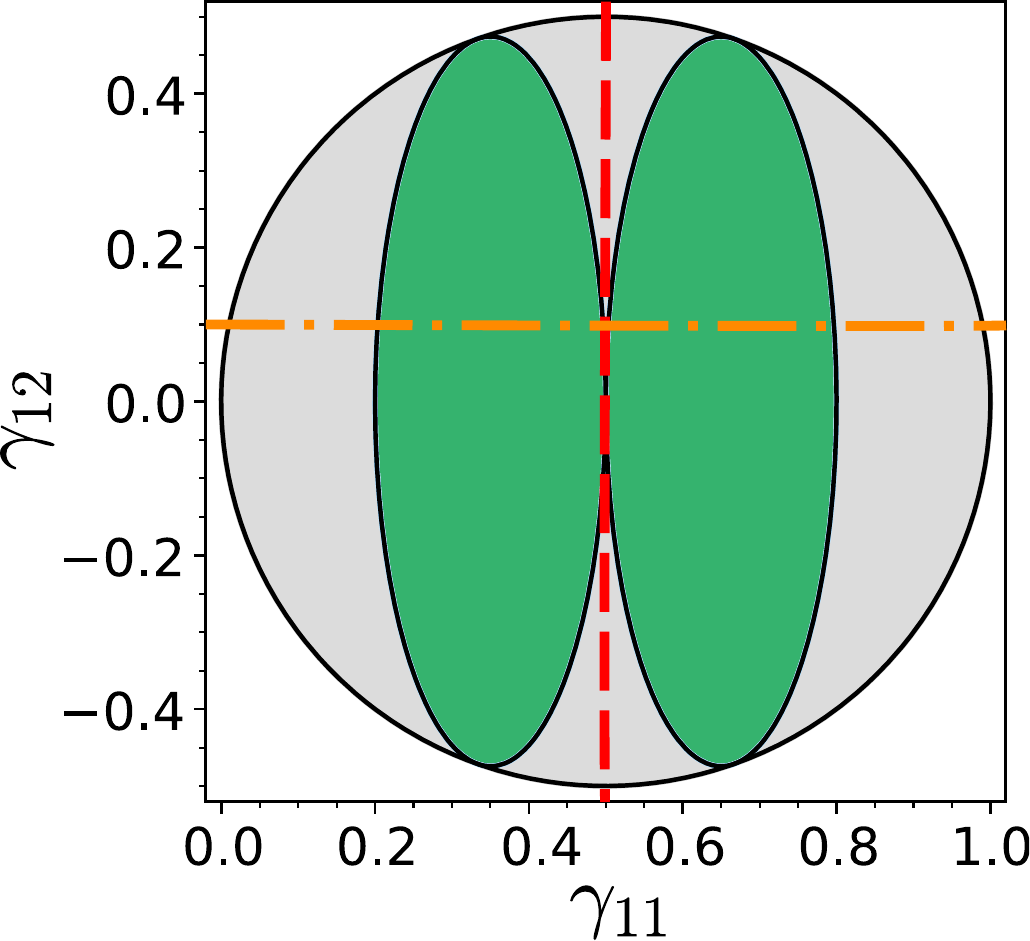}}&
\thead{\includegraphics[width=0.24\linewidth]{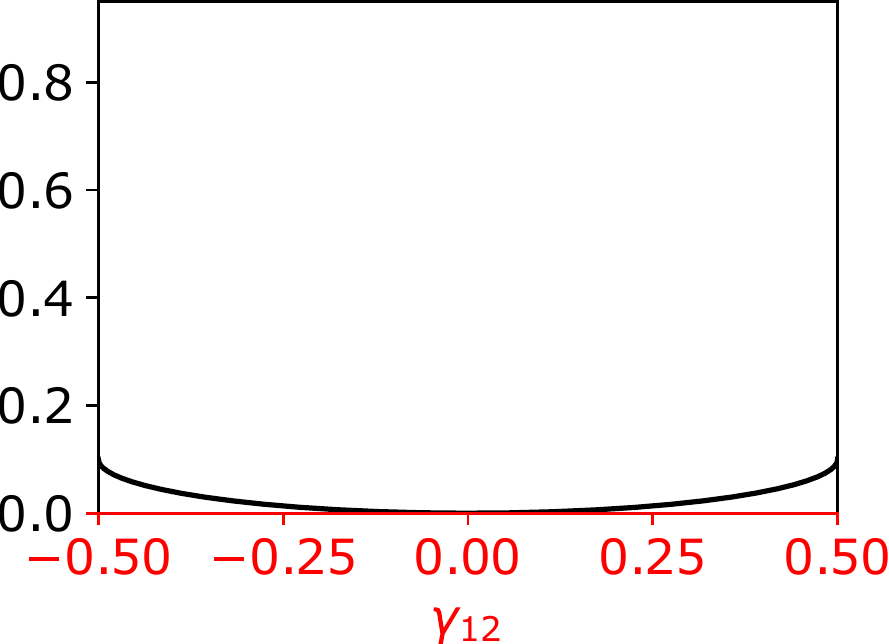}}&
\thead{\includegraphics[width=0.24\linewidth]{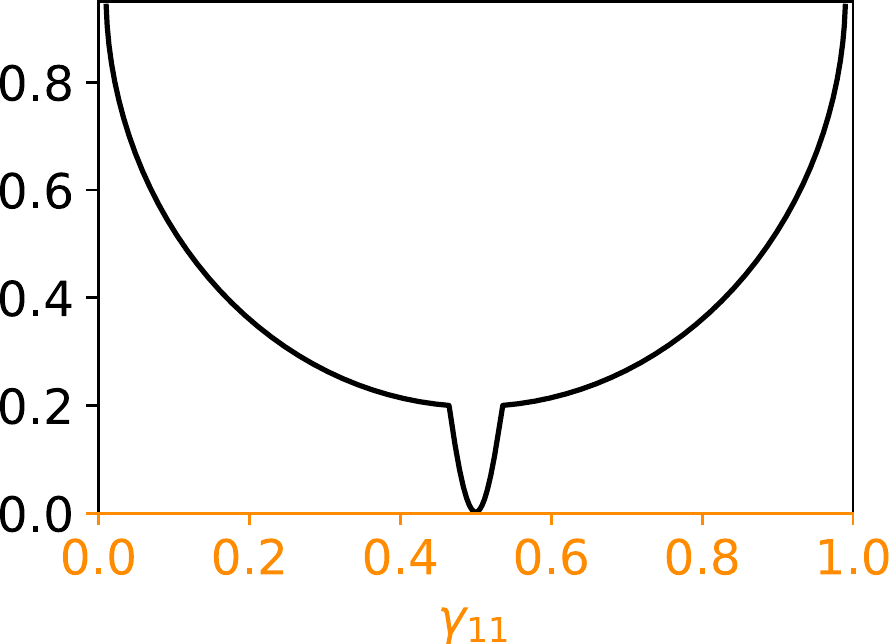}}\\
\thead{\includegraphics[width=0.24\linewidth]{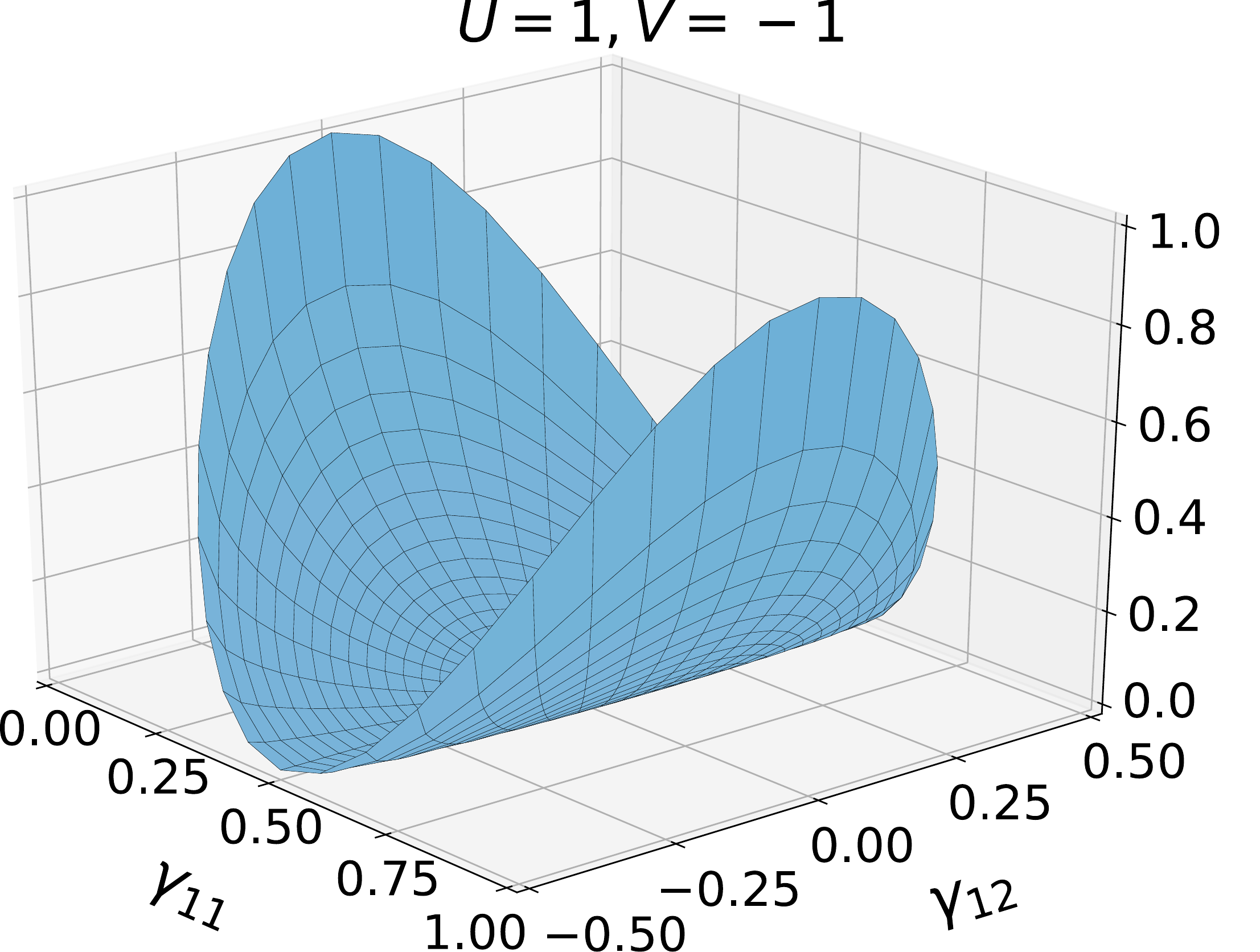}}&
\thead{\includegraphics[width=0.2\linewidth]{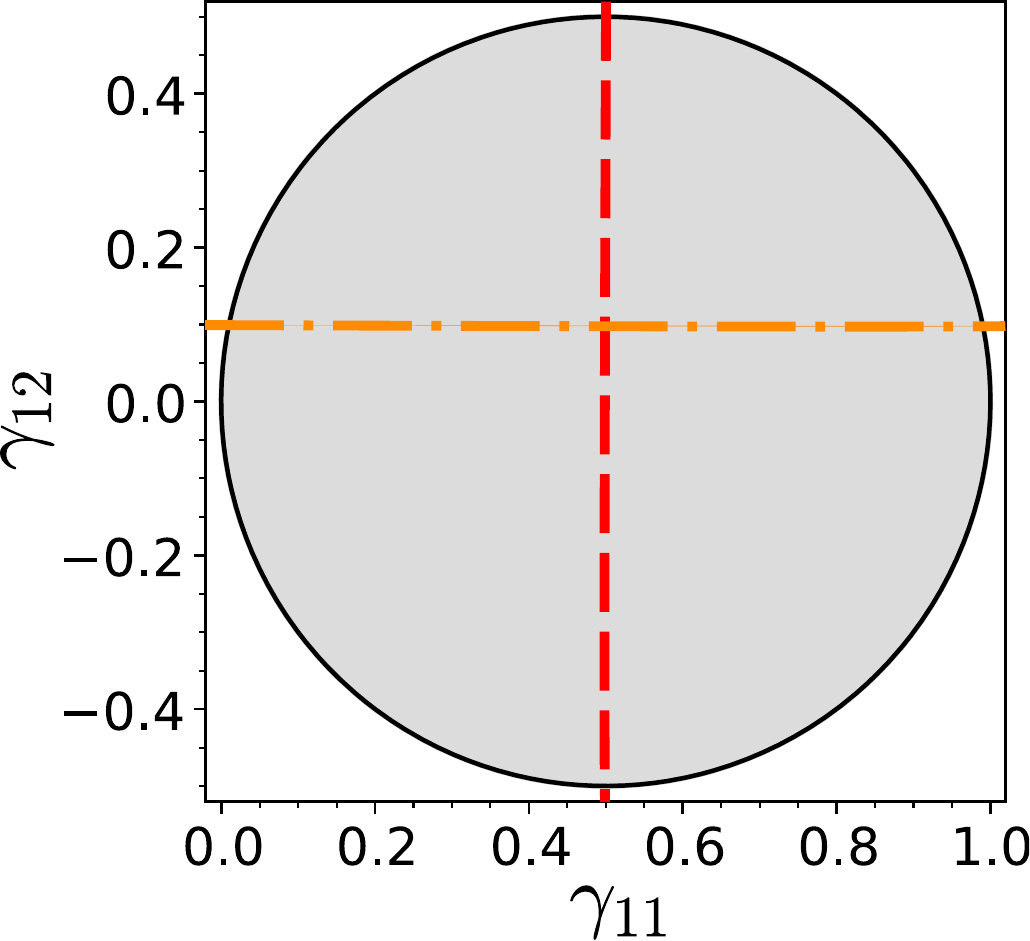}}&
\thead{\includegraphics[width=0.24\linewidth]{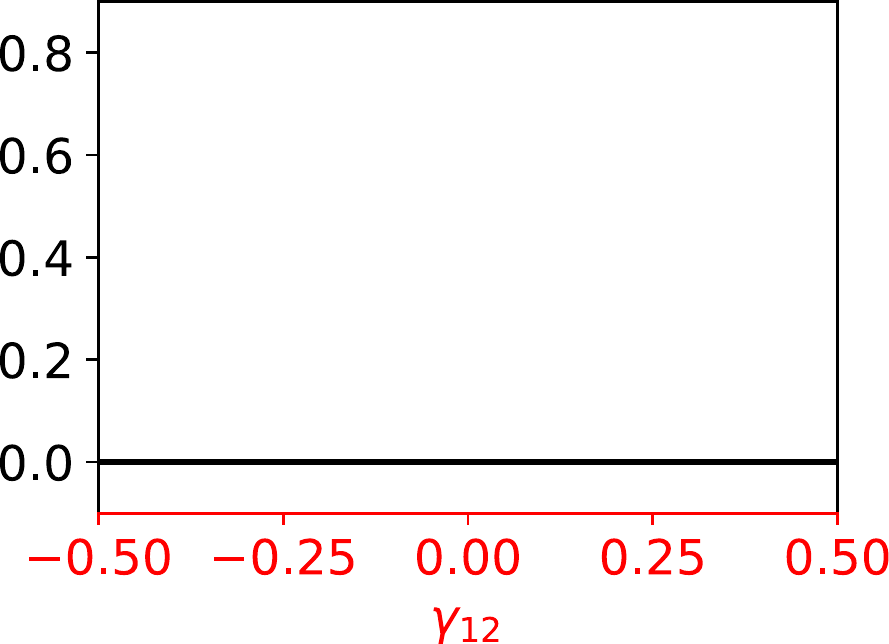}}&
\thead{\includegraphics[width=0.24\linewidth]{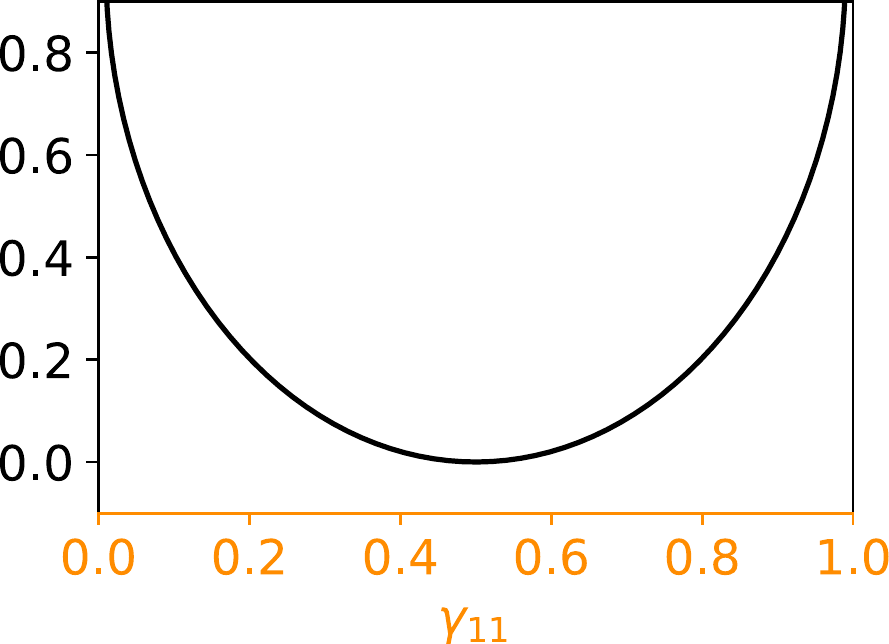}}\\
\thead{\includegraphics[width=0.24\linewidth]{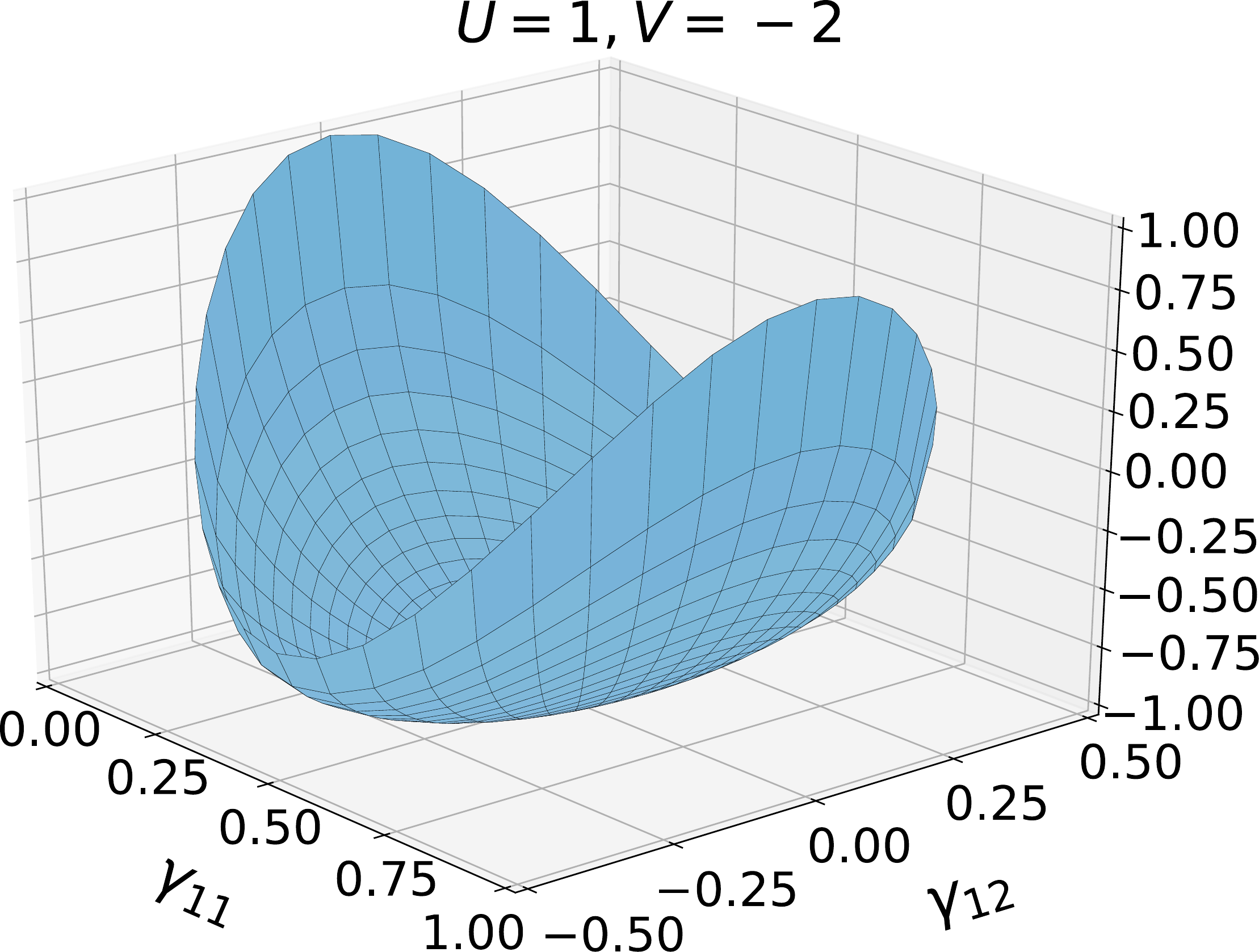}}&
\thead{\includegraphics[width=0.2\linewidth]{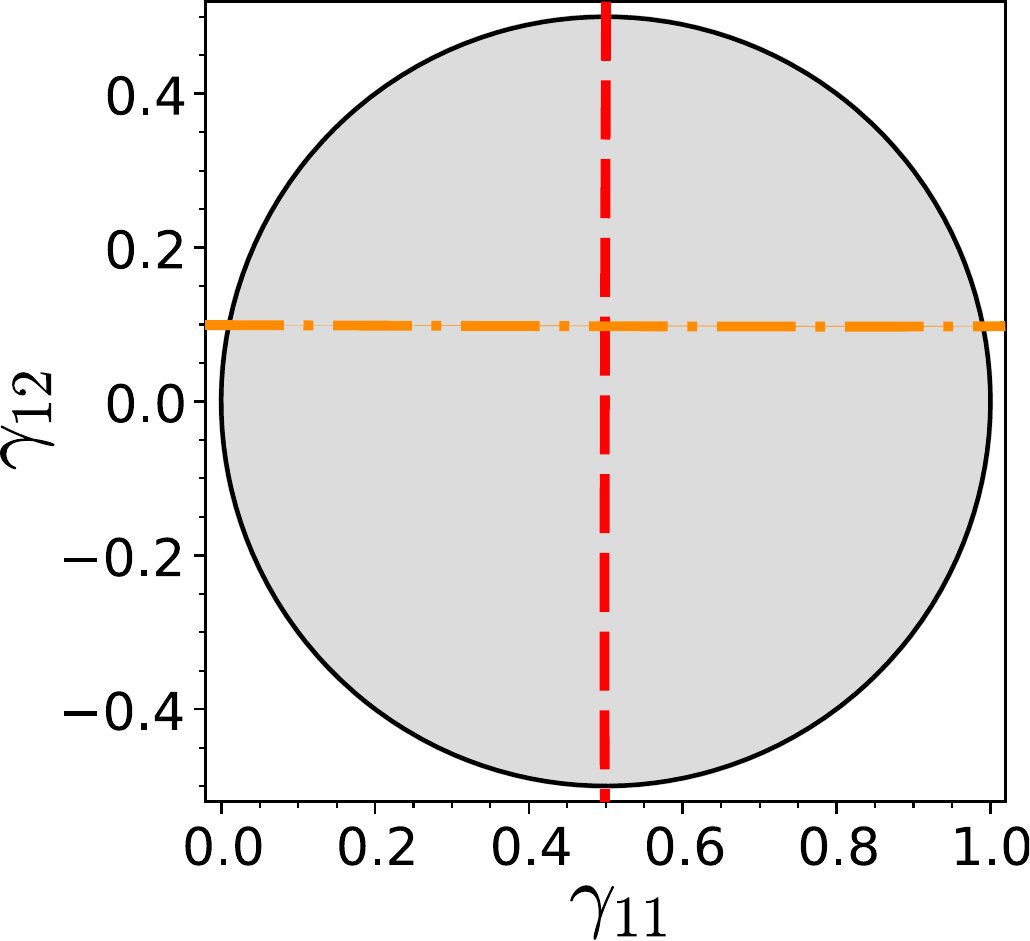}}&
\thead{\includegraphics[width=0.24\linewidth]{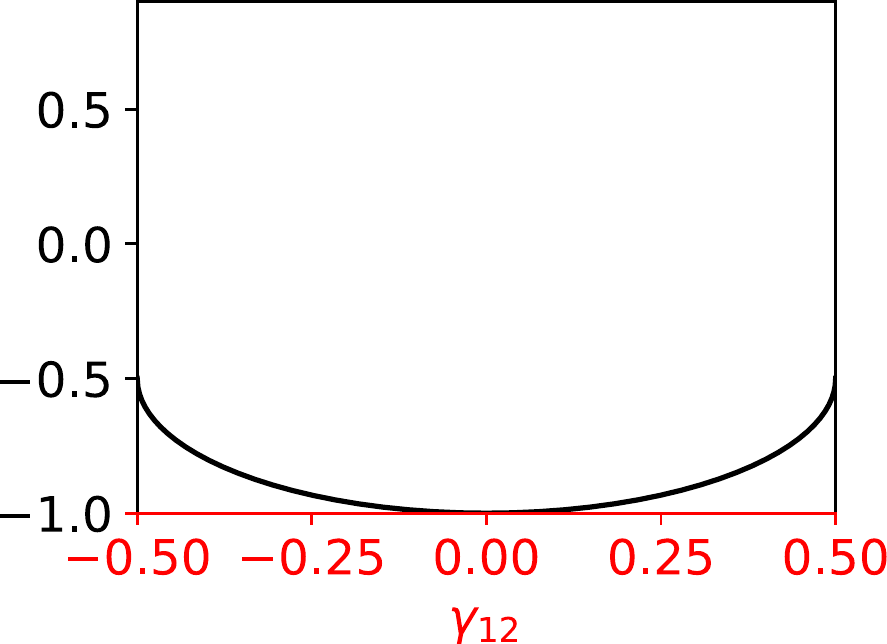}}&
\thead{\includegraphics[width=0.24\linewidth]{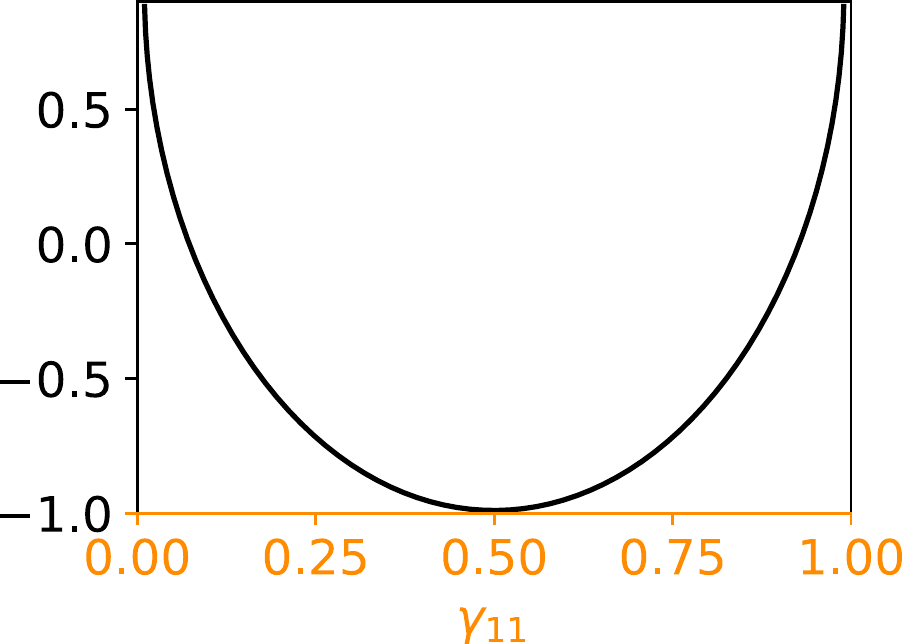}}
\end{tabular}
\caption{Illustration of the universal functional $\mathcal{F}^{(p)}_\RR$ and next to it the corresponding pure state $v$-representable 1RDMs (grey) as well as two two-dimensional slices of $\mathcal{F}^{(p)}_\RR$ for different values of $U, V$ ($X=0$). The 1RDMs that are not pure state $v$-representable are marked in green. The dashed lines in the second column depict the values of $\gamma_{12}$ and $\gamma_{11}$ that were fixed in the respective plots in the third and fourth column. \label{fig:FRR_arbitrary}}
\end{table*}

\subsection{v-representability in the generalized Hubbard dimer\label{sec:vrepr-gen}}

In this section, we solve the pure state $v$-representability problem for the generalized Hubbard dimer.
According to Table \ref{fig:FRR_arbitrary}, the functional $\mathcal{F}_\RR^{(p)}$ is not convex for most pairs of $U$ and $V$. To further illustrate the pure state $v$-representability we present next to each functional a plot of its domain indicating the 1RDMs $\gamma$ which are not pure state $v$-representable in green. The pure state $v$-representable 1RDMs are shown in grey.
As shown by numerical means in Ref.~\onlinecite{Neck2001}, the non-$v$-representable regions depend on the interaction $W$ and thus change as a function of the free parameters $U, V$.
Recall that we set $X=0$ in Table \ref{fig:FRR_arbitrary} since this will not affect any results or insights.
In particular, a non-vanishing $X$ does not modify the leading order of the exchange force discussed in Sec.~\ref{sec:force} since the respective term in $\mathcal{F}_{\RR}^{(p)}$ is linear in $R$. The two-dimensional slices of $\mathcal{F}_\RR^{(p)}$ in Table \ref{fig:FRR_arbitrary} were obtained from $\mathcal{F}_\RR^{(p)}$ by fixing $\gamma_{11}=0.5$ in the third column and $\gamma_{12}=0.1$ in the fourth column (counting from the left hand side). Despite the similarity to the schematic illustration in Fig.~\ref{fig:F_schematic}, it is in general \textit{not} possible to infer $v$-representability of a 1RDM $\gamma$ from a lower-dimensional slice of the functional. This manifests itself in the fact that $v$-representability is indeed a \textit{global} property of the universal functional.

For the equation of the ellipses restricting the non-$v$-representable 1RDMs $\gamma$, we obtain for $U>V,X=0$ (see Appendix \ref{app:ellipses}),
\begin{equation}
\frac{\gamma_{12}^2 }{2\left(1-\frac{V}{U}\right)}+ \frac{\left(|\gamma_{11}-\frac{1}{2}| - \frac{1}{4}\sqrt{1-\left(\frac{V}{U}\right)^2} \right)^2}{1 - \left(\frac{V}{U}\right)^2 } =\frac{1}{16}
\end{equation}
and in the case of $U<V$ ($X=0$) we have
\begin{eqnarray}
&& \frac{\left(3-\frac{U}{V}\right)^2\left(|\gamma_{12}| -\frac{1}{2}\sqrt{\frac{2\left(1-\frac{U}{V}\right)}{\left(3-\frac{U}{V}\right)^2}} \right)^2}{2\left(1-\frac{U}{V}\right)}  + \frac{\left(3-\frac{U}{V}\right)\left(\gamma_{11}-\frac{1}{2}\right)^2}{4\left(1-\frac{U}{V}\right)} \nonumber\\
&&= \frac{1}{16} \,.
\end{eqnarray}
There are two main differences with respect to the ordinary Hubbard dimer discussed in Sec.~\ref{sec:HubbardDimer} (compare also Fig.~\ref{fig:vrepr-dimer} and Table \ref{fig:FRR_arbitrary}). First, the ellipses restricting the set of non-$v$-representable 1RDMs $\gamma$ can change in size and move inside the disk such that they do not touch its boundary anymore. In addition, these ellipses can be rotated by $90$ degrees. Second, they can touch the boundary at four instead of two or zero points depending on the two parameters $U$ and $V$. As we will prove in Sec.~\ref{sec:force}, these are the only boundary points where the exchange force vanishes.


\subsection{Exchange force \label{sec:force}}


Based on the analytic expression for the universal functional obtained in Eq.~\eqref{eq:FRR_polar}, we prove in the following the existence of a fermionic exchange force \cite{Schilling2019} close to the boundary of the domain $\mathcal{P}_2^1$ of $\mathcal{F}_\RR^{(p)}$ for an arbitrary isotropic, reflection symmetric pair interaction \eqref{eq:W_generic}. Taking the derivative of $\mathcal{F}_\RR^{(p)}$ with respect to the distance $R$, as introduced in Fig.~\ref{fig:disk}, yields to leading order for small $R$,
\begin{equation}\label{eq:dFdD}
\frac{\partial \mathcal{F}_\RR^{(p)}}{\partial R} = -\frac{|\mathrm{sin}^2(\varphi)(V-U) - 2V|}{2}\frac{1}{\sqrt{R}} + \mathcal{O}(R^0)\,.
\end{equation}
The overall minus sign of the leading term ensures that the exchange force is always \textit{repulsive}.
Thus, we indeed find as expected that the gradient of the universal functional diverges repulsively at the boundary of the set $\mathcal{P}^1_N$. Since $\mathcal{F}_\RR^{(e)}=\mathrm{conv}(\mathcal{F}_\RR^{(p)})$ the same holds true for the ensemble functional $\mathcal{F}_\RR^{(e)}$. These findings therefore confirm the existence of the fermionic exchange force also in systems without translational symmetry.

Intriguingly, the prefactor of the $1/\sqrt{R}$ divergence for $R\to 0$ contains crucial information about the microscopic details and thus provides insights into the system-specific properties: By considering different angles $\varphi$ one could apparently extract the values of the two coupling parameters $U,V$. In will be one of the promising future challenges to understand how this key finding generalized to larger systems, with an emphasis on the Coulomb interaction.

Moreover, as it has been explained in Sec.~\ref{sec:v-rep-general}, the fermionic exchange force implies that whenever $\partial\mathcal{F}/\partial\gamma$ diverges in \eqref{eq:dFdD}, the corresponding 1RDMs on the boundary $\partial\mathcal{P}^1_N/ \partial\mathcal{E}^1_N$ are not pure/ensemble state $v$-representable.
For the Hubbard dimer with on-site interaction discussed in Sec.~\ref{sec:HubbardDimer}, the leading order term in Eq.~\eqref{eq:dFdD} reduces to $\partial \mathcal{F}_\RR^{(p)}/\partial R = -U\mathrm{sin}^2(\varphi)/2\sqrt{R} + \mathcal{O}(R^0)$.
Furthermore, we observe that for specific choices of $U$ and $V$ for the generalized Hubbard dimer we can find an angle $\varphi$ such that the prefactor in front of the $1/\sqrt{R}$ divergence vanishes. Solving $\mathrm{sin}^2(\varphi)(V-U) - 2V=0$ for the angle $\varphi$ yields in total four solutions,
\begin{equation}\label{eq:varphi1k}
\varphi = \pm \arcsin\left(\sqrt{\frac{2V}{V-U}}\right) + 2\pi m \,,\quad m\in \ZZ
\end{equation}
and
\begin{equation}\label{eq:varphi2k}
\varphi = \pi \pm \arcsin\left(\sqrt{\frac{2V}{V-U}}\right) + 2\pi m \,,\quad m\in \ZZ\,.
\end{equation}
Thus, it is possible to reach the boundary of $\mathcal{P}_2^1$ on either zero, two or four points depending on the values of $U$ and $V$ (c.f.~Eqs.~\eqref{eq:varphi1k} and \eqref{eq:varphi2k}). Equivalently, the gradient of the universal functional does not diverge at those points. Whenever a solution for $\varphi\in \RR$ exists the following must hold
\begin{equation}
|V|\leq |U|\,.
\end{equation}
For $U\geq 0$ and $V\geq 0$ this can only be satisfied for $V=0$ which in turn leads to $\varphi =0, \pi$, in agreement with the result for the ordinary Hubbard dimer discussed in Sec.~\ref{sec:HubbardDimer}. For $U\geq 0$ and negative $V$ we obtain the restriction $V\in [-U, 0]$ and in this case there are four points where the prefactor in Eq.~\eqref{eq:dFdD} vanishes. The same holds for $U\leq 0$ and $V\in[0, |U|]$, whereas for $U\leq 0$ and $V< 0$ we obtain no valid solution for $\varphi$.

Last but not least, it is worth noticing that the prefactor in front of the $1/\sqrt{R}$ divergence in \eqref{eq:dFdD} can only vanish for some boundary points if one of the ellipses describing the subsets of non-pure state $v$-representable 1RDMs touches it. Thus, those touch points are indeed pure state $v$-representable in the sense that they can be obtained as ground state 1RDMs of an, in this case degenerate, Hamiltonian $H(h)=h+W$. The same holds true in the context of ensemble $v$-representability.

\section{Summary and conclusions}
Our work has advanced the foundation of one-particle reduced density matrix functional theory (1RDMFT) by refining, relating and illustrating some of its fundamental features and underlying concepts.

In the first part, we have formalised the \emph{scope} of a functional theory by identifying it with an affine space of Hamiltonians $H(h)= h +W$ of interest. Addressing the ground state problem exclusively for that class of systems --- as it is indeed done in each scientific subfield --- leads immediately in virtue of the Rayleigh-Ritz variational principle to a universal functional. This more general perspective on functional theory has the advantage that the functional variable can be identified in a concise manner through the Riesz representation theorem. It is given by the unique Riesz vector, i.e., the simplest possible reduced state that still allows one to calculate the expectation value of any $h$. In particular, this reasoning also explains how the functional variable could be simplified if the one-particle Hamiltonian $h$ exhibits further symmetries, or more generally, is restricted to a subspace. Due to its practical relevance, we applied these fundamental considerations to Hamiltonians with (conventional) time-reversal symmetry. This means nothing else than that the scope of the 1RDMFT is restricted to  real-valued matrices $h$. Following our proposed  paradigm of irreducibility based on Riesz' representation theorem this offers the opportunity to restrict the functional variable from the complex-valued 1RDM $\gt$ to its real part $\g \equiv \Re(\gt)$. In that case, one could even further reduce 1RDMFT by restricting the constrained search formalism to real-valued $N$-particle quantum states. These options and the choice between Levy/pure and Valone/ensemble 1RDMFT yields in total six equivalent universal functionals which are all listed and characterized in Fig.~\ref{fig:choices}. Most importantly, all these functionals are related to each other in concise mathematical terms according to Fig.~\ref{fig:functionals-general}.

In complete analogy to the functional theory, also the notion of $v$-representability is a relative concept. As it is illustrated in Fig.~\ref{fig:v-rep}, it refers as well to the underlying scope, variable, and the choice between pure/ensemble and real/complex $N$-particle quantum states. Last but not least, in Sec.~\ref{sec:v-rep-general} we exploited the geometric interpretation of the Legendre-Fenchel transformation to relate the notion of $v$-representability to the form of the corresponding universal functional. To be more specific, the comparison of a universal pure and ensemble functional identifies the non-pure state $v$-representable 1RDMs in the interior of the domain, while generic points on the boundary are expected to be never $v$-representable due to the fermionic exchange force.

Due to the rigorous and more universal character of our approach, various definitions, insights and findings could in principle also be translated into the context of density functional theory (DFT).  When restricting the affine space of one-particle Hamiltonians to $h_{t}(v)\equiv t+v$ with \emph{fixed} kinetic energy operator $t$ and variable external potential $v$, our approach identifies immediately the particle density as the natural variable and thus establishes DFT. It is worth noticing, however, that one of the conceptual facets of our work on 1RDMFT does not appear in DFT: Since the particle density is always real-valued by definition, the natural variable is unambiguous and the choice of referring to complex or real numbers would therefore affect only the functional but not its variable. In that sense, such considerations could complement related studies in DFT on $v$-representability, and in particular the potential-density mapping\cite{UK02,Penz22, PL23}.

In the second part of our work, we then discussed and illustrated all these conceptual aspects for the ordinary Hubbard dimer model and a generalization thereof. In particular, the latter allowed us to systematically explore and confirm the striking dependence of various fundamental features on the pair-interaction $W$. For this, we first derived by analytical means closed formulas for all six universal functionals for the Hubbard dimer (c.f Fig.~\ref{fig:choices}) and revealed concise relations among them (c.f Fig.~\ref{fig:functionals}). In particular, we proved the equivalence of the two functionals $\mathcal{F}^{(p)}_\CC(\g)$ and $\mathcal{F}^{(e)}_\RR(\g)$ (see Sec.~\ref{sec:functional-onsite-complex}), a relation that was conjectured by numerical means in Ref.~\onlinecite{Cohen2016}. Since our proof is merely based on the geometry of quantum states and does not refer to any specific interaction, it is equally valid for the generalized Hubbard dimer in Sec.~\ref{sec:generic}.
This result leads to an important insight: adding `unnecessary' degrees of freedom in the constrained search formalism with pure states  simulates a certain degree of mixedness. Indeed, since the interaction $W$ is assumed to not depend on the extra degrees of freedom, one can trace them out which in turn leads to a mixed state.
Moreover, according to Sec.~\ref{sec:v-rep-general}, the comparison of all six functionals then allowed us to solve each variant of the $v$-representability problem. For instance, since $\mathcal{F}^{(p)}_\CC(\g)$ was found to be convex, all $\g$ are complex-pure state $v$-representable, while the same is not true for the full complex-valued 1RDM $\gt$.
%

For the generalized dimer, we could derive closed formulas for the four universal functionals which depend on the reduced variable $\g \equiv \Re(\gt)$, in particular $\mathcal{F}^{(p)}_\RR(\g)$. All six universal functionals obey the same relations as for the ordinary dimer (c.f Fig.~\ref{fig:functionals}). The corresponding $v$-representability problems could therefore be solved again in a straightforward manner and we confirmed conclusively by analytical means the strong influence of the pair interaction $W$ on their solution. Intriguingly, the sets of non-pure state $v$-representable 1RDMs were found to rotate and change in size.

Last but not least, the closed formulas of the universal functionals allowed us to conclusively confirm  the existence of the fermionic exchange force also for systems without translational symmetry. In particular, the prefactor of its universal diverging behaviour at the boundary of the domain depends on $W$. This crucial observation, also in combination with our other findings on the $v$-representability problem, raises the following far-reaching questions in the context of larger quantum systems: (i) Which information about the system ($W$) does the diverging fermionic exchange force provide and would it be possible to experimentally access it? (ii) How does the position, shape and topological structure of the set of non-$v$-representable 1RDMs reflect crucial features of the quantum system?

\begin{acknowledgments}
We thank E.K.U.\hspace{0.5mm}Gross for inspiring discussions, C.L.\hspace{0.5mm}Benavides-Riveros for helpful comments on the manuscript and D.P.\hspace{0.5mm}Kooi for bringing Ref.~\onlinecite{V-bachelor} to our attention. We acknowledge financial support from the Deutsche Forschungsgemeinschaft (Grant SCHI 1476/1-1) (A.Y.C, J.L and C.S.), the Munich Center for Quantum Science and Technology (C.S.) and the International Max Planck Research School for Quantum Science and Technology (IMPRS-QST) (J.L.). The project/research is also part of the Munich Quantum Valley, which is supported by the Bavarian state government with funds from the Hightech Agenda Bayern Plus.
\end{acknowledgments}

\appendix
\section{Proof of $\widetilde{\mathcal{F}}_\CC^{(p)}(\tilde{\gamma}) = \mathcal{F}_\RR^{(p)}(\tilde{\gamma}_{11},|\tilde{\gamma}_{12}|)$ for on-site interaction\label{app:Ftilde}}
In this section, we prove the relation $\widetilde{\mathcal{F}}_\CC^{(p)}(\tilde{\gamma})=\mathcal{F}_\RR^{(p)}(\tilde{\gamma}_{11},|\tilde{\gamma}_{12}|)$ for the Hubbard dimer with on-site interaction. The reference basis in the singlet spin sector consists of the three orthonormal states $\ket{\Phi_1} = c_{1\uparrow}^\dagger c_{1\downarrow}^\dagger\ket{0}$, $\ket{\Phi_2} = c_{2\uparrow}^\dagger c_{2\downarrow}^\dagger\ket{0}$ and $\ket{\Phi_3} = (c_{1\uparrow}^\dagger c_{2\downarrow}^\dagger - c_{1\downarrow}^\dagger c_{2\uparrow}^\dagger)\ket{0}/\sqrt{2}$, where $\ket{0}$ denotes the vacuum state.
Then, the functional
\begin{equation}
\widetilde{\mathcal{F}}^{(p)}_\CC(\gt) = \min_{\ket{\Psi}\mapsto\gt} \bra{\Psi}W\ket{\Psi}
\end{equation}
is obtained by minimizing the expectation value of the interaction $W$ over all $2$-fermion wave functions
\begin{equation}\label{Psicomplex}
\ket{\Psi} = a\ket{\Phi_1} + b\ket{\Phi_2} + c\ket{\Phi_3}
\end{equation}
with $a, b, c\in \CC$. Since
\begin{equation}\label{eq:Wvsc2}
\bra{\Psi}W\ket{\Psi} = U(1 - |c|^2)
\end{equation}
is invariant under a change of the global phase of the state $\ket{\Psi}$, we assume w.l.o.g.~$c\in \RR$. Therefore, we are left with a minimization of $\bra{\Psi}W\ket{\Psi}$ over all $\ket{\Psi}$ involving five free parameters, namely the moduli $|a|, |b|$, the phases $\varphi_a, \varphi_b$ and the real parameter $c$, under the three constraints
\begin{eqnarray}
1 &=& |a|^2 + |b|^2 + c^2 \label{eq:cond1} \\
\gt_{11} &=& |a|^2 + \frac{c^2}{2} \label{eq:cond2}\\
\gt_{12} &=& \frac{c}{\sqrt{2}}\left(a + b^*\right)\,, \label{eq:cond3}
\end{eqnarray}
where $b^*$ is the complex conjugate of $b$. In the next step, we make use of the fact that the universal functional $\widetilde{\mathcal{F}}^{(p)}_\CC$ does not dependent on the complex phase of $\gt_{12}$.
This follows directly from the fact that any complex phase of $\gt_{12}$ could be absorbed into $a$ and $b$ in Eq.~\eqref{eq:cond3}, while such a change of $a$ or $b$ does neither affect the other two constraints \eqref{eq:cond1}, \eqref{eq:cond2} not the interaction energy \eqref{eq:Wvsc2}.
Consequently, the third condition \eqref{eq:cond3} can be reduced to
\begin{eqnarray}\label{eq:g12s_phi}
&&|\gt_{12}|^2 = \frac{c^2}{2}\left(|a|^2 + |b|^2 +2\Re(ab)\right)  \\
&&= \frac{c^2}{2}\left(1 - c^2 + \mathrm{cos}(\varphi_{ab})\sqrt{4\gt_{11}(1 - \gt_{11})+ c^2(c^2 - 2)}\right)\nonumber\,,
\end{eqnarray}
where we used Eqs.~\eqref{eq:cond1} and \eqref{eq:cond2} in the second line and introduced the new variable $\varphi_{ab} = \varphi_a + \varphi_b$. Eq.\eqref{eq:g12s_phi} can be rewritten as follows,
\begin{eqnarray}\label{eq:c2-expanded}
&&c^8\sin^2(\varphi_{ab}) - 2c^6\sin^2(\varphi_{ab}) - c^4\left(4\gt_{11}(1 - \gt_{11})\cos^2(\varphi_{ab})\right.\nonumber \\
&&\left. - 4|\gt_{12}|^2 - 1\right) - 4c^2|\gt_{12}|^2 + 4|\gt_{12}|^4 =0\,.
\end{eqnarray}
Therefore, we have to find the $c^2$ which minimizes $\bra{\Psi}W\ket{\Psi} = U(1 - c^2)$. This is achieved by first differentiating Eq.~\eqref{eq:c2-expanded} with respect to $\varphi_{ab}$, which leads to
\begin{widetext}
\begin{equation}\label{eq:dc2dphi}
\frac{\mathrm{d} c^2}{\mathrm{d}\varphi_{ab}} = \frac{c^2\sin(\varphi_{ab})\left(c^4 - 2c^2 + 4\gt_{11}(1 - \gt_{11})\right)}{\cos(\varphi_{ab})\left(2c^4 - 3c^2 + 4\gt_{11}(1 - \gt_{11})\right) + (1 - 2c^2)\sqrt{4\gt_{11}(1 - \gt_{11}) + c^2(c^2 - 2)}}\,,
\end{equation}
\end{widetext}
and, second, using the above expression to determine the extremal points of $c^2$ as a function of $\varphi_{ab}$. The right hand side of Eq.~\eqref{eq:dc2dphi} is equal to zero for $\varphi_{ab} = 0, \pi (+ 2\pi m, m \in \ZZ)$. For both solutions for $\varphi_{ab}$, Eq.~\eqref{eq:dc2dphi} yields
\begin{equation}\label{eq:c2pm}
c^2_\pm = \frac{|\gt_{12}|^2\left(1 \pm \sqrt{1 - 4\left[|\gt_{12}|^2 + \left(\gt_{11} - \frac{1}{2}\right)^2\right]} \right)}{2\left[|\gt_{12}|^2 + \left(\gt_{11} - \frac{1}{2}\right)^2\right]}\,.
\end{equation}
Thus, for both $\varphi_{ab} = 0, \pi $ the solution $c^2_+$ minimizes the expectation value of the interaction for $U\geq 0$ and the solution $c^2_-$ for $U\leq 0$. This then immediately leads to
\begin{widetext}
\begin{eqnarray}
\widetilde{\mathcal{F}}_{\CC}^{(p)}(\gt) = U\frac{(\gt_{11}-\tfrac{1}{2})^2+\tfrac{1}{2}|\gt_{12}|^2[1 - \mathrm{sgn}(U)\sqrt{1-4(\gt_{11}-\tfrac{1}{2})^2-4|\gt_{12}|^2}]}{(\gt_{11}-\tfrac{1}{2})^2+|\gt_{12}|^2}\,.
\end{eqnarray}
\end{widetext}
By comparing the above result to the well-known expression for $\mathcal{F}^{(p)}_\RR(\g)$\cite{Pastor2011a, Cohen2016} (see Eq.~\eqref{eq:FRR_1} in the main text), we observe that
\begin{equation}
\widetilde{\mathcal{F}}^{(p)}_\CC(\gt) = \mathcal{F}^{(p)}_\RR(\gt_{11}, |\gt_{12}|)\,.
\end{equation}
It remains to check whether $\varphi_{ab} = 0, \pi $ indeed corresponds to a maximal solution for $c^2$ or not. Evaluating the second derivative of $c^2$ with respect to $\varphi_{ab}$ at $\varphi_{ab} = 0, \pi$ and inserting Eq.~\eqref{eq:c2pm} yields
\begin{equation}\label{eq:d2c2}
\frac{\mathrm{d}^2 c^2}{\mathrm{d}\varphi_{ab}^2} = \mp \frac{c_\pm^4\left[4\gt_{11}(1 - \gt_{11}) + c_\pm^2(c_\pm^2 - 2)\right]}{\sqrt{1 - 4\left[|\gt_{12}|^2 + \left(\gt_{11} - \frac{1}{2}\right)^2\right]}}\,.
\end{equation}
In order to determine the sign of the right hand side in Eq.~\eqref{eq:d2c2} it is sufficient to evaluate
\begin{equation}\label{eq:sign}
\mathrm{sgn}\left[\mp \left(4\gt_{11}(1 - \gt_{11}) + c_\pm^2(c_\pm^2 - 2)\right)\right]
\end{equation}
for all points $(\gt_{11}, |\gt_{12}|)$ which satisfy $(\gt_{11} - 1/2)^2 + |\gt_{12}|^2 \leq 1/4$. In the following, we focus on $c_+$, i.e.~the solution which minimizes $U(1 - c^2)$ for $U\geq 0$. The calculation for $U\leq 0$ follows analogously. Then, to check the sign of \eqref{eq:sign}, we only need to determine the minimum of $f(\gt_{11}, |\gt_{12}|) = 4\gt_{11}(1 - \gt_{11}) + c_+^2(c_+^2 - 2)$ which is equal to zero. Consequently, the right hand side of Eq.~\eqref{eq:d2c2} is always negative for all possible points $(\gt_{11}, |\gt_{12}|)$ in the domain of $\widetilde{\mathcal{F}}^{(p)}_\CC$.

\section{Derivation of $\mathcal{F}_{\RR}^{(p)}(\gamma)$ for generic interactions \label{app:FR_gen}}

We derive the pure universal functional $\mathcal{F}^{(p)}_{\RR}$ for a generic isotropic, reflection symmetric interaction
\begin{eqnarray}
W &=& U\left(\ket{\Phi_1}\!\bra{\Phi_1} + \ket{\Phi_2}\!\bra{\Phi_2}\right)  + V\left(\ket{\Phi_1}\!\bra{\Phi_2} + \mathrm{h.c.}\right) \nonumber \\
& \quad &+ X\left(\ket{\Phi_1}\!\bra{\Phi_3} + \ket{\Phi_2}\!\bra{\Phi_3} + \mathrm{h.c.}\right)\,,
\end{eqnarray}
where $U, V, X\in \RR$.
It follows that the expectation value of this interaction $W$ with a general pure state $\ket{\Psi} = a\ket{\Phi_1} + b\ket{\Phi_2} + c\ket{\Phi_3}$ with $a,b,c\in \RR$
is given by
\begin{eqnarray}\label{eq:WPsi}
\bra{\Psi}W\ket{\Psi} &=& U\left(1 - c^2\right) + 2Vab+ 2\sqrt{2}X\gamma_{12} \\
&=& U- V+2\sqrt{2}X\gamma_{12} + c^2(V-U) + \frac{2V\gamma_{12}^2}{c^2}\,. \nonumber
\end{eqnarray}
Thus, $\bra{\Psi}W\ket{\Psi}$ is a function of $c^2$ only.
The universal functional $\mathcal{F}^{(p)}_{\RR}$ then follows from minimizing $\bra{\Psi}W\ket{\Psi}$ \eqref{eq:WPsi} under the constraints $a^2 + b^2 + c^2 = 1$, $\gamma_{11} = a^2 + c^2/2$ and $\gamma_{12} = c(a + b)/\sqrt{2}$. Inverting these three constraints for $c^2$ yields the two solutions
\begin{equation}
c^2_{\pm} = \frac{\gamma_{12}^2\left(\frac{1}{2}\pm\sqrt{\frac{1}{4} - \left[|\gamma_{12}|^2 + \left(\gamma_{11} - \frac{1}{2}\right)^2\right]}\right)}{\left(\gamma_{11} - \frac{1}{2}\right)^2 + \gamma_{12}^2}\,.
\end{equation}
Thus, the minimization over the free parameter $c$ in $\mathcal{F}^{(p)}_{\RR}$ reduces to
\begin{eqnarray}
\mathcal{F}^{(p)}_{\RR}(\g) &=& U - V + 2\sqrt{2}X\gamma_{12} \\
&& +\min_{c^2_{\pm}}\left(c^2_{\pm}(V-U) + c^2_{\mp}\frac{2V\left[\left(\gamma_{11}-\frac{1}{2}\right)^2 + \gamma_{12}^2\right]}{\gamma_{12}^2}\right)\,.\nonumber
\end{eqnarray}
This minimization can be executed and we eventually arrive at
\begin{widetext}
\begin{equation}\label{eq:FR_gen}
\mathcal{F}_\RR^{(p)}(\gamma) = U + 2\sqrt{2}X\gamma_{12}+ \frac{(V-U)\gamma_{12}^2}{2((\gamma_{11}-\frac{1}{2})^2 + \gamma_{12}^2)} - \sqrt{\frac{1}{4} - \left(\left(\gamma_{11}-\frac{1}{2}\right)^2 + \gamma_{12}^2\right)}\left\vert \frac{\gamma_{12}^2(V-U)}{(\gamma_{11}-\frac{1}{2})^2 + \gamma_{12}^2} - 2V\right\vert\,.
\end{equation}
\end{widetext}
The above equation can be rewritten in terms of the polar coordinates $\gamma_{11} = [1+(1-2R)\mathrm{cos}(\varphi)]/2$, $\gamma_{12} = (1-2R)\mathrm{sin}(\varphi)/2$, where $R$ denotes the distance of a 1RDM to the boundary of the disk describing the set $\mathcal{P}^1_N$ (see also Fig.~\ref{fig:disk}) and $\varphi$ is the polar angle. Using the new variables $R$ and $\varphi$, Eq.~\eqref{eq:FR_gen} reduces to the more compact expression
\begin{eqnarray}
&&\mathcal{F}_\RR^{(p)}(\gamma) =U + \sqrt{2}X(1-2R)\mathrm{sin}(\varphi) + \frac{1}{2}(V-U)\mathrm{sin}^2(\varphi) \nonumber \\
&&- \frac{1}{2}\sqrt{1-(1-2R)^2}\left\vert (V-U)\mathrm{sin}^2(\varphi) - 2V\right\vert\,.
\end{eqnarray}

\section{Non-$v$-representable 1RDMs for the generalized Hubbard dimer \label{app:ellipses}}

In this section, we derive the equations for the ellipses surrounding the 1RDMs which are not pure state $v$-representable in the generalized Hubbard dimer. As shown below, in this setting it is indeed sufficient to investigate the Hessian to answer this question since for the not pure state $v$-representable 1RDMs the functional is locally not convex. It is important to notice that for more complicated geometries of the sets of not pure state $v$-representable 1RDMs the Hessian will be in general not sufficient to find a solution to the pure state $v$-representability problem.
The Hessian is a symmetric matrix that describes the second derivative of the energy functional and takes the form $H_{ij}=\partial_i\partial_jf(x_i,x_j)$. For the universal functional $\mathcal{F}^{(p)}_\RR$ this matrix's eigenvalues take the form
\begin{widetext}
\begin{equation}
	\lambda_\pm =  \frac{1}{2}\left(\partial_{\gamma_{11}}^2\mathcal{F}^{(p)}_\RR + \partial_{\gamma_{12}}^2\mathcal{F}^{(p)}_\RR\right)  \pm \frac{1}{2}\sqrt{ \left(\partial_{\gamma_{11}}^2\mathcal{F}^{(p)}_\RR + \partial_{\gamma_{12}}^2\mathcal{F}^{(p)}_\RR\right)^2- 4 \left( \partial^2_{\gamma_{11}}\mathcal{F}^{(p)}_\RR\partial^2_{\gamma_{12}}\mathcal{F}^{(p)}_\RR - (\partial_{\gamma_{11}}\partial_{\gamma_{12}}\mathcal{F}^{(p)}_\RR)^2 \right)}\,.
	\label{eigen_hessian}
\end{equation}
\end{widetext}
We are only interested in the eigenvalue's sign to distinguish between the 1RDMs, which are pure state $v$-representable and those which are not. As we will see below, here this information can be obtained by evaluating the sign of the Hessian's determinant
\begin{equation}\label{hessian_det}
	\partial^2_{x}\mathcal{F}^{(p)}_\RR\partial^2_{y}\mathcal{F}^{(p)}_\RR- \left(\partial_{x}\partial_{y}\mathcal{F}^{(p)}_\RR\right)^2\,,
\end{equation}
where we introduced $x=\gamma_{11}-\frac{1}{2}$ and $y=\gamma_{12}$ and finding the values of $(x,y)$ where this goes to zero. Plugging the functional $\mathcal{F}^{(p)}_\RR$ into Eq.~\eqref{hessian_det} yields
\begin{widetext}
\begin{equation}
	\begin{aligned}
	&(xy)^2\Bigg(2(y^2-x^2)(\xi-2\sqrt{\tfrac{1}{4}-x^2-y^2}) + \frac{(x^2+y^2)}{\sqrt{\tfrac{1}{4}-x^2-y^2}} \Big( 2 (x^2+y^2) + \frac{(x^2+y^2)(y^2+\tfrac{2V}{U-V}(x^2-y^2)}{(\tfrac{1}{4}-x^2-y^2)} \Big)  \Bigg)^2 = \\
	&\Bigg(y^2(3x^2-y^2)(\xi-2\sqrt{\tfrac{1}{4}-x^2-y^2}) + \frac{(x^2+y^2)}{\sqrt{\tfrac{1}{4}-x^2-y^2}} \Big( -4x^2y^2+ \frac{(x^2+y^2)(\tfrac{1}{4}-y^2)(y^2+\tfrac{2V}{U-V}(x^2-y^2)}{(\tfrac{1}{4}-x^2-y^2)} \Big)  \Bigg)\\
	&\Bigg(x^2(x^2-3y^2)(\xi-2\sqrt{\tfrac{1}{4}-x^2-y^2}) + \frac{(x^2+y^2)}{\sqrt{\tfrac{1}{4}-x^2-y^2}} \Big(4x^2y^2+ \frac{(x^2+y^2)(\frac{1}{4}-x^2)(y^2+\tfrac{2V}{U-V}(x^2-y^2)}{(\tfrac{1}{4}-x^2-y^2)} \Big)  \Bigg)\,,
	\end{aligned}
\end{equation}
where $\xi = \text{sign}[-2V(x^2+y^2)+y^2(V-U)]$. This can be expanded and simplified to
\begin{equation}
	\begin{aligned}
		&\Big( \tfrac{1}{2} - \xi\sqrt{\tfrac{1}{4}-x^2-y^2} - (x^2+y^2) \Big) \Big( 1 +\frac{(x^2-y^2)(x^2+y^2)^2}{4x^2y^2(\tfrac{1}{4}-x^2-y^2)} \Big( \frac{y^2}{x^2+y^2} + \frac{2V}{U-V} \Big)  \Big) = \\
		&\frac{(x^2+y^2)^3}{4x^2y^2(\tfrac{1}{4}-x^2-y^2)} \Big( \frac{y^2}{x^2+y^2} + \frac{2V}{U-V} \Big)\Big( y^2 + \frac{2V}{U-V} (x^2+y^2) -(x^2-y^2)\Big)\,.
	\end{aligned}
\end{equation}
The above expression can be further simplified to
\begin{equation}
	\begin{aligned}
		1 - 4 (x^2+y^2) = \Bigg( 2(x^2+y^2) -1 + \frac{2(x^2+y^2)^2 \Big((x^2+y^2)(1+\tfrac{2V}{U-V}) -y^2 \Big)}{(x^4-y^4)+ \frac{x^2y^2\big( 1-4(x^2+y^2) \big)}{y^2+ \tfrac{2V}{U-V}(x^2+y^2)} } \Bigg)^2\,.
	\end{aligned}
\end{equation}
\end{widetext}
Solving and expanding the above equation yields elliptic solutions which take different forms when $U>V$ and when $U<V$. For the case where  $U>V$ the solution takes the form
\begin{equation}
	\Big( \frac{1}{4} \Big)^2 = \frac{1}{2(1-\frac{V}{U})} y^2 + \frac{1}{(1 - (\frac{V}{U})^2 )}\Bigg(|x| - \frac{1}{4}\sqrt{1-\Big(\frac{V}{U}\Big)^2} \Bigg)^2
\end{equation}
and when $V<U$ the solution takes the form
\begin{equation}
	\Big( \frac{1}{4} \Big)^2 = \frac{(3-\frac{U}{V})^2}{8(1-\frac{U}{V})} \Bigg(|y| -\frac{1}{4}\sqrt{\frac{8(1-\frac{U}{V})}{(3-\frac{U}{V})^2}} \Bigg)^2 + \frac{3-\frac{U}{V}}{4(1-\frac{U}{V})}x^2\,.
\end{equation}
Due to the geometry of these ellipses, the above two expressions finally describe the two ellipses restricting the set of 1RDMs which are not pure state $v$-representable for the generalized Hubbard dimer.

\bibliography{Refs}

\begin{thebibliography}{89}%
\makeatletter
\providecommand \@ifxundefined [1]{%
 \@ifx{#1\undefined}
}%
\providecommand \@ifnum [1]{%
 \ifnum #1\expandafter \@firstoftwo
 \else \expandafter \@secondoftwo
 \fi
}%
\providecommand \@ifx [1]{%
 \ifx #1\expandafter \@firstoftwo
 \else \expandafter \@secondoftwo
 \fi
}%
\providecommand \natexlab [1]{#1}%
\providecommand \enquote  [1]{``#1''}%
\providecommand \bibnamefont  [1]{#1}%
\providecommand \bibfnamefont [1]{#1}%
\providecommand \citenamefont [1]{#1}%
\providecommand \href@noop [0]{\@secondoftwo}%
\providecommand \href [0]{\begingroup \@sanitize@url \@href}%
\providecommand \@href[1]{\@@startlink{#1}\@@href}%
\providecommand \@@href[1]{\endgroup#1\@@endlink}%
\providecommand \@sanitize@url [0]{\catcode `\\12\catcode `\$12\catcode
  `\&12\catcode `\#12\catcode `\^12\catcode `\_12\catcode `\%12\relax}%
\providecommand \@@startlink[1]{}%
\providecommand \@@endlink[0]{}%
\providecommand \url  [0]{\begingroup\@sanitize@url \@url }%
\providecommand \@url [1]{\endgroup\@href {#1}{\urlprefix }}%
\providecommand \urlprefix  [0]{URL }%
\providecommand \Eprint [0]{\href }%
\providecommand \doibase [0]{https://doi.org/}%
\providecommand \selectlanguage [0]{\@gobble}%
\providecommand \bibinfo  [0]{\@secondoftwo}%
\providecommand \bibfield  [0]{\@secondoftwo}%
\providecommand \translation [1]{[#1]}%
\providecommand \BibitemOpen [0]{}%
\providecommand \bibitemStop [0]{}%
\providecommand \bibitemNoStop [0]{.\EOS\space}%
\providecommand \EOS [0]{\spacefactor3000\relax}%
\providecommand \BibitemShut  [1]{\csname bibitem#1\endcsname}%
\let\auto@bib@innerbib\@empty
\bibitem [{\citenamefont {Hohenberg}\ and\ \citenamefont {Kohn}(1964)}]{HK}%
  \BibitemOpen
  \bibfield  {author} {\bibinfo {author} {\bibfnamefont {P.}~\bibnamefont
  {Hohenberg}}\ and\ \bibinfo {author} {\bibfnamefont {W.}~\bibnamefont
  {Kohn}},\ }\bibfield  {title} {\enquote {\bibinfo {title} {Inhomogeneous
  electron gas},}\ }\href {https://doi.org/10.1103/PhysRev.136.B864} {\bibfield
   {journal} {\bibinfo  {journal} {Phys. Rev.}\ }\textbf {\bibinfo {volume}
  {136}},\ \bibinfo {pages} {B864} (\bibinfo {year} {1964})}\BibitemShut
  {NoStop}%
\bibitem [{\citenamefont {Gross}\ and\ \citenamefont {Dreizler}(2013)}]{GD95}%
  \BibitemOpen
  \bibfield  {author} {\bibinfo {author} {\bibfnamefont {E.}~\bibnamefont
  {Gross}}\ and\ \bibinfo {author} {\bibfnamefont {R.}~\bibnamefont
  {Dreizler}},\ }\href {https://www.springer.com/la/book/9781475799774} {\emph
  {\bibinfo {title} {Density functional theory}}},\ \bibinfo {series} {Nato
  Science Series B}, Vol.\ \bibinfo {volume} {337}\ (\bibinfo  {publisher}
  {Springer Science \& Business Media},\ \bibinfo {year} {2013})\BibitemShut
  {NoStop}%
\bibitem [{\citenamefont {Capelle}, \citenamefont {Ullrich},\ and\
  \citenamefont {Vignale}(2007)}]{Capelle07}%
  \BibitemOpen
  \bibfield  {author} {\bibinfo {author} {\bibfnamefont {K.}~\bibnamefont
  {Capelle}}, \bibinfo {author} {\bibfnamefont {C.~A.}\ \bibnamefont
  {Ullrich}},\ and\ \bibinfo {author} {\bibfnamefont {G.}~\bibnamefont
  {Vignale}},\ }\bibfield  {title} {\enquote {\bibinfo {title} {Degenerate
  ground states and nonunique potentials: Breakdown and restoration of density
  functionals},}\ }\href {https://doi.org/10.1103/PhysRevA.76.012508}
  {\bibfield  {journal} {\bibinfo  {journal} {Phys. Rev. A}\ }\textbf {\bibinfo
  {volume} {76}},\ \bibinfo {pages} {012508} (\bibinfo {year}
  {2007})}\BibitemShut {NoStop}%
\bibitem [{\citenamefont {Gilbert}(1975)}]{Gilbert}%
  \BibitemOpen
  \bibfield  {author} {\bibinfo {author} {\bibfnamefont {T.~L.}\ \bibnamefont
  {Gilbert}},\ }\bibfield  {title} {\enquote {\bibinfo {title} {{Hohenberg-Kohn
  theorem for nonlocal ex\-ter\-nal potentials}},}\ }\href
  {https://link.aps.org/doi/10.1103/PhysRevB.12.2111} {\bibfield  {journal}
  {\bibinfo  {journal} {Phys. Rev. B}\ }\textbf {\bibinfo {volume} {12}},\
  \bibinfo {pages} {2111} (\bibinfo {year} {1975})}\BibitemShut {NoStop}%
\bibitem [{\citenamefont {Levy}(1979)}]{LE79}%
  \BibitemOpen
  \bibfield  {author} {\bibinfo {author} {\bibfnamefont {M.}~\bibnamefont
  {Levy}},\ }\bibfield  {title} {\enquote {\bibinfo {title} {Universal
  variational functionals of electron densities, first-order density matrices,
  and natural spin-orbitals and solution of the v-representability problem},}\
  }\href {http://www.pnas.org/content/76/12/6062} {\bibfield  {journal}
  {\bibinfo  {journal} {Proc. Natl. Acad. Sci. U.S.A}\ }\textbf {\bibinfo
  {volume} {76}},\ \bibinfo {pages} {6062} (\bibinfo {year}
  {1979})}\BibitemShut {NoStop}%
\bibitem [{\citenamefont {Valone}(1980)}]{V80}%
  \BibitemOpen
  \bibfield  {author} {\bibinfo {author} {\bibfnamefont {S.~M.}\ \bibnamefont
  {Valone}},\ }\bibfield  {title} {\enquote {\bibinfo {title} {Consequences of
  extending 1-matrix energy functionals from pure–state representable to all
  ensemble representable 1-matrices},}\ }\href
  {https://doi.org/10.1063/1.440249} {\bibfield  {journal} {\bibinfo  {journal}
  {J. Chem. Phys.}\ }\textbf {\bibinfo {volume} {73}},\ \bibinfo {pages} {1344}
  (\bibinfo {year} {1980})}\BibitemShut {NoStop}%
\bibitem [{\citenamefont {Klyachko}(2006)}]{KL06}%
  \BibitemOpen
  \bibfield  {author} {\bibinfo {author} {\bibfnamefont {A.}~\bibnamefont
  {Klyachko}},\ }\bibfield  {title} {\enquote {\bibinfo {title} {Quantum
  marginal problem and {N}-representability},}\ }\href
  {http://stacks.iop.org/1742-6596/36/i=1/a=014} {\bibfield  {journal}
  {\bibinfo  {journal} {J. Phys. Conf. Ser.}\ }\textbf {\bibinfo {volume}
  {36}},\ \bibinfo {pages} {72} (\bibinfo {year} {2006})}\BibitemShut {NoStop}%
\bibitem [{\citenamefont {Altunbulak}\ and\ \citenamefont
  {Klyachko}(2008)}]{AK08}%
  \BibitemOpen
  \bibfield  {author} {\bibinfo {author} {\bibfnamefont {M.}~\bibnamefont
  {Altunbulak}}\ and\ \bibinfo {author} {\bibfnamefont {A.}~\bibnamefont
  {Klyachko}},\ }\bibfield  {title} {\enquote {\bibinfo {title} {The {P}auli
  principle revisited},}\ }\href {https://doi.org/10.1007/s00220-008-0552-z}
  {\bibfield  {journal} {\bibinfo  {journal} {Commun. Math. Phys.}\ }\textbf
  {\bibinfo {volume} {282}},\ \bibinfo {pages} {287} (\bibinfo {year}
  {2008})}\BibitemShut {NoStop}%
\bibitem [{\citenamefont {Klyachko}(2009)}]{Kly}%
  \BibitemOpen
  \bibfield  {author} {\bibinfo {author} {\bibfnamefont {A.}~\bibnamefont
  {Klyachko}},\ }\bibfield  {title} {\enquote {\bibinfo {title} {The {P}auli
  exclusion principle and beyond},}\ }\href {http://arxiv.org/abs/0904.2009}
  {\bibfield  {journal} {\bibinfo  {journal} {arXiv:0904.2009}\ } (\bibinfo
  {year} {2009})}\BibitemShut {NoStop}%
\bibitem [{\citenamefont {Schilling}(2018)}]{Schilling2018}%
  \BibitemOpen
  \bibfield  {author} {\bibinfo {author} {\bibfnamefont {C.}~\bibnamefont
  {Schilling}},\ }\bibfield  {title} {\enquote {\bibinfo {title}
  {{Communication: Relating the pure and ensemble density matrix
  functional}},}\ }\href {https://doi.org/10.1063/1.5080088} {\bibfield
  {journal} {\bibinfo  {journal} {J. Chem. Phys.}\ }\textbf {\bibinfo {volume}
  {149}},\ \bibinfo {pages} {231102} (\bibinfo {year} {2018})}\BibitemShut
  {NoStop}%
\bibitem [{\citenamefont {Cioslowski}\ and\ \citenamefont
  {Pernal}(2004)}]{CP04}%
  \BibitemOpen
  \bibfield  {author} {\bibinfo {author} {\bibfnamefont {J.}~\bibnamefont
  {Cioslowski}}\ and\ \bibinfo {author} {\bibfnamefont {K.}~\bibnamefont
  {Pernal}},\ }\bibfield  {title} {\enquote {\bibinfo {title} {Size versus
  volume extensivity of a new class of density matrix functionals},}\ }\href
  {https://doi.org/10.1063/1.1738411} {\bibfield  {journal} {\bibinfo
  {journal} {J. Chem. Phys.}\ }\textbf {\bibinfo {volume} {120}},\ \bibinfo
  {pages} {10364--10367} (\bibinfo {year} {2004})}\BibitemShut {NoStop}%
\bibitem [{\citenamefont {Cioslowski}(2005)}]{C05}%
  \BibitemOpen
  \bibfield  {author} {\bibinfo {author} {\bibfnamefont {J.}~\bibnamefont
  {Cioslowski}},\ }\bibfield  {title} {\enquote {\bibinfo {title} {New
  constraints upon the electron-electron repulsion energy functional of the
  one-electron reduced density matrix},}\ }\href
  {https://doi.org/10.1063/1.2074527} {\bibfield  {journal} {\bibinfo
  {journal} {J. Chem. Phys.}\ }\textbf {\bibinfo {volume} {123}},\ \bibinfo
  {pages} {164106} (\bibinfo {year} {2005})}\BibitemShut {NoStop}%
\bibitem [{\citenamefont {Rohr}\ and\ \citenamefont {Pernal}(2011)}]{RP11}%
  \BibitemOpen
  \bibfield  {author} {\bibinfo {author} {\bibfnamefont {D.~R.}\ \bibnamefont
  {Rohr}}\ and\ \bibinfo {author} {\bibfnamefont {K.}~\bibnamefont {Pernal}},\
  }\bibfield  {title} {\enquote {\bibinfo {title} {Open-shell reduced density
  matrix functional theory},}\ }\href {https://doi.org/10.1063/1.3624609}
  {\bibfield  {journal} {\bibinfo  {journal} {J. Chem. Phys.}\ }\textbf
  {\bibinfo {volume} {135}},\ \bibinfo {pages} {074104} (\bibinfo {year}
  {2011})}\BibitemShut {NoStop}%
\bibitem [{\citenamefont {Pernal}(2012)}]{P12}%
  \BibitemOpen
  \bibfield  {author} {\bibinfo {author} {\bibfnamefont {K.}~\bibnamefont
  {Pernal}},\ }\bibfield  {title} {\enquote {\bibinfo {title} {Excitation
  energies from range-separated time-dependent density and density matrix
  functional theory},}\ }\href {https://doi.org/10.1063/1.4712019} {\bibfield
  {journal} {\bibinfo  {journal} {J. Chem. Phys.}\ }\textbf {\bibinfo {volume}
  {136}},\ \bibinfo {pages} {184105} (\bibinfo {year} {2012})}\BibitemShut
  {NoStop}%
\bibitem [{\citenamefont {Wang}\ and\ \citenamefont {Knowles}(2015)}]{WK15}%
  \BibitemOpen
  \bibfield  {author} {\bibinfo {author} {\bibfnamefont {J.}~\bibnamefont
  {Wang}}\ and\ \bibinfo {author} {\bibfnamefont {P.~J.}\ \bibnamefont
  {Knowles}},\ }\bibfield  {title} {\enquote {\bibinfo {title} {Nonuniqueness
  of algebraic first-order density-matrix functionals},}\ }\href
  {https://doi.org/10.1103/PhysRevA.92.012520} {\bibfield  {journal} {\bibinfo
  {journal} {Phys. Rev. A}\ }\textbf {\bibinfo {volume} {92}},\ \bibinfo
  {pages} {012520} (\bibinfo {year} {2015})}\BibitemShut {NoStop}%
\bibitem [{\citenamefont {Baldsiefen}, \citenamefont {Cangi},\ and\
  \citenamefont {Gross}(2015)}]{BCG15}%
  \BibitemOpen
  \bibfield  {author} {\bibinfo {author} {\bibfnamefont {T.}~\bibnamefont
  {Baldsiefen}}, \bibinfo {author} {\bibfnamefont {A.}~\bibnamefont {Cangi}},\
  and\ \bibinfo {author} {\bibfnamefont {E.~K.~U.}\ \bibnamefont {Gross}},\
  }\bibfield  {title} {\enquote {\bibinfo {title}
  {Reduced-density-matrix-functional theory at finite temperature: Theoretical
  foundations},}\ }\href {https://doi.org/10.1103/PhysRevA.92.052514}
  {\bibfield  {journal} {\bibinfo  {journal} {Phys. Rev. A}\ }\textbf {\bibinfo
  {volume} {92}},\ \bibinfo {pages} {052514} (\bibinfo {year}
  {2015})}\BibitemShut {NoStop}%
\bibitem [{\citenamefont {Giesbertz}, \citenamefont {Uimonen},\ and\
  \citenamefont {van Leeuwen}(2018)}]{GUL18}%
  \BibitemOpen
  \bibfield  {author} {\bibinfo {author} {\bibfnamefont {K.~J.~H.}\
  \bibnamefont {Giesbertz}}, \bibinfo {author} {\bibfnamefont {A.-M.}\
  \bibnamefont {Uimonen}},\ and\ \bibinfo {author} {\bibfnamefont
  {R.}~\bibnamefont {van Leeuwen}},\ }\bibfield  {title} {\enquote {\bibinfo
  {title} {Approximate energy functionals for one-body reduced density matrix
  functional theory from many-body perturbation theory},}\ }\href
  {https://link.springer.com/article/10.1140/epjb/e2018-90279-1} {\bibfield
  {journal} {\bibinfo  {journal} {Eur. Phys. J. B}\ }\textbf {\bibinfo {volume}
  {91}},\ \bibinfo {pages} {1} (\bibinfo {year} {2018})}\BibitemShut {NoStop}%
\bibitem [{\citenamefont {Schilling}\ and\ \citenamefont
  {Schilling}(2019)}]{Schilling2019}%
  \BibitemOpen
  \bibfield  {author} {\bibinfo {author} {\bibfnamefont {C.}~\bibnamefont
  {Schilling}}\ and\ \bibinfo {author} {\bibfnamefont {R.}~\bibnamefont
  {Schilling}},\ }\bibfield  {title} {\enquote {\bibinfo {title} {Diverging
  exchange force and form of the exact density matrix functional},}\ }\href
  {https://doi.org/10.1103/PhysRevLett.122.013001} {\bibfield  {journal}
  {\bibinfo  {journal} {Phys. Rev. Lett.}\ }\textbf {\bibinfo {volume} {122}},\
  \bibinfo {pages} {013001} (\bibinfo {year} {2019})}\BibitemShut {NoStop}%
\bibitem [{\citenamefont {Gritsenko}, \citenamefont {Wang},\ and\ \citenamefont
  {Knowles}(2019)}]{GWK19}%
  \BibitemOpen
  \bibfield  {author} {\bibinfo {author} {\bibfnamefont {O.~V.}\ \bibnamefont
  {Gritsenko}}, \bibinfo {author} {\bibfnamefont {J.}~\bibnamefont {Wang}},\
  and\ \bibinfo {author} {\bibfnamefont {P.~J.}\ \bibnamefont {Knowles}},\
  }\bibfield  {title} {\enquote {\bibinfo {title} {Symmetry dependence and
  universality of practical algebraic functionals in density-matrix-functional
  theory},}\ }\href {https://doi.org/10.1103/PhysRevA.99.042516} {\bibfield
  {journal} {\bibinfo  {journal} {Phys. Rev. A}\ }\textbf {\bibinfo {volume}
  {99}},\ \bibinfo {pages} {042516} (\bibinfo {year} {2019})}\BibitemShut
  {NoStop}%
\bibitem [{\citenamefont {Cioslowski}, \citenamefont {Mih\'{a}lka},\ and\
  \citenamefont {Szabados}(2019)}]{Cioslowski19}%
  \BibitemOpen
  \bibfield  {author} {\bibinfo {author} {\bibfnamefont {J.}~\bibnamefont
  {Cioslowski}}, \bibinfo {author} {\bibfnamefont {Z.~E.}\ \bibnamefont
  {Mih\'{a}lka}},\ and\ \bibinfo {author} {\bibfnamefont {A.}~\bibnamefont
  {Szabados}},\ }\bibfield  {title} {\enquote {\bibinfo {title} {Bilinear
  constraints upon the correlation contribution to the electron–electron
  repulsion energy as a functional of the one-electron reduced density
  matrix},}\ }\href {https://doi.org/10.1021/acs.jctc.9b00443} {\bibfield
  {journal} {\bibinfo  {journal} {J. Chem. Theory Comput.}\ }\textbf {\bibinfo
  {volume} {15}},\ \bibinfo {pages} {4862--4872} (\bibinfo {year}
  {2019})}\BibitemShut {NoStop}%
\bibitem [{\citenamefont {Cioslowski}(2020{\natexlab{a}})}]{C20b}%
  \BibitemOpen
  \bibfield  {author} {\bibinfo {author} {\bibfnamefont {J.}~\bibnamefont
  {Cioslowski}},\ }\bibfield  {title} {\enquote {\bibinfo {title} {Off-diagonal
  derivative discontinuities in the reduced density matrices of electronic
  systems},}\ }\href {https://doi.org/10.1063/5.0023955} {\bibfield  {journal}
  {\bibinfo  {journal} {J. Chem. Phys.}\ }\textbf {\bibinfo {volume} {153}},\
  \bibinfo {pages} {154108} (\bibinfo {year} {2020}{\natexlab{a}})}\BibitemShut
  {NoStop}%
\bibitem [{\citenamefont {Cioslowski}(2020{\natexlab{b}})}]{C20a}%
  \BibitemOpen
  \bibfield  {author} {\bibinfo {author} {\bibfnamefont {J.}~\bibnamefont
  {Cioslowski}},\ }\bibfield  {title} {\enquote {\bibinfo {title} {One-electron
  reduced density matrix functional theory of spin-polarized systems},}\ }\href
  {https://doi.org/10.1021/acs.jctc.9b01155} {\bibfield  {journal} {\bibinfo
  {journal} {J. Chem. Theory Comput.}\ }\textbf {\bibinfo {volume} {16}},\
  \bibinfo {pages} {1578} (\bibinfo {year} {2020}{\natexlab{b}})}\BibitemShut
  {NoStop}%
\bibitem [{\citenamefont {Giesbertz}(2020)}]{G20}%
  \BibitemOpen
  \bibfield  {author} {\bibinfo {author} {\bibfnamefont {K.~J.~H.}\
  \bibnamefont {Giesbertz}},\ }\bibfield  {title} {\enquote {\bibinfo {title}
  {Implications of the unitary invariance and symmetry restrictions on the
  development of proper approximate one-body reduced-density-matrix
  functionals},}\ }\href {https://doi.org/10.1103/PhysRevA.102.052814}
  {\bibfield  {journal} {\bibinfo  {journal} {Phys. Rev. A}\ }\textbf {\bibinfo
  {volume} {102}},\ \bibinfo {pages} {052814} (\bibinfo {year}
  {2020})}\BibitemShut {NoStop}%
\bibitem [{\citenamefont {Cioslowski}(2020{\natexlab{c}})}]{C153-20}%
  \BibitemOpen
  \bibfield  {author} {\bibinfo {author} {\bibfnamefont {J.}~\bibnamefont
  {Cioslowski}},\ }\bibfield  {title} {\enquote {\bibinfo {title} {Construction
  of explicitly correlated one-electron reduced density matrices},}\ }\href
  {https://doi.org/10.1063/5.0031195} {\bibfield  {journal} {\bibinfo
  {journal} {J. Chem. Phys.}\ }\textbf {\bibinfo {volume} {153}},\ \bibinfo
  {pages} {224109} (\bibinfo {year} {2020}{\natexlab{c}})}\BibitemShut
  {NoStop}%
\bibitem [{\citenamefont {Macia{\.{z}}ek}(2021)}]{M21}%
  \BibitemOpen
  \bibfield  {author} {\bibinfo {author} {\bibfnamefont {T.}~\bibnamefont
  {Macia{\.{z}}ek}},\ }\bibfield  {title} {\enquote {\bibinfo {title}
  {Repulsively diverging gradient of the density functional in the reduced
  density matrix functional theory},}\ }\href
  {https://doi.org/10.1088/1367-2630/ac309c} {\bibfield  {journal} {\bibinfo
  {journal} {New J. Phys.}\ }\textbf {\bibinfo {volume} {23}},\ \bibinfo
  {pages} {113006} (\bibinfo {year} {2021})}\BibitemShut {NoStop}%
\bibitem [{\citenamefont {Schilling}\ and\ \citenamefont
  {Pittalis}(2021)}]{Schilling21}%
  \BibitemOpen
  \bibfield  {author} {\bibinfo {author} {\bibfnamefont {C.}~\bibnamefont
  {Schilling}}\ and\ \bibinfo {author} {\bibfnamefont {S.}~\bibnamefont
  {Pittalis}},\ }\bibfield  {title} {\enquote {\bibinfo {title} {Ensemble
  reduced density matrix functional theory for excited states and hierarchical
  generalization of {P}auli's exclusion principle},}\ }\href
  {https://doi.org/10.1103/PhysRevLett.127.023001} {\bibfield  {journal}
  {\bibinfo  {journal} {Phys. Rev. Lett.}\ }\textbf {\bibinfo {volume} {127}},\
  \bibinfo {pages} {023001} (\bibinfo {year} {2021})}\BibitemShut {NoStop}%
\bibitem [{\citenamefont {Liebert}\ \emph {et~al.}(2022)\citenamefont
  {Liebert}, \citenamefont {Castillo}, \citenamefont {Labb\'e},\ and\
  \citenamefont {Schilling}}]{LCLS21}%
  \BibitemOpen
  \bibfield  {author} {\bibinfo {author} {\bibfnamefont {J.}~\bibnamefont
  {Liebert}}, \bibinfo {author} {\bibfnamefont {F.}~\bibnamefont {Castillo}},
  \bibinfo {author} {\bibfnamefont {J.-P.}\ \bibnamefont {Labb\'e}},\ and\
  \bibinfo {author} {\bibfnamefont {C.}~\bibnamefont {Schilling}},\ }\bibfield
  {title} {\enquote {\bibinfo {title} {Foundation of one-particle reduced
  density matrix functional theory for excited states},}\ }\href
  {https://doi.org/10.1021/acs.jctc.1c00561} {\bibfield  {journal} {\bibinfo
  {journal} {J. Chem. Theory Comput.}\ }\textbf {\bibinfo {volume} {18}},\
  \bibinfo {pages} {124--140} (\bibinfo {year} {2022})}\BibitemShut {NoStop}%
\bibitem [{\citenamefont {Di~Sabatino}\ \emph {et~al.}(2022)\citenamefont
  {Di~Sabatino}, \citenamefont {Koskelo}, \citenamefont {Berger},\ and\
  \citenamefont {Romaniello}}]{DSKBR22}%
  \BibitemOpen
  \bibfield  {author} {\bibinfo {author} {\bibfnamefont {S.}~\bibnamefont
  {Di~Sabatino}}, \bibinfo {author} {\bibfnamefont {J.}~\bibnamefont
  {Koskelo}}, \bibinfo {author} {\bibfnamefont {J.~A.}\ \bibnamefont
  {Berger}},\ and\ \bibinfo {author} {\bibfnamefont {P.}~\bibnamefont
  {Romaniello}},\ }\bibfield  {title} {\enquote {\bibinfo {title} {Introducing
  screening in one-body density matrix functionals: Impact on charged
  excitations of model systems via the extended {K}oopmans' theorem},}\ }\href
  {https://doi.org/10.1103/PhysRevB.105.235123} {\bibfield  {journal} {\bibinfo
   {journal} {Phys. Rev. B}\ }\textbf {\bibinfo {volume} {105}},\ \bibinfo
  {pages} {235123} (\bibinfo {year} {2022})}\BibitemShut {NoStop}%
\bibitem [{\citenamefont {Senjean}\ \emph {et~al.}(2022)\citenamefont
  {Senjean}, \citenamefont {Yalouz}, \citenamefont {Nakatani},\ and\
  \citenamefont {Fromager}}]{SYNF22}%
  \BibitemOpen
  \bibfield  {author} {\bibinfo {author} {\bibfnamefont {B.}~\bibnamefont
  {Senjean}}, \bibinfo {author} {\bibfnamefont {S.}~\bibnamefont {Yalouz}},
  \bibinfo {author} {\bibfnamefont {N.}~\bibnamefont {Nakatani}},\ and\
  \bibinfo {author} {\bibfnamefont {E.}~\bibnamefont {Fromager}},\ }\bibfield
  {title} {\enquote {\bibinfo {title} {Reduced density matrix functional theory
  from an ab initio seniority-zero wave function: Exact and approximate
  formulations along adiabatic connection paths},}\ }\href
  {https://doi.org/10.1103/PhysRevA.106.032203} {\bibfield  {journal} {\bibinfo
   {journal} {Phys. Rev. A}\ }\textbf {\bibinfo {volume} {106}},\ \bibinfo
  {pages} {032203} (\bibinfo {year} {2022})}\BibitemShut {NoStop}%
\bibitem [{\citenamefont {Sutter}\ and\ \citenamefont
  {Giesbertz}(2023)}]{SG22}%
  \BibitemOpen
  \bibfield  {author} {\bibinfo {author} {\bibfnamefont {S.~M.}\ \bibnamefont
  {Sutter}}\ and\ \bibinfo {author} {\bibfnamefont {K.~J.~H.}\ \bibnamefont
  {Giesbertz}},\ }\bibfield  {title} {\enquote {\bibinfo {title} {One-body
  reduced density-matrix functional theory for the canonical ensemble},}\
  }\href {https://doi.org/10.1103/PhysRevA.107.022210} {\bibfield  {journal}
  {\bibinfo  {journal} {Phys. Rev. A}\ }\textbf {\bibinfo {volume} {107}},\
  \bibinfo {pages} {022210} (\bibinfo {year} {2023})}\BibitemShut {NoStop}%
\bibitem [{\citenamefont {Gibney}, \citenamefont {Boyn},\ and\ \citenamefont
  {Mazziotti}(2022{\natexlab{a}})}]{GBM22}%
  \BibitemOpen
  \bibfield  {author} {\bibinfo {author} {\bibfnamefont {D.}~\bibnamefont
  {Gibney}}, \bibinfo {author} {\bibfnamefont {J.-N.}\ \bibnamefont {Boyn}},\
  and\ \bibinfo {author} {\bibfnamefont {D.~A.}\ \bibnamefont {Mazziotti}},\
  }\bibfield  {title} {\enquote {\bibinfo {title} {Density functional theory
  transformed into a one-electron reduced-density-matrix functional theory for
  the capture of static correlation},}\ }\href
  {https://doi.org/10.1021/acs.jpclett.2c00083} {\bibfield  {journal} {\bibinfo
   {journal} {J. Phys. Chem. Lett.}\ }\textbf {\bibinfo {volume} {13}},\
  \bibinfo {pages} {1382--1388} (\bibinfo {year}
  {2022}{\natexlab{a}})}\BibitemShut {NoStop}%
\bibitem [{\citenamefont {Gibney}, \citenamefont {Boyn},\ and\ \citenamefont
  {Mazziotti}(2022{\natexlab{b}})}]{GBM22-2}%
  \BibitemOpen
  \bibfield  {author} {\bibinfo {author} {\bibfnamefont {D.}~\bibnamefont
  {Gibney}}, \bibinfo {author} {\bibfnamefont {J.-N.}\ \bibnamefont {Boyn}},\
  and\ \bibinfo {author} {\bibfnamefont {D.~A.}\ \bibnamefont {Mazziotti}},\
  }\bibfield  {title} {\enquote {\bibinfo {title} {Comparison of density-matrix
  corrections to density functional theory},}\ }\href
  {https://doi.org/10.1021/acs.jctc.2c00625} {\bibfield  {journal} {\bibinfo
  {journal} {J. Chem. Theory Comput.}\ }\textbf {\bibinfo {volume} {18}},\
  \bibinfo {pages} {6600--6607} (\bibinfo {year}
  {2022}{\natexlab{b}})}\BibitemShut {NoStop}%
\bibitem [{\citenamefont {Kamil}\ \emph {et~al.}(2016)\citenamefont {Kamil},
  \citenamefont {Schade}, \citenamefont {Pruschke},\ and\ \citenamefont
  {Bl\"ochl}}]{Kamil2016}%
  \BibitemOpen
  \bibfield  {author} {\bibinfo {author} {\bibfnamefont {E.}~\bibnamefont
  {Kamil}}, \bibinfo {author} {\bibfnamefont {R.}~\bibnamefont {Schade}},
  \bibinfo {author} {\bibfnamefont {T.}~\bibnamefont {Pruschke}},\ and\
  \bibinfo {author} {\bibfnamefont {P.~E.}\ \bibnamefont {Bl\"ochl}},\
  }\bibfield  {title} {\enquote {\bibinfo {title} {Reduced density-matrix
  functionals applied to the {H}ubbard dimer},}\ }\href
  {https://doi.org/10.1103/PhysRevB.93.085141} {\bibfield  {journal} {\bibinfo
  {journal} {Phys. Rev. B}\ }\textbf {\bibinfo {volume} {93}},\ \bibinfo
  {pages} {085141} (\bibinfo {year} {2016})}\BibitemShut {NoStop}%
\bibitem [{\citenamefont {Baldsiefen}\ \emph {et~al.}(2017)\citenamefont
  {Baldsiefen}, \citenamefont {Cangi}, \citenamefont {Eich},\ and\
  \citenamefont {Gross}}]{BCEG17}%
  \BibitemOpen
  \bibfield  {author} {\bibinfo {author} {\bibfnamefont {T.}~\bibnamefont
  {Baldsiefen}}, \bibinfo {author} {\bibfnamefont {A.}~\bibnamefont {Cangi}},
  \bibinfo {author} {\bibfnamefont {F.~G.}\ \bibnamefont {Eich}},\ and\
  \bibinfo {author} {\bibfnamefont {E.~K.~U.}\ \bibnamefont {Gross}},\
  }\bibfield  {title} {\enquote {\bibinfo {title} {Exchange-correlation
  approximations for reduced-density-matrix-functional theory at finite
  temperature: Capturing magnetic phase transitions in the homogeneous electron
  gas},}\ }\href {https://doi.org/10.1103/PhysRevA.96.062508} {\bibfield
  {journal} {\bibinfo  {journal} {Phys. Rev. A}\ }\textbf {\bibinfo {volume}
  {96}},\ \bibinfo {pages} {062508} (\bibinfo {year} {2017})}\BibitemShut
  {NoStop}%
\bibitem [{\citenamefont {Schade}, \citenamefont {Kamil},\ and\ \citenamefont
  {Bl{\"o}chl}(2017)}]{SKB17}%
  \BibitemOpen
  \bibfield  {author} {\bibinfo {author} {\bibfnamefont {R.}~\bibnamefont
  {Schade}}, \bibinfo {author} {\bibfnamefont {E.}~\bibnamefont {Kamil}},\ and\
  \bibinfo {author} {\bibfnamefont {P.}~\bibnamefont {Bl{\"o}chl}},\ }\bibfield
   {title} {\enquote {\bibinfo {title} {Reduced density-matrix functionals from
  many-particle theory},}\ }\href {https://doi.org/10.1140/epjst/e2017-70046-0}
  {\bibfield  {journal} {\bibinfo  {journal} {Eur. Phys. J. Special Topics}\
  }\textbf {\bibinfo {volume} {226}},\ \bibinfo {pages} {2677} (\bibinfo {year}
  {2017})}\BibitemShut {NoStop}%
\bibitem [{\citenamefont {Schade}\ and\ \citenamefont {Bl\"ochl}(2018)}]{SB18}%
  \BibitemOpen
  \bibfield  {author} {\bibinfo {author} {\bibfnamefont {R.}~\bibnamefont
  {Schade}}\ and\ \bibinfo {author} {\bibfnamefont {P.~E.}\ \bibnamefont
  {Bl\"ochl}},\ }\bibfield  {title} {\enquote {\bibinfo {title} {Adaptive
  cluster approximation for reduced density-matrix functional theory},}\ }\href
  {https://doi.org/10.1103/PhysRevB.97.245131} {\bibfield  {journal} {\bibinfo
  {journal} {Phys. Rev. B}\ }\textbf {\bibinfo {volume} {97}},\ \bibinfo
  {pages} {245131} (\bibinfo {year} {2018})}\BibitemShut {NoStop}%
\bibitem [{\citenamefont {M\"uller}, \citenamefont {T\"ows},\ and\
  \citenamefont {Pastor}(2018)}]{MTP18}%
  \BibitemOpen
  \bibfield  {author} {\bibinfo {author} {\bibfnamefont {T.~S.}\ \bibnamefont
  {M\"uller}}, \bibinfo {author} {\bibfnamefont {W.}~\bibnamefont {T\"ows}},\
  and\ \bibinfo {author} {\bibfnamefont {G.~M.}\ \bibnamefont {Pastor}},\
  }\bibfield  {title} {\enquote {\bibinfo {title} {Exploiting the links between
  ground-state correlations and independent-fermion entropy in the {H}ubbard
  model},}\ }\href {https://doi.org/10.1103/PhysRevB.98.045135} {\bibfield
  {journal} {\bibinfo  {journal} {Phys. Rev. B}\ }\textbf {\bibinfo {volume}
  {98}},\ \bibinfo {pages} {045135} (\bibinfo {year} {2018})}\BibitemShut
  {NoStop}%
\bibitem [{\citenamefont {Mitxelena}, \citenamefont {Mayorga},\ and\
  \citenamefont {Piris}(2018)}]{Mitxelena18}%
  \BibitemOpen
  \bibfield  {author} {\bibinfo {author} {\bibfnamefont {I.}~\bibnamefont
  {Mitxelena}}, \bibinfo {author} {\bibfnamefont {M.}~\bibnamefont {Mayorga}},\
  and\ \bibinfo {author} {\bibfnamefont {M.}~\bibnamefont {Piris}},\ }\bibfield
   {title} {\enquote {\bibinfo {title} {Phase dilemma in natural orbital
  functional theory from the {N}-representability perspective},}\ }\href
  {https://doi.org/10.1140/epjb/e2018-90078-8} {\bibfield  {journal} {\bibinfo
  {journal} {The European Physical Journal B}\ }\textbf {\bibinfo {volume}
  {91}},\ \bibinfo {pages} {109} (\bibinfo {year} {2018})}\BibitemShut
  {NoStop}%
\bibitem [{\citenamefont {Piris}(2019)}]{Piris19}%
  \BibitemOpen
  \bibfield  {author} {\bibinfo {author} {\bibfnamefont {M.}~\bibnamefont
  {Piris}},\ }\bibfield  {title} {\enquote {\bibinfo {title} {Natural orbital
  functional for multiplets},}\ }\href
  {https://doi.org/10.1103/PhysRevA.100.032508} {\bibfield  {journal} {\bibinfo
   {journal} {Phys. Rev. A}\ }\textbf {\bibinfo {volume} {100}},\ \bibinfo
  {pages} {032508} (\bibinfo {year} {2019})}\BibitemShut {NoStop}%
\bibitem [{\citenamefont {Mitxelena}\ and\ \citenamefont
  {Piris}(2020)}]{Mitxelena20}%
  \BibitemOpen
  \bibfield  {author} {\bibinfo {author} {\bibfnamefont {I.}~\bibnamefont
  {Mitxelena}}\ and\ \bibinfo {author} {\bibfnamefont {M.}~\bibnamefont
  {Piris}},\ }\bibfield  {title} {\enquote {\bibinfo {title} {Analytic
  gradients for spin multiplets in natural orbital functional theory},}\ }\href
  {https://doi.org/10.1063/5.0012897} {\bibfield  {journal} {\bibinfo
  {journal} {The Journal of Chemical Physics}\ }\textbf {\bibinfo {volume}
  {153}},\ \bibinfo {pages} {044101} (\bibinfo {year} {2020})}\BibitemShut
  {NoStop}%
\bibitem [{\citenamefont {Piris}\ and\ \citenamefont
  {Mitxelena}(2021)}]{Piris21-1}%
  \BibitemOpen
  \bibfield  {author} {\bibinfo {author} {\bibfnamefont {M.}~\bibnamefont
  {Piris}}\ and\ \bibinfo {author} {\bibfnamefont {I.}~\bibnamefont
  {Mitxelena}},\ }\bibfield  {title} {\enquote {\bibinfo {title} {{DoNOF}: An
  open-source implementation of natural-orbital-functional-based methods for
  quantum chemistry},}\ }\href
  {https://doi.org/https://doi.org/10.1016/j.cpc.2020.107651} {\bibfield
  {journal} {\bibinfo  {journal} {Comput. Phys. Commun.}\ }\textbf {\bibinfo
  {volume} {259}},\ \bibinfo {pages} {107651} (\bibinfo {year}
  {2021})}\BibitemShut {NoStop}%
\bibitem [{\citenamefont {Piris}(2021)}]{Piris21-2}%
  \BibitemOpen
  \bibfield  {author} {\bibinfo {author} {\bibfnamefont {M.}~\bibnamefont
  {Piris}},\ }\bibfield  {title} {\enquote {\bibinfo {title} {Global natural
  orbital functional: Towards the complete description of the electron
  correlation},}\ }\href {https://doi.org/10.1103/PhysRevLett.127.233001}
  {\bibfield  {journal} {\bibinfo  {journal} {Phys. Rev. Lett.}\ }\textbf
  {\bibinfo {volume} {127}},\ \bibinfo {pages} {233001} (\bibinfo {year}
  {2021})}\BibitemShut {NoStop}%
\bibitem [{\citenamefont {Su}(2021)}]{Su21a}%
  \BibitemOpen
  \bibfield  {author} {\bibinfo {author} {\bibfnamefont {N.~Q.}\ \bibnamefont
  {Su}},\ }\bibfield  {title} {\enquote {\bibinfo {title} {Unity of
  {K}ohn-{S}ham density-functional theory and reduced-density-matrix-functional
  theory},}\ }\href {https://doi.org/10.1103/PhysRevA.104.052809} {\bibfield
  {journal} {\bibinfo  {journal} {Phys. Rev. A}\ }\textbf {\bibinfo {volume}
  {104}},\ \bibinfo {pages} {052809} (\bibinfo {year} {2021})}\BibitemShut
  {NoStop}%
\bibitem [{\citenamefont {Yao}, \citenamefont {Fang},\ and\ \citenamefont
  {Su}(2021)}]{YFS21}%
  \BibitemOpen
  \bibfield  {author} {\bibinfo {author} {\bibfnamefont {Y.-F.}\ \bibnamefont
  {Yao}}, \bibinfo {author} {\bibfnamefont {W.-H.}\ \bibnamefont {Fang}},\ and\
  \bibinfo {author} {\bibfnamefont {N.~Q.}\ \bibnamefont {Su}},\ }\bibfield
  {title} {\enquote {\bibinfo {title} {Handling ensemble {N}-representability
  constraint in explicit-by-implicit manner},}\ }\href
  {https://doi.org/10.1021/acs.jpclett.1c01835} {\bibfield  {journal} {\bibinfo
   {journal} {J. Phys. Chem. Lett.}\ }\textbf {\bibinfo {volume} {12}},\
  \bibinfo {pages} {6788--6793} (\bibinfo {year} {2021})}\BibitemShut {NoStop}%
\bibitem [{\citenamefont {Kooi}(2022)}]{Kooi22}%
  \BibitemOpen
  \bibfield  {author} {\bibinfo {author} {\bibfnamefont {D.~P.}\ \bibnamefont
  {Kooi}},\ }\bibfield  {title} {\enquote {\bibinfo {title} {Efficient bosonic
  and fermionic {S}inkhorn algorithms for non-interacting ensembles in one-body
  reduced density matrix functional theory in the canonical ensemble},}\ }\href
  {https://arxiv.org/abs/2205.15058} {\bibfield  {journal} {\bibinfo  {journal}
  {arXiv:2205.15058}\ } (\bibinfo {year} {2022})}\BibitemShut {NoStop}%
\bibitem [{\citenamefont {Schade}\ \emph {et~al.}(2022)\citenamefont {Schade},
  \citenamefont {Bauer}, \citenamefont {Tamoev}, \citenamefont {Mazur},
  \citenamefont {Plessl},\ and\ \citenamefont {K\"uhne}}]{SBTMPK22}%
  \BibitemOpen
  \bibfield  {author} {\bibinfo {author} {\bibfnamefont {R.}~\bibnamefont
  {Schade}}, \bibinfo {author} {\bibfnamefont {C.}~\bibnamefont {Bauer}},
  \bibinfo {author} {\bibfnamefont {K.}~\bibnamefont {Tamoev}}, \bibinfo
  {author} {\bibfnamefont {L.}~\bibnamefont {Mazur}}, \bibinfo {author}
  {\bibfnamefont {C.}~\bibnamefont {Plessl}},\ and\ \bibinfo {author}
  {\bibfnamefont {T.~D.}\ \bibnamefont {K\"uhne}},\ }\bibfield  {title}
  {\enquote {\bibinfo {title} {Parallel quantum chemistry on noisy
  intermediate-scale quantum computers},}\ }\href
  {https://doi.org/10.1103/PhysRevResearch.4.033160} {\bibfield  {journal}
  {\bibinfo  {journal} {Phys. Rev. Res.}\ }\textbf {\bibinfo {volume} {4}},\
  \bibinfo {pages} {033160} (\bibinfo {year} {2022})}\BibitemShut {NoStop}%
\bibitem [{\citenamefont {Rodríguez-Mayorga}, \citenamefont {Giesbertz},\ and\
  \citenamefont {Visscher}(2022)}]{RMGV22}%
  \BibitemOpen
  \bibfield  {author} {\bibinfo {author} {\bibfnamefont {M.}~\bibnamefont
  {Rodríguez-Mayorga}}, \bibinfo {author} {\bibfnamefont {K.~J.}\ \bibnamefont
  {Giesbertz}},\ and\ \bibinfo {author} {\bibfnamefont {L.}~\bibnamefont
  {Visscher}},\ }\bibfield  {title} {\enquote {\bibinfo {title} {Relativistic
  reduced density matrix functional theory},}\ }\href
  {https://doi.org/10.21468/SciPostChem.1.2.004} {\bibfield  {journal}
  {\bibinfo  {journal} {SciPost Chem.}\ }\textbf {\bibinfo {volume} {1}},\
  \bibinfo {pages} {004} (\bibinfo {year} {2022})}\BibitemShut {NoStop}%
\bibitem [{\citenamefont {Lew-Yee}\ and\ \citenamefont {del
  Campo}(2022)}]{LYDC22}%
  \BibitemOpen
  \bibfield  {author} {\bibinfo {author} {\bibfnamefont {J.~F.~H.}\
  \bibnamefont {Lew-Yee}}\ and\ \bibinfo {author} {\bibfnamefont {J.~M.}\
  \bibnamefont {del Campo}},\ }\bibfield  {title} {\enquote {\bibinfo {title}
  {Charge delocalization error in {P}iris natural orbital functionals},}\
  }\href {https://doi.org/10.1063/5.0102310} {\bibfield  {journal} {\bibinfo
  {journal} {J. Chem. Phys.}\ }\textbf {\bibinfo {volume} {157}},\ \bibinfo
  {pages} {104113} (\bibinfo {year} {2022})}\BibitemShut {NoStop}%
\bibitem [{\citenamefont {Lemke}, \citenamefont {Kussmann},\ and\ \citenamefont
  {Ochsenfeld}(2022)}]{LKO22}%
  \BibitemOpen
  \bibfield  {author} {\bibinfo {author} {\bibfnamefont {Y.}~\bibnamefont
  {Lemke}}, \bibinfo {author} {\bibfnamefont {J.}~\bibnamefont {Kussmann}},\
  and\ \bibinfo {author} {\bibfnamefont {C.}~\bibnamefont {Ochsenfeld}},\
  }\bibfield  {title} {\enquote {\bibinfo {title} {Efficient integral-direct
  methods for self-consistent reduced density matrix functional theory
  calculations on central and graphics processing units},}\ }\href
  {https://doi.org/10.1021/acs.jctc.2c00231} {\bibfield  {journal} {\bibinfo
  {journal} {J. Chem. Theory Comput.}\ }\textbf {\bibinfo {volume} {18}},\
  \bibinfo {pages} {4229--4244} (\bibinfo {year} {2022})}\BibitemShut {NoStop}%
\bibitem [{\citenamefont {Ai}, \citenamefont {Fang},\ and\ \citenamefont
  {Su}(2022)}]{AFS22}%
  \BibitemOpen
  \bibfield  {author} {\bibinfo {author} {\bibfnamefont {W.}~\bibnamefont
  {Ai}}, \bibinfo {author} {\bibfnamefont {W.-H.}\ \bibnamefont {Fang}},\ and\
  \bibinfo {author} {\bibfnamefont {N.~Q.}\ \bibnamefont {Su}},\ }\bibfield
  {title} {\enquote {\bibinfo {title} {Functional-based description of
  electronic dynamic and strong correlation: Old issues and new insights},}\
  }\href {https://doi.org/10.1021/acs.jpclett.2c00084} {\bibfield  {journal}
  {\bibinfo  {journal} {J. Phys. Chem. Lett.}\ }\textbf {\bibinfo {volume}
  {13}},\ \bibinfo {pages} {1744--1751} (\bibinfo {year} {2022})}\BibitemShut
  {NoStop}%
\bibitem [{\citenamefont {Di~Sabatino}\ \emph {et~al.}(2023)\citenamefont
  {Di~Sabatino}, \citenamefont {Koskelo}, \citenamefont {Berger},\ and\
  \citenamefont {Romaniello}}]{DSKBR23}%
  \BibitemOpen
  \bibfield  {author} {\bibinfo {author} {\bibfnamefont {S.}~\bibnamefont
  {Di~Sabatino}}, \bibinfo {author} {\bibfnamefont {J.}~\bibnamefont
  {Koskelo}}, \bibinfo {author} {\bibfnamefont {J.~A.}\ \bibnamefont
  {Berger}},\ and\ \bibinfo {author} {\bibfnamefont {P.}~\bibnamefont
  {Romaniello}},\ }\bibfield  {title} {\enquote {\bibinfo {title} {Screened
  extended {K}oopmans' theorem: Photoemission at weak and strong
  correlation},}\ }\href {https://doi.org/10.1103/PhysRevB.107.035111}
  {\bibfield  {journal} {\bibinfo  {journal} {Phys. Rev. B}\ }\textbf {\bibinfo
  {volume} {107}},\ \bibinfo {pages} {035111} (\bibinfo {year}
  {2023})}\BibitemShut {NoStop}%
\bibitem [{\citenamefont {L\'opez-Sandoval}\ and\ \citenamefont
  {Pastor}(2000)}]{Pastor2000}%
  \BibitemOpen
  \bibfield  {author} {\bibinfo {author} {\bibfnamefont {R.}~\bibnamefont
  {L\'opez-Sandoval}}\ and\ \bibinfo {author} {\bibfnamefont {G.~M.}\
  \bibnamefont {Pastor}},\ }\bibfield  {title} {\enquote {\bibinfo {title}
  {Density-matrix functional theory of the {H}ubbard model: An exact numerical
  study},}\ }\href {https://doi.org/10.1103/PhysRevB.61.1764} {\bibfield
  {journal} {\bibinfo  {journal} {Phys. Rev. B}\ }\textbf {\bibinfo {volume}
  {61}},\ \bibinfo {pages} {1764--1772} (\bibinfo {year} {2000})}\BibitemShut
  {NoStop}%
\bibitem [{\citenamefont {Van~Neck}\ \emph {et~al.}(2001)\citenamefont
  {Van~Neck}, \citenamefont {Waroquier}, \citenamefont {Peirs}, \citenamefont
  {Van~Speybroeck},\ and\ \citenamefont {Dewulf}}]{Neck2001}%
  \BibitemOpen
  \bibfield  {author} {\bibinfo {author} {\bibfnamefont {D.}~\bibnamefont
  {Van~Neck}}, \bibinfo {author} {\bibfnamefont {M.}~\bibnamefont {Waroquier}},
  \bibinfo {author} {\bibfnamefont {K.}~\bibnamefont {Peirs}}, \bibinfo
  {author} {\bibfnamefont {V.}~\bibnamefont {Van~Speybroeck}},\ and\ \bibinfo
  {author} {\bibfnamefont {Y.}~\bibnamefont {Dewulf}},\ }\bibfield  {title}
  {\enquote {\bibinfo {title} {$v$-representability of one-body density
  matrices},}\ }\href {https://doi.org/10.1103/PhysRevA.64.042512} {\bibfield
  {journal} {\bibinfo  {journal} {Phys. Rev. A}\ }\textbf {\bibinfo {volume}
  {64}},\ \bibinfo {pages} {042512} (\bibinfo {year} {2001})}\BibitemShut
  {NoStop}%
\bibitem [{\citenamefont {L\'opez-Sandoval}\ and\ \citenamefont
  {Pastor}(2002)}]{Lopez2002}%
  \BibitemOpen
  \bibfield  {author} {\bibinfo {author} {\bibfnamefont {R.}~\bibnamefont
  {L\'opez-Sandoval}}\ and\ \bibinfo {author} {\bibfnamefont {G.~M.}\
  \bibnamefont {Pastor}},\ }\bibfield  {title} {\enquote {\bibinfo {title}
  {Density-matrix functional theory of strongly correlated lattice fermions},}\
  }\href {https://doi.org/10.1103/PhysRevB.66.155118} {\bibfield  {journal}
  {\bibinfo  {journal} {Phys. Rev. B}\ }\textbf {\bibinfo {volume} {66}},\
  \bibinfo {pages} {155118} (\bibinfo {year} {2002})}\BibitemShut {NoStop}%
\bibitem [{\citenamefont {Requist}\ and\ \citenamefont
  {Pankratov}(2008)}]{Requist2008}%
  \BibitemOpen
  \bibfield  {author} {\bibinfo {author} {\bibfnamefont {R.}~\bibnamefont
  {Requist}}\ and\ \bibinfo {author} {\bibfnamefont {O.}~\bibnamefont
  {Pankratov}},\ }\bibfield  {title} {\enquote {\bibinfo {title} {Generalized
  {K}ohn-{S}ham system in one-matrix functional theory},}\ }\href
  {https://doi.org/10.1103/PhysRevB.77.235121} {\bibfield  {journal} {\bibinfo
  {journal} {Phys. Rev. B}\ }\textbf {\bibinfo {volume} {77}},\ \bibinfo
  {pages} {235121} (\bibinfo {year} {2008})}\BibitemShut {NoStop}%
\bibitem [{\citenamefont {Sauban\`ere}\ and\ \citenamefont
  {Pastor}(2011)}]{Pastor2011a}%
  \BibitemOpen
  \bibfield  {author} {\bibinfo {author} {\bibfnamefont {M.}~\bibnamefont
  {Sauban\`ere}}\ and\ \bibinfo {author} {\bibfnamefont {G.~M.}\ \bibnamefont
  {Pastor}},\ }\bibfield  {title} {\enquote {\bibinfo {title} {Density-matrix
  functional study of the {H}ubbard model on one- and two-dimensional bipartite
  lattices},}\ }\href {https://doi.org/10.1103/PhysRevB.84.035111} {\bibfield
  {journal} {\bibinfo  {journal} {Phys. Rev. B}\ }\textbf {\bibinfo {volume}
  {84}},\ \bibinfo {pages} {035111} (\bibinfo {year} {2011})}\BibitemShut
  {NoStop}%
\bibitem [{\citenamefont {T\"ows}\ and\ \citenamefont
  {Pastor}(2011)}]{Pastor2011b}%
  \BibitemOpen
  \bibfield  {author} {\bibinfo {author} {\bibfnamefont {W.}~\bibnamefont
  {T\"ows}}\ and\ \bibinfo {author} {\bibfnamefont {G.~M.}\ \bibnamefont
  {Pastor}},\ }\bibfield  {title} {\enquote {\bibinfo {title} {Lattice density
  functional theory of the single-impurity {A}nderson model: {D}evelopment and
  applications},}\ }\href {https://doi.org/10.1103/PhysRevB.83.235101}
  {\bibfield  {journal} {\bibinfo  {journal} {Phys. Rev. B}\ }\textbf {\bibinfo
  {volume} {83}},\ \bibinfo {pages} {235101} (\bibinfo {year}
  {2011})}\BibitemShut {NoStop}%
\bibitem [{\citenamefont {Fuks}\ \emph {et~al.}(2013)\citenamefont {Fuks},
  \citenamefont {Farzanehpour}, \citenamefont {Tokatly}, \citenamefont {Appel},
  \citenamefont {Kurth},\ and\ \citenamefont {Rubio}}]{Fuks2013}%
  \BibitemOpen
  \bibfield  {author} {\bibinfo {author} {\bibfnamefont {J.~I.}\ \bibnamefont
  {Fuks}}, \bibinfo {author} {\bibfnamefont {M.}~\bibnamefont {Farzanehpour}},
  \bibinfo {author} {\bibfnamefont {I.~V.}\ \bibnamefont {Tokatly}}, \bibinfo
  {author} {\bibfnamefont {H.}~\bibnamefont {Appel}}, \bibinfo {author}
  {\bibfnamefont {S.}~\bibnamefont {Kurth}},\ and\ \bibinfo {author}
  {\bibfnamefont {A.}~\bibnamefont {Rubio}},\ }\bibfield  {title} {\enquote
  {\bibinfo {title} {Time-dependent exchange-correlation functional for a
  {H}ubbard dimer: Quantifying nonadiabatic effects},}\ }\href
  {https://doi.org/10.1103/PhysRevA.88.062512} {\bibfield  {journal} {\bibinfo
  {journal} {Phys. Rev. A}\ }\textbf {\bibinfo {volume} {88}},\ \bibinfo
  {pages} {062512} (\bibinfo {year} {2013})}\BibitemShut {NoStop}%
\bibitem [{\citenamefont {Fuks}\ and\ \citenamefont {Maitra}(2014)}]{Fuks2014}%
  \BibitemOpen
  \bibfield  {author} {\bibinfo {author} {\bibfnamefont {J.~I.}\ \bibnamefont
  {Fuks}}\ and\ \bibinfo {author} {\bibfnamefont {N.~T.}\ \bibnamefont
  {Maitra}},\ }\bibfield  {title} {\enquote {\bibinfo {title} {Challenging
  adiabatic time-dependent density functional theory with a {H}ubbard dimer:
  the case of time-resolved long-range charge transfer},}\ }\href
  {https://doi.org/10.1039/C4CP00118D} {\bibfield  {journal} {\bibinfo
  {journal} {Phys. Chem. Chem. Phys.}\ }\textbf {\bibinfo {volume} {16}},\
  \bibinfo {pages} {14504--14513} (\bibinfo {year} {2014})}\BibitemShut
  {NoStop}%
\bibitem [{\citenamefont {Carrascal}\ \emph {et~al.}(2015)\citenamefont
  {Carrascal}, \citenamefont {Ferrer}, \citenamefont {C.},\ and\ \citenamefont
  {Burke}}]{Carrascal2015}%
  \BibitemOpen
  \bibfield  {author} {\bibinfo {author} {\bibfnamefont {D.~J.}\ \bibnamefont
  {Carrascal}}, \bibinfo {author} {\bibfnamefont {J.}~\bibnamefont {Ferrer}},
  \bibinfo {author} {\bibfnamefont {S.~J.}\ \bibnamefont {C.}},\ and\ \bibinfo
  {author} {\bibfnamefont {K.}~\bibnamefont {Burke}},\ }\bibfield  {title}
  {\enquote {\bibinfo {title} {The {H}ubbard dimer: a density functional case
  study of a many-body problem},}\ }\href
  {https://doi.org/10.1088/0953-8984/27/39/393001} {\bibfield  {journal}
  {\bibinfo  {journal} {Journal of Physics: Condensed Matter}\ }\textbf
  {\bibinfo {volume} {27}},\ \bibinfo {pages} {393001} (\bibinfo {year}
  {2015})}\BibitemShut {NoStop}%
\bibitem [{\citenamefont {Cohen}\ and\ \citenamefont
  {Mori-S\'anchez}(2016)}]{Cohen2016}%
  \BibitemOpen
  \bibfield  {author} {\bibinfo {author} {\bibfnamefont {A.~J.}\ \bibnamefont
  {Cohen}}\ and\ \bibinfo {author} {\bibfnamefont {P.}~\bibnamefont
  {Mori-S\'anchez}},\ }\bibfield  {title} {\enquote {\bibinfo {title}
  {Landscape of an exact energy functional},}\ }\href
  {https://doi.org/10.1103/PhysRevA.93.042511} {\bibfield  {journal} {\bibinfo
  {journal} {Phys. Rev. A}\ }\textbf {\bibinfo {volume} {93}},\ \bibinfo
  {pages} {042511} (\bibinfo {year} {2016})}\BibitemShut {NoStop}%
\bibitem [{\citenamefont {Deur}, \citenamefont {Mazouin},\ and\ \citenamefont
  {Fromager}(2017)}]{Deur17}%
  \BibitemOpen
  \bibfield  {author} {\bibinfo {author} {\bibfnamefont {K.}~\bibnamefont
  {Deur}}, \bibinfo {author} {\bibfnamefont {L.}~\bibnamefont {Mazouin}},\ and\
  \bibinfo {author} {\bibfnamefont {E.}~\bibnamefont {Fromager}},\ }\bibfield
  {title} {\enquote {\bibinfo {title} {Exact ensemble density functional theory
  for excited states in a model system: Investigating the weight dependence of
  the correlation energy},}\ }\href
  {https://doi.org/10.1103/PhysRevB.95.035120} {\bibfield  {journal} {\bibinfo
  {journal} {Phys. Rev. B}\ }\textbf {\bibinfo {volume} {95}},\ \bibinfo
  {pages} {035120} (\bibinfo {year} {2017})}\BibitemShut {NoStop}%
\bibitem [{\citenamefont {Deur}\ \emph {et~al.}(2018)\citenamefont {Deur},
  \citenamefont {Mazouin}, \citenamefont {Senjean},\ and\ \citenamefont
  {Fromager}}]{Deur18}%
  \BibitemOpen
  \bibfield  {author} {\bibinfo {author} {\bibfnamefont {K.}~\bibnamefont
  {Deur}}, \bibinfo {author} {\bibfnamefont {L.}~\bibnamefont {Mazouin}},
  \bibinfo {author} {\bibfnamefont {B.}~\bibnamefont {Senjean}},\ and\ \bibinfo
  {author} {\bibfnamefont {E.}~\bibnamefont {Fromager}},\ }\bibfield  {title}
  {\enquote {\bibinfo {title} {Exploring weight-dependent density-functional
  approximations for ensembles in the {H}ubbard dimer},}\ }\href
  {https://doi.org/10.1140/epjb/e2018-90124-7} {\bibfield  {journal} {\bibinfo
  {journal} {Eur. Phys. J. B}\ }\textbf {\bibinfo {volume} {91}},\ \bibinfo
  {pages} {162} (\bibinfo {year} {2018})}\BibitemShut {NoStop}%
\bibitem [{\citenamefont {Deur}\ and\ \citenamefont {Fromager}(2019)}]{Deur19}%
  \BibitemOpen
  \bibfield  {author} {\bibinfo {author} {\bibfnamefont {K.}~\bibnamefont
  {Deur}}\ and\ \bibinfo {author} {\bibfnamefont {E.}~\bibnamefont
  {Fromager}},\ }\bibfield  {title} {\enquote {\bibinfo {title} {Ground and
  excited energy levels can be extracted exactly from a single ensemble
  density-functional theory calculation},}\ }\href
  {https://doi.org/10.1063/1.5084312} {\bibfield  {journal} {\bibinfo
  {journal} {J. Chem. Phys.}\ }\textbf {\bibinfo {volume} {150}},\ \bibinfo
  {pages} {094106} (\bibinfo {year} {2019})}\BibitemShut {NoStop}%
\bibitem [{\citenamefont {Di~Sabatino}, \citenamefont {Verdozzi},\ and\
  \citenamefont {Romaniello}(2021)}]{DVR21}%
  \BibitemOpen
  \bibfield  {author} {\bibinfo {author} {\bibfnamefont {S.}~\bibnamefont
  {Di~Sabatino}}, \bibinfo {author} {\bibfnamefont {C.}~\bibnamefont
  {Verdozzi}},\ and\ \bibinfo {author} {\bibfnamefont {P.}~\bibnamefont
  {Romaniello}},\ }\bibfield  {title} {\enquote {\bibinfo {title} {Time
  dependent reduced density matrix functional theory at strong correlation:
  insights from a two-site {A}nderson impurity model},}\ }\href
  {https://doi.org/10.1039/D1CP01742J} {\bibfield  {journal} {\bibinfo
  {journal} {Phys. Chem. Chem. Phys.}\ }\textbf {\bibinfo {volume} {23}},\
  \bibinfo {pages} {16730--16738} (\bibinfo {year} {2021})}\BibitemShut
  {NoStop}%
\bibitem [{\citenamefont {Verzijl}(2022)}]{V-bachelor}%
  \BibitemOpen
  \bibfield  {author} {\bibinfo {author} {\bibfnamefont {H.}~\bibnamefont
  {Verzijl}},\ }\emph {\bibinfo {title} {The effects of spin restrictions on
  the landscape of the exact electron-electron interaction functional in the
  {H}ubbard dimer}},\ \href
  {https://scripties.uba.uva.nl/search?id=record_30119} {\bibinfo {type}
  {Bachelor's thesis}},\ \bibinfo  {school} {Vrije Universiteit Amsterdam}
  (\bibinfo {year} {2022})\BibitemShut {NoStop}%
\bibitem [{\citenamefont {Krein}\ and\ \citenamefont
  {Milman}(1940)}]{Krein1940}%
  \BibitemOpen
  \bibfield  {author} {\bibinfo {author} {\bibfnamefont {M.}~\bibnamefont
  {Krein}}\ and\ \bibinfo {author} {\bibfnamefont {D.}~\bibnamefont {Milman}},\
  }\bibfield  {title} {\enquote {\bibinfo {title} {On extreme points of regular
  convex sets},}\ }\href {http://eudml.org/doc/219061} {\bibfield  {journal}
  {\bibinfo  {journal} {Studia Math.}\ }\textbf {\bibinfo {volume} {9}},\
  \bibinfo {pages} {133--1Dre38} (\bibinfo {year} {1940})}\BibitemShut
  {NoStop}%
\bibitem [{\citenamefont {Lieb}(1983)}]{L83}%
  \BibitemOpen
  \bibfield  {author} {\bibinfo {author} {\bibfnamefont {E.~H.}\ \bibnamefont
  {Lieb}},\ }\bibfield  {title} {\enquote {\bibinfo {title} {Density
  functionals for coulomb systems},}\ }\href
  {https://doi.org/10.1002/qua.560240302} {\bibfield  {journal} {\bibinfo
  {journal} {Int. J. Quantum Chem.}\ }\textbf {\bibinfo {volume} {24}},\
  \bibinfo {pages} {243} (\bibinfo {year} {1983})}\BibitemShut {NoStop}%
\bibitem [{Note1()}]{Note1}%
  \BibitemOpen
  \bibinfo {note} {Although most applications of 1RDMFT in quantum chemistry so
  far restrict to time-reversal symmetric Hamiltonians, there are quite a few
  relevant systems which break that symmetry. The latter include the systems
  with external magnetic fields and velocity dependent forces in general\cite
  {TCPH22}, and chiral cavities\cite {Huebner21}.}\BibitemShut {Stop}%
\bibitem [{\citenamefont {Haake}, \citenamefont {Gnutzmann},\ and\
  \citenamefont {Ku\'s}(2019)}]{HGK19}%
  \BibitemOpen
  \bibfield  {author} {\bibinfo {author} {\bibfnamefont {F.}~\bibnamefont
  {Haake}}, \bibinfo {author} {\bibfnamefont {S.}~\bibnamefont {Gnutzmann}},\
  and\ \bibinfo {author} {\bibfnamefont {M.}~\bibnamefont {Ku\'s}},\ }\href
  {https://doi.org/10.1007/978-3-319-97580-1} {\emph {\bibinfo {title} {Quantum
  Signatures of Chaos}}}\ (\bibinfo  {publisher} {Springer Cham},\ \bibinfo
  {year} {2019})\BibitemShut {NoStop}%
\bibitem [{\citenamefont {Smith}(1966)}]{Smith66}%
  \BibitemOpen
  \bibfield  {author} {\bibinfo {author} {\bibfnamefont {D.~W.}\ \bibnamefont
  {Smith}},\ }\bibfield  {title} {\enquote {\bibinfo {title}
  {$n$-representability problem for fermion density matrices. ii. the
  first-order density matrix with $n$ even},}\ }\href
  {https://doi.org/10.1103/PhysRev.147.896} {\bibfield  {journal} {\bibinfo
  {journal} {Phys. Rev.}\ }\textbf {\bibinfo {volume} {147}},\ \bibinfo {pages}
  {896--898} (\bibinfo {year} {1966})}\BibitemShut {NoStop}%
\bibitem [{\citenamefont {Chakraborty}\ and\ \citenamefont
  {Mazziotti}(2014)}]{CM14}%
  \BibitemOpen
  \bibfield  {author} {\bibinfo {author} {\bibfnamefont {R.}~\bibnamefont
  {Chakraborty}}\ and\ \bibinfo {author} {\bibfnamefont {D.~A.}\ \bibnamefont
  {Mazziotti}},\ }\bibfield  {title} {\enquote {\bibinfo {title} {Generalized
  pauli conditions on the spectra of one-electron reduced density matrices of
  atoms and molecules},}\ }\href {https://doi.org/10.1103/PhysRevA.89.042505}
  {\bibfield  {journal} {\bibinfo  {journal} {Phys. Rev. A}\ }\textbf {\bibinfo
  {volume} {89}},\ \bibinfo {pages} {042505} (\bibinfo {year}
  {2014})}\BibitemShut {NoStop}%
\bibitem [{\citenamefont {Schilling}(2015)}]{CS15}%
  \BibitemOpen
  \bibfield  {author} {\bibinfo {author} {\bibfnamefont {C.}~\bibnamefont
  {Schilling}},\ }\bibfield  {title} {\enquote {\bibinfo {title} {Quasipinning
  and its relevance for $n$-fermion quantum states},}\ }\href
  {https://doi.org/10.1103/PhysRevA.91.022105} {\bibfield  {journal} {\bibinfo
  {journal} {Phys. Rev. A}\ }\textbf {\bibinfo {volume} {91}},\ \bibinfo
  {pages} {022105} (\bibinfo {year} {2015})}\BibitemShut {NoStop}%
\bibitem [{\citenamefont {Kohn}(1983)}]{Kohn83}%
  \BibitemOpen
  \bibfield  {author} {\bibinfo {author} {\bibfnamefont {W.}~\bibnamefont
  {Kohn}},\ }\bibfield  {title} {\enquote {\bibinfo {title}
  {$v$-representability and density functional theory},}\ }\href
  {https://doi.org/10.1103/PhysRevLett.51.1596} {\bibfield  {journal} {\bibinfo
   {journal} {Phys. Rev. Lett.}\ }\textbf {\bibinfo {volume} {51}},\ \bibinfo
  {pages} {1596} (\bibinfo {year} {1983})}\BibitemShut {NoStop}%
\bibitem [{\citenamefont {Helgaker}\ and\ \citenamefont
  {Teale}(2022)}]{Helgaker22}%
  \BibitemOpen
  \bibfield  {author} {\bibinfo {author} {\bibfnamefont {T.}~\bibnamefont
  {Helgaker}}\ and\ \bibinfo {author} {\bibfnamefont {A.~M.}\ \bibnamefont
  {Teale}},\ }\bibfield  {title} {\enquote {\bibinfo {title} {Lieb variation
  principle in density-functional theory},}\ }\href
  {https://arxiv.org/abs/2204.12216} {\bibfield  {journal} {\bibinfo  {journal}
  {arXiv:2204.12216}\ } (\bibinfo {year} {2022})}\BibitemShut {NoStop}%
\bibitem [{\citenamefont {Penz}\ and\ \citenamefont {van
  Leeuwen}(2021)}]{Penz22}%
  \BibitemOpen
  \bibfield  {author} {\bibinfo {author} {\bibfnamefont {M.}~\bibnamefont
  {Penz}}\ and\ \bibinfo {author} {\bibfnamefont {R.}~\bibnamefont {van
  Leeuwen}},\ }\bibfield  {title} {\enquote {\bibinfo {title}
  {Density-functional theory on graphs},}\ }\href
  {https://doi.org/10.1063/5.0074249} {\bibfield  {journal} {\bibinfo
  {journal} {J. Chem. Phys.}\ }\textbf {\bibinfo {volume} {155}},\ \bibinfo
  {pages} {244111} (\bibinfo {year} {2021})}\BibitemShut {NoStop}%
\bibitem [{\citenamefont {Benavides-Riveros}\ \emph {et~al.}(2020)\citenamefont
  {Benavides-Riveros}, \citenamefont {Wolff}, \citenamefont {Marques},\ and\
  \citenamefont {Schilling}}]{Benavides20}%
  \BibitemOpen
  \bibfield  {author} {\bibinfo {author} {\bibfnamefont {C.~L.}\ \bibnamefont
  {Benavides-Riveros}}, \bibinfo {author} {\bibfnamefont {J.}~\bibnamefont
  {Wolff}}, \bibinfo {author} {\bibfnamefont {M.~A.~L.}\ \bibnamefont
  {Marques}},\ and\ \bibinfo {author} {\bibfnamefont {C.}~\bibnamefont
  {Schilling}},\ }\bibfield  {title} {\enquote {\bibinfo {title} {Reduced
  density matrix functional theory for bosons},}\ }\href
  {https://doi.org/10.1103/PhysRevLett.124.180603} {\bibfield  {journal}
  {\bibinfo  {journal} {Phys. Rev. Lett.}\ }\textbf {\bibinfo {volume} {124}},\
  \bibinfo {pages} {180603} (\bibinfo {year} {2020})}\BibitemShut {NoStop}%
\bibitem [{\citenamefont {Liebert}\ and\ \citenamefont
  {Schilling}(2021)}]{LS21}%
  \BibitemOpen
  \bibfield  {author} {\bibinfo {author} {\bibfnamefont {J.}~\bibnamefont
  {Liebert}}\ and\ \bibinfo {author} {\bibfnamefont {C.}~\bibnamefont
  {Schilling}},\ }\bibfield  {title} {\enquote {\bibinfo {title} {Functional
  theory for {B}ose-{E}instein condensates},}\ }\href
  {https://doi.org/10.1103/PhysRevResearch.3.013282} {\bibfield  {journal}
  {\bibinfo  {journal} {Phys. Rev. Research}\ }\textbf {\bibinfo {volume}
  {3}},\ \bibinfo {pages} {013282} (\bibinfo {year} {2021})}\BibitemShut
  {NoStop}%
\bibitem [{\citenamefont {Schilling}, \citenamefont {Benavides-Riveros},\ and\
  \citenamefont {Vrana}(2017)}]{PhysRevA.96.052312}%
  \BibitemOpen
  \bibfield  {author} {\bibinfo {author} {\bibfnamefont {C.}~\bibnamefont
  {Schilling}}, \bibinfo {author} {\bibfnamefont {C.~L.}\ \bibnamefont
  {Benavides-Riveros}},\ and\ \bibinfo {author} {\bibfnamefont
  {P.}~\bibnamefont {Vrana}},\ }\bibfield  {title} {\enquote {\bibinfo {title}
  {Reconstructing quantum states from single-party information},}\ }\href
  {https://doi.org/10.1103/PhysRevA.96.052312} {\bibfield  {journal} {\bibinfo
  {journal} {Phys. Rev. A}\ }\textbf {\bibinfo {volume} {96}},\ \bibinfo
  {pages} {052312} (\bibinfo {year} {2017})}\BibitemShut {NoStop}%
\bibitem [{\citenamefont {Schilling}\ \emph {et~al.}(2020)\citenamefont
  {Schilling}, \citenamefont {Benavides-Riveros}, \citenamefont {Lopes},
  \citenamefont {Macia{\.{z}}ek},\ and\ \citenamefont
  {Sawicki}}]{Schilling_2020}%
  \BibitemOpen
  \bibfield  {author} {\bibinfo {author} {\bibfnamefont {C.}~\bibnamefont
  {Schilling}}, \bibinfo {author} {\bibfnamefont {C.~L.}\ \bibnamefont
  {Benavides-Riveros}}, \bibinfo {author} {\bibfnamefont {A.}~\bibnamefont
  {Lopes}}, \bibinfo {author} {\bibfnamefont {T.}~\bibnamefont
  {Macia{\.{z}}ek}},\ and\ \bibinfo {author} {\bibfnamefont {A.}~\bibnamefont
  {Sawicki}},\ }\bibfield  {title} {\enquote {\bibinfo {title} {Implications of
  pinned occupation numbers for natural orbital expansions: I. generalizing the
  concept of active spaces},}\ }\href
  {https://doi.org/10.1088/1367-2630/ab64b0} {\bibfield  {journal} {\bibinfo
  {journal} {New J. Phys.}\ }\textbf {\bibinfo {volume} {22}},\ \bibinfo
  {pages} {023001} (\bibinfo {year} {2020})}\BibitemShut {NoStop}%
\bibitem [{\citenamefont {Macia{\.{z}}ek}\ \emph {et~al.}(2020)\citenamefont
  {Macia{\.{z}}ek}, \citenamefont {Sawicki}, \citenamefont {Gross},
  \citenamefont {Lopes},\ and\ \citenamefont {Schilling}}]{MSGLS20-2}%
  \BibitemOpen
  \bibfield  {author} {\bibinfo {author} {\bibfnamefont {T.}~\bibnamefont
  {Macia{\.{z}}ek}}, \bibinfo {author} {\bibfnamefont {A.}~\bibnamefont
  {Sawicki}}, \bibinfo {author} {\bibfnamefont {D.}~\bibnamefont {Gross}},
  \bibinfo {author} {\bibfnamefont {A.}~\bibnamefont {Lopes}},\ and\ \bibinfo
  {author} {\bibfnamefont {C.}~\bibnamefont {Schilling}},\ }\bibfield  {title}
  {\enquote {\bibinfo {title} {Implications of pinned occupation numbers for
  natural orbital expansions. ii: rigorous derivation and extension to
  non-fermionic systems},}\ }\href {https://doi.org/10.1088/1367-2630/ab64b1}
  {\bibfield  {journal} {\bibinfo  {journal} {New J. Phys.}\ }\textbf {\bibinfo
  {volume} {22}},\ \bibinfo {pages} {023002} (\bibinfo {year}
  {2020})}\BibitemShut {NoStop}%
\bibitem [{Note2()}]{Note2}%
  \BibitemOpen
  \bibinfo {note} {Although we will distinguish in the following carefully
  between different variants of 1RDMFT according to Sec.~\ref
  {sec:real-vs-complex} we will safely continue using the non-specific symbols
  $\protect \mathcal {P}^1_N, \protect \mathcal {E}^1_N$. It will namely be
  clear from the context to which variant they refer to.}\BibitemShut {Stop}%
\bibitem [{Note3()}]{Note3}%
  \BibitemOpen
  \bibinfo {note} {References in the literature usually discuss repulsive
  interactions with $U\geq 0$ \cite {Pastor2011a, Cohen2016}. Nevertheless, the
  derivation of $\protect \mathcal {F}_{\protect \mathbbm {R}}^{(p)}$ can be
  extended to attractive interactions in a straightforward manner yielding the
  result in Eq.~\protect \eqref {eq:FRR_1}}\BibitemShut {NoStop}%
\bibitem [{\citenamefont {Rockafellar}(1997)}]{R97}%
  \BibitemOpen
  \bibfield  {author} {\bibinfo {author} {\bibfnamefont {R.~T.}\ \bibnamefont
  {Rockafellar}},\ }\href {https://press.princeton.edu/titles/1815.html} {\emph
  {\bibinfo {title} {Convex {A}nalysis}}}\ (\bibinfo  {publisher} {Princeton
  university press},\ \bibinfo {year} {1997})\BibitemShut {NoStop}%
\bibitem [{\citenamefont {Kienesberger}(ming)}]{Kienesberger-MA}%
  \BibitemOpen
  \bibfield  {author} {\bibinfo {author} {\bibfnamefont {L.}~\bibnamefont
  {Kienesberger}},\ }\emph {\bibinfo {title} {The curse of universality in
  functional theory}},\ \href@noop {} {Master's thesis},\ \bibinfo  {school}
  {Ludwig-Maximilans-Universit\"at M\"unchen} (\bibinfo {year}
  {upcoming})\BibitemShut {NoStop}%
\bibitem [{\citenamefont {Ullrich}\ and\ \citenamefont {Kohn}(2002)}]{UK02}%
  \BibitemOpen
  \bibfield  {author} {\bibinfo {author} {\bibfnamefont {C.~A.}\ \bibnamefont
  {Ullrich}}\ and\ \bibinfo {author} {\bibfnamefont {W.}~\bibnamefont {Kohn}},\
  }\bibfield  {title} {\enquote {\bibinfo {title} {Degeneracy in density
  functional theory: Topology in the $v$ and $n$ spaces},}\ }\href
  {https://doi.org/10.1103/PhysRevLett.89.156401} {\bibfield  {journal}
  {\bibinfo  {journal} {Phys. Rev. Lett.}\ }\textbf {\bibinfo {volume} {89}},\
  \bibinfo {pages} {156401} (\bibinfo {year} {2002})}\BibitemShut {NoStop}%
\bibitem [{\citenamefont {Penz}\ and\ \citenamefont {van
  Leeuwen}(2023)}]{PL23}%
  \BibitemOpen
  \bibfield  {author} {\bibinfo {author} {\bibfnamefont {M.}~\bibnamefont
  {Penz}}\ and\ \bibinfo {author} {\bibfnamefont {R.}~\bibnamefont {van
  Leeuwen}},\ }\bibfield  {title} {\enquote {\bibinfo {title} {Geometry of
  degeneracy in potential and density space},}\ }\href
  {https://doi.org/10.22331/q-2023-02-09-918} {\bibfield  {journal} {\bibinfo
  {journal} {{Quantum}}\ }\textbf {\bibinfo {volume} {7}},\ \bibinfo {pages}
  {918} (\bibinfo {year} {2023})}\BibitemShut {NoStop}%
\bibitem [{\citenamefont {Tellgren}\ \emph {et~al.}(2022)\citenamefont
  {Tellgren}, \citenamefont {Culpitt}, \citenamefont {Peters},\ and\
  \citenamefont {Helgaker}}]{TCPH22}%
  \BibitemOpen
  \bibfield  {author} {\bibinfo {author} {\bibfnamefont {E.}~\bibnamefont
  {Tellgren}}, \bibinfo {author} {\bibfnamefont {T.}~\bibnamefont {Culpitt}},
  \bibinfo {author} {\bibfnamefont {L.}~\bibnamefont {Peters}},\ and\ \bibinfo
  {author} {\bibfnamefont {T.}~\bibnamefont {Helgaker}},\ }\bibfield  {title}
  {\enquote {\bibinfo {title} {Molecular vibrations in the presence of
  velocity-dependent forces},}\ }\href {https://arxiv.org/abs/2212.10246}
  {\bibfield  {journal} {\bibinfo  {journal} {arXiv:2212.10246}\ } (\bibinfo
  {year} {2022})}\BibitemShut {NoStop}%
\bibitem [{\citenamefont {H\"ubener}\ \emph {et~al.}(2021)\citenamefont
  {H\"ubener}, \citenamefont {De~Giovannini}, \citenamefont {Sch\"afer},
  \citenamefont {Andberger}, \citenamefont {Ruggenthaler}, \citenamefont
  {Faist},\ and\ \citenamefont {Rubio}}]{Huebner21}%
  \BibitemOpen
  \bibfield  {author} {\bibinfo {author} {\bibfnamefont {H.}~\bibnamefont
  {H\"ubener}}, \bibinfo {author} {\bibfnamefont {U.}~\bibnamefont
  {De~Giovannini}}, \bibinfo {author} {\bibfnamefont {C.}~\bibnamefont
  {Sch\"afer}}, \bibinfo {author} {\bibfnamefont {J.}~\bibnamefont
  {Andberger}}, \bibinfo {author} {\bibfnamefont {M.}~\bibnamefont
  {Ruggenthaler}}, \bibinfo {author} {\bibfnamefont {J.}~\bibnamefont
  {Faist}},\ and\ \bibinfo {author} {\bibfnamefont {A.}~\bibnamefont {Rubio}},\
  }\bibfield  {title} {\enquote {\bibinfo {title} {Engineering quantum
  materials with chiral optical cavities},}\ }\href
  {https://doi.org/10.1038/s41563-020-00801-7} {\bibfield  {journal} {\bibinfo
  {journal} {Nat. Mater.}\ }\textbf {\bibinfo {volume} {20}},\ \bibinfo {pages}
  {438–442} (\bibinfo {year} {2021})}\BibitemShut {NoStop}%
\end{thebibliography}%
\end{document}